\documentclass[12pt, draftclsnofoot, onecolumn]{IEEEtran}
\usepackage{amsmath, amssymb, amsthm, bm}
\usepackage{graphicx}
\usepackage{color, colortbl} 
\usepackage[dvipsnames,svgnames,x11names]{xcolor}
\makeatother
\usepackage{tabularx, boldline, multirow}
\usepackage[boxed,ruled,vlined]{algorithm2e}
\usepackage{enumerate}
\usepackage[font=footnotesize]{caption,subcaption}
\usepackage{hyperref, cite}
\usepackage{url,comment}
\usepackage[normalem]{ulem}
\usepackage[shortlabels]{enumitem}
\usepackage{pgfplots}
\usepackage{tikz}
\usepackage{mathtools}
\usepackage{accents}
\usepackage{balance}
\usepackage{acronym}
\usepackage{arydshln}
\usepackage{scalerel}

\usetikzlibrary{calc}
\makeatletter
\newcommand{\gettikzxy}[3]{%
  \tikz@scan@one@point\pgfutil@firstofone#1\relax
  \edef#2{\the\pgf@x}%
  \edef#3{\the\pgf@y}%
}

\newcommand*\numcircledmod[1]{\raisebox{.5pt}{\textcircled{\raisebox{-.9pt} {#1}}}}

\newtheorem{rem}{Remark}
\acrodef{crb}[CRB]{Cram\'er-Rao bound}
\acrodef{awgn}[AWGN]{additive white Gaussian noise}
\acrodef{bs}[BS]{base station}
\acrodef{bp}[BP]{belief propagation}
\acrodef{1d}[1D]{one-dimensional}
\acrodef{3d}[3D]{three-dimensional}
\acrodef{5g}[5G]{fifth generation}
\acrodef{6g}[6G]{sixth generation}
\acrodef{ue}[UE]{user equipment}
\acrodef{iue}[IUE]{image user equipment}
\acrodef{los}[LOS]{line-of-sight}
\acrodef{aoa}[AoA]{angle-of-arrival}
\acrodefplural{aoa}[AoAs]{angle-of-arrival}
\acrodef{aod}[AoD]{angle-of-departure}
\acrodefplural{aod}[AoDs]{angle-of-departures}
\acrodef{toa}[ToA]{time-of-arrival}
\acrodefplural{toa}[ToAs]{time-of-arrival}
\acrodef{tdoa}[TDoA]{time-difference-of-arrival}
\acrodef{ris}[RIS]{reconfigurable intelligent surface}
\acrodefplural{ris}[RISs]{reconfigurable intelligent surfaces}
\acrodef{isac}[ISAC]{integrated sensing and communications}
\acrodef{nris}[NRIS]{non-RIS}
\acrodef{cp}[CP]{cyclic prefix}
\acrodef{tx}[Tx]{transmitter}
\acrodef{rx}[Rx]{receiver}
\acrodefplural{rx}[Rxs]{receivers}
\acrodef{psd}[PSD]{power spectral density}
\acrodef{crlb}[CRLB]{Cram\'er-Rao lower bounds}
\acrodefplural{crlb}[CRLBs]{Cram\'er-Rao lower bounds}
\acrodef{rss}[RSS]{received signal strength}
\acrodef{los}[LOS]{line-of-sight}
\acrodef{nlos}[NLOS]{non line-of-sight}
\acrodef{fov}[FoV]{field-of-view}
\acrodef{dft}[DFT]{discrete Fourier transform}
\acrodef{fft}[FFT]{fast Fourier transform}
\acrodef{fim}[FIM]{Fisher information matrix}
\acrodefplural{fim}[FIMs]{Fisher information matrix}
\acrodef{efim}[EFIM]{equivalent Fisher information matrix}
\acrodef{upa}[UPA]{uniform planar array}
\acrodefplural{upa}[UPAs]{uniform planar arrays}
\acrodef{peb}[PEB]{position error bound}
\acrodef{slam}[SLAM]{simultaneous localization and mapping}
\acrodef{snr}[SNR]{signal-to-noise ratio}
\acrodefplural{snr}[SNRs]{signal-to-noise ratio}
\acrodef{sre}[SRE]{smart radio environment}
\acrodefplural{sre}[SRE]{smart radio environments}
\acrodef{mpmb}[MPMB]{marginal Poisson multi-Bernoulli}
\acrodef{mb}[MB]{multi-Bernoulli}
\acrodef{pmb}[PMB]{Poisson multi-Bernoulli}
\acrodef{pmbm}[PMBM]{Poisson multi-Bernoulli mixture}
\acrodef{mimo}[MIMO]{multiple-input  multiple-output}
\acrodef{rfs}[RFS]{random finite set}
\acrodef{rfid}[RFID]{radio-frequency identification}
\acrodef{siso}[SISO]{single-input single-output}
\acrodef{miso}[MISO]{multiple-input single-output}
\acrodef{ici}[ICI]{inter-carrier interference}
\acrodef{iid}[iid]{independent and identically distributed}
\acrodef{ml}[ML]{maximum likelihood}
\acrodef{pdf}[PDF]{probability density function}
\acrodef{cdf}[CDF]{cumulative distribution function}
\acrodef{ofdm}[OFDM]{orthogonal frequency-division multiplexing}
\acrodef{qos}[QoS]{Quality of Service}
\acrodef{sp}[SP]{scattering point}
\acrodefplural{sp}[SPs]{scattering points}
\acrodef{ula}[ULA]{uniform linear array}
\acrodef{va}[VA]{virtual anchor}
\acrodefplural{va}[VAs]{virtual anchors}
\acrodef{ls}[LS]{large surface}
\acrodefplural{ls}[LSs]{large surfaces}
\acrodef{rp}[RP]{reflection point}
\acrodefplural{ls}[RPs]{reflection point}
\acrodef{eb}[EB]{error bound}
\acrodefplural{eb}[EBs]{error bounds}
\acrodef{peb}[PEB]{position error bound}
\acrodefplural{peb}[PEBs]{position error bound}
\acrodef{heb}[HEB]{heading error bound}
\acrodefplural{heb}[HEBs]{heading error bound}
\acrodef{seb}[SEB]{speed error bound}
\acrodefplural{seb}[SEBs]{speed error bound}
\acrodef{mae}[MAE]{mean absolute error}
\acrodefplural{mae}[MAEs]{mean absolute error}
\acrodef{ab}[AB]{arbitrary beam}
\acrodef{cb}[CB]{conventional beam}
\acrodef{gospa}[GOSPA]{generalized optimal subpattern assignment}
\acrodef{mae}[MAE]{mean absolute error}
\acrodef{ppp}[PPP]{Poisson point process}
\acrodef{mmwave}[mmWave]{millimeter-wave}

\begin{document}
\bstctlcite{IEEEexample:BSTcontrol}
\title{RIS-Enabled and Access-Point-Free \\Simultaneous Radio Localization and Mapping}
\author{
\IEEEauthorblockN{
Hyowon Kim, \IEEEmembership{Member, IEEE}, Hui Chen, \IEEEmembership{Member, IEEE}, \\Musa~Furkan~Keskin,
\IEEEmembership{Member, IEEE},
Yu~Ge, \IEEEmembership{Student Member, IEEE}, 
Kamran~Keykhosravi,  \IEEEmembership{Member,~IEEE}, 
George C. Alexandropoulos, \IEEEmembership{Senior Member, IEEE}, 
Sunwoo~Kim,~\IEEEmembership{Senior~Member,~IEEE}, 
and \\Henk Wymeersch, \IEEEmembership{Senior Member, IEEE}
}

\thanks{H. Kim, H. Chen, M. F. Keskin, Y. Ge, and H. Wymeersch are with the Department of Electrical Engineering, Chalmers University of Technology, 412 58 Gothenburg, Sweden (emails: \{hyowon, hui.chen, furkan, yuge, henkw\}@chalmers.se).}
\thanks{K. Keykhosravi is with Ericsson Research, Ericsson AB, Gothenburg, Sweden (e-mail:kamran.keykhosravi@ericsson.com).}
\thanks{G. C. Alexandropoulos is with the Department of Informatics and Telecommunications, National and Kapodistrian University of Athens, 15784 Athens, Greece (e-mail: alexandg@di.uoa.gr).}
\thanks{S. Kim is with the Department of Electronic Engineering, Hanyang University, 04763 Seoul, South Korea (email: remero@hanyang.ac.kr).}
\thanks{This work was supported in part by the EU H2020 RISE-6G project under grant 101017011, the Chalmers Area of Advance Transport 6G-Cities project, and Basic Science Research Program through the National Research Foundation of Korea (2022R1A6A3A03068510).}
}

\maketitle
\vspace{-1cm}
\begin{abstract}
    In the upcoming sixth generation~(6G) of wireless communication systems, 
    reconfigurable intelligent surfaces~(RISs) are regarded as one of the promising technological enablers, which can provide programmable signal propagation.
    Therefore, simultaneous radio localization and mapping~(SLAM) with RISs appears as an emerging research direction within the 6G ecosystem.
    In this paper, we propose a novel framework of RIS-enabled radio SLAM for wireless operation without the intervention of access points (APs).
    We first design the RIS phase profiles leveraging prior information for the user equipment~(UE), such that they uniformly illuminate the angular sector where the UE is probabilistically located.
    Second, we modify the marginal Poisson multi-Bernoulli SLAM filter and estimate the UE state and landmarks, which enables efficient mapping of the radio propagation environment.
    Third, we derive the theoretical Cram\'er-Rao lower bounds on the estimators for the channel parameters and the UE state.
    We finally evaluate the performance of the proposed method under scenarios with a limited number of transmissions, taking into account the channel coherence time.
    Our results demonstrate that the RIS enables solving the radio SLAM problem with zero APs, 
    and that the consideration of the Doppler shift contributes to improving the UE speed estimates.
\end{abstract}
\vspace{-2mm}
\IEEEpeerreviewmaketitle

\vspace{-1mm}
\section{Introduction}\label{sec:Introduction}
\vspace{-1mm}
    \Ac{slam} has been a highly active research topic in wireless communications~\cite{Hyowon_TWC2020,Yu_EK-PMBM_JSAC2021,Hyowon_MPMB_TVT2022,HenkGlobecom2018,HyowonAsilomar2018,Erik_BPSLAM_TWC2019,Rico_BPSLAM_JSTSP2019,yassin2018mosaic,palacios2018communication} as well as in robotics~\cite{Durrant2006SLAM1,Durrant2006SLAM2}.
    According to the late advances in wireless systems design, spatial information has been attracting attention in location-aware communications~\cite{Taranto_LocAC_SPM2014,Koivisto_LocAC_CM2017} for \ac{5g} systems, as well as in \ac{isac}~\cite{An_ISAC_Survey2022,Fan_ISAC_JSAC2022,Sundeep_RISISAC_2022}, a concept intended for beyond \ac{5g} and \ac{6g}.
    In \ac{5g}, large bandwidths and multiple antennas are considered at \ac{mmwave} carrier frequencies, hence, high-resolvable \ac{toa}, \ac{aod}, and \ac{aoa} for each resolvable signal propagation path are feasible~\cite{HenkGlobecom2018,TR2022}.
    This has been lately motivating various studies on radio \ac{slam}~\cite{Hyowon_TWC2020,Yu_EK-PMBM_JSAC2021,Hyowon_MPMB_TVT2022,HenkGlobecom2018,HyowonAsilomar2018,Erik_BPSLAM_TWC2019,Rico_BPSLAM_JSTSP2019,yassin2018mosaic,palacios2018communication}.
    
    In line with the \ac{5g} developments, radio \ac{slam} methods have been devised~\cite{Hyowon_TWC2020,Yu_EK-PMBM_JSAC2021,Hyowon_MPMB_TVT2022,HenkGlobecom2018,HyowonAsilomar2018,Erik_BPSLAM_TWC2019,Rico_BPSLAM_JSTSP2019,yassin2018mosaic,palacios2018communication}, enabling estimating the \ac{ue} state while simultaneously mapping the landmarks in the propagation environment.
    In~\cite{Hyowon_TWC2020,Yu_EK-PMBM_JSAC2021,Hyowon_MPMB_TVT2022},  \acp{rfs} were adopted for modeling the unknown position and number of landmarks, as well as the uncertain data association between landmarks and measurements with missed detections and false alarms.
    Depending on the case-by-case scenario, we approximate the \ac{rfs} density for modeling a set of landmarks and can utilize the desired \ac{rfs}.
    By avoiding the assignment problem for data association, the trade-off between the performance and computational complexity was balanced in~\cite{Hyowon_TWC2020}.
    An efficient implementation using the extended Kalman filter was developed in~\cite{Yu_EK-PMBM_JSAC2021}, by computing the marginalized \ac{slam} density together with data association.
    From the \ac{rfs}-\ac{slam} density, which can handle the \ac{slam} problem in a theoretically optimal manner, a series of \ac{slam} densities was derived by marginalizing out either multiple hypotheses or nuisance variables~\cite{Hyowon_MPMB_TVT2022}.
    This implementation was computationally efficient and its connection to the \ac{bp}-based \ac{slam} filter was showcased~\cite{Erik_BPSLAM_TWC2019}.
    In \ac{bp}-\ac{slam}~\cite{HenkGlobecom2018,HyowonAsilomar2018,Erik_BPSLAM_TWC2019,Rico_BPSLAM_JSTSP2019}, the landmark position was modeled by a vector, and the marginalized densities, corresponding to individual \ac{ue} state and landmarks, were computed via \ac{bp}.
    With the known number of landmarks, the \ac{slam} problem was formulated as a factor graph~\cite{HenkGlobecom2018,HyowonAsilomar2018} and multiple types of landmarks were considered in~\cite{HyowonAsilomar2018}. For the case of unknown number of landmarks, the factor graph including data association was proposed in~\cite{Erik_BPSLAM_TWC2019}, and the authors in~\cite{Rico_BPSLAM_JSTSP2019} additionally considered the clock bias in the \ac{ue} state estimation problem.
    Non-Bayesian methods using geometry relations were recently studied in~\cite{yassin2018mosaic,palacios2018communication}. Considering known number of landmarks and data association, the \ac{slam} problem was formulated and solved in~\cite{yassin2018mosaic}. The authors in~\cite{palacios2018communication} considered a single \ac{bs} with an unknown position to the SLAM system.
    
    Complementary to SLAM, \ac{isac} and \acp{ris} have recently emerged as very promising tehcnologies for the future \ac{6g} wireless communications, providing the necessary measurement data for high-precision SLAM.
    ISAC is recently attracting increasing attention according to which, sensing and communication functionalities can be delivered simultaneously in a given propagation environment~\cite{An_ISAC_Survey2022,Fan_ISAC_JSAC2022,Sundeep_RISISAC_2022}, while \acp{ris} \cite{huang2019reconfigurable,Tsinghua_RIS_Tutorial} can offer programmable signal propagation \cite{Emil_RIS_SPM2022}, enabling various wireless applications \cite{RISE6G_COMMAG}. 
    For example, \acp{ris} have been proposed as a new paradigm in radio localization~\cite{Henk_RISSLAM_VTM2020,Zhang_RISSL_Proc2022}, rendering location estimation feasible even in cases where it is not with conventional nodes \cite{Keykhosravi2022infeasible}.
    Leveraging its radio propagation control capability, an \ac{ris} can provide a reference point with reliable geometric measurements, generating the desired signal path and power.
    In a \ac{siso} system, channel parameter estimation and the \ac{crlb} on the \ac{ue} state estimation were studied in~\cite{Kamran_RISMobility_JSTSP2022}.
    The \acp{crlb} on the position and heading\footnote{The heading is the direction of the agent facing in the angular domain, also known as orientation.} were derived in a \ac{mimo} system in~\cite{Ahmed_RISLoc_TSP2021}, and localization with an array-of-subarray-based THz systems was discussed in~\cite{chen2022tutorial}.
    The authors in~\cite{Davide_RISLoc_TWC2021} considered a blocked \ac{los} path and studied \ac{ris}-aided localization, and furthermore, user localization and \ac{aoa} estimation were simultaneously addressed for \ac{mimo} systems~\cite{Boyu_RISLoc_JSTSP2022}.
    Without a \ac{bs}, self-localization  was proposed in an \ac{ris}-equipped \ac{siso} wireless system in~\cite{Kamran_RISloc_ICC2022}.
    In sensing and radar applications aided by reflective \acp{ris}, the receiver side was used to sense the signal reflected by the \ac{ris} elements, which enabled coverage extension~\cite{Aubry_RIS-Sensing_2022} and \ac{snr} enhancement~\cite{Buzzi_RIS-Sensing_ICASSP2022}. 
    The unknown \ac{ris} position and orientation were addressed by bistatic sensing in \cite{Reza_SensingRIS_2022}.
    Considering the \ac{ris}-aided sensing and localization together, several \ac{slam} cases were introduced in \cite{Zhang_RISSL_Proc2022}, 
    and a \ac{slam} algorithm was developed for the case where multiple \acp{ris} are considered as the unknown controllable passive objects \cite{Yang_MetaSLAM_TWC2022}.
    In joint radar and communication systems, \ac{ris}-aided sensing helped to improve the performance of radar detection and alleviate mutual interference~\cite{He_RIS-Sensing_JSAC20222}.
    Very recently, various applications and benefits of \acp{ris} for sensing functionalities and for \ac{isac} systems were discussed \cite{Sundeep_RISISAC_2022}.
    However, in spite of the latter \ac{ris}-based advantages, signals reaching a passive object via and \ac{ris} suffer from severe path loss.

    The RIS capability to adjust the phase shifts of its individual elements provides additional degrees of freedom that can boost the localization performance in \ac{ris}-aided wireless systems~\cite{Henk_RISSLAM_VTM2020}.
%
In~\cite{loc_awareness_JSAC_2022,RIS_beam_training_TWC_2021}, heuristic designs of directional \ac{ris} phase profiles were presented with the objective of \ac{snr} maximization in the presence of prior knowledge about the \ac{ue} location.
    Localization-optimal designs via the \ac{crlb} minimization for a single-antenna receiver were studied in~\cite{RIS_Loc_JSTSP_2022}, leading to considerable accuracy improvements over standard directional and \ac{dft}-based phase profiles, at a cost of increased transmission delays. 
    To generate the arbitrary beampatterns, \ac{ris} configuration optimization was introduced in~\cite{Mustafa_ArbitraryBeam_6G2022}.
    For multi-objects sensing using the round-trip signals via the RIS, the \ac{ris} configurations and beamforming vectors were jointly designed in~\cite{Kaitao_RISSensing_TC2022}.
    The authors in \cite{Vahid_BeamChannelEst_CL2022} considered the \ac{ue} position for \ac{ris}-based illumination, with the goal to reduce the \ac{ris} reconfiguration overhead, offering a large illumination area.
    To reduce the beam training overhead while sustaining the beam's quality, a hierarchical \ac{ris} phase profile design was devised in
    \cite{George_NFBeamManange_2022}.
    In \cite{Kamran_RISMobility_JSTSP2022}, the Doppler effect, induced by mobility in various relevant applications, was considered~\cite{Gosan_Highspeed_WC2020}.
    While the impacts of this effect and its exploitation have been extensively studied in the context of traditional localization \cite{Han_Doppler_TIT2015, kakkavas2019performance,Hui_Doppler_Globecom2022}, it has received only limited treatment in RIS-aided localization. 
    In the absence of 
    prior information about the \ac{ue} state,
    random phase configurations (either random element phases or random directions) can be employed 
    \cite{Davide_RISLoc_TWC2021,Kamran_RISMobility_JSTSP2022}, providing an \ac{snr} gain over time. Moreover, 
    temporal RIS phase coding offers an extra degree of freedom that allows for the separation of controlled (i.e., through the \ac{ris}) and uncontrolled paths (i.e., due to other reflectors and scatterers in the signal propagation environment)~\cite{Kamran_RISMobility_JSTSP2022,Davide_RISLoc_TWC2021}, thereby facilitating subsequent processing. 
    In \ac{ris}-enabled \ac{slam}, sensing is carried out by the \ac{ue}, while the \ac{ris} contributes in the estimation of the \ac{ue} state, and only indirectly improves the mapping/sensing performance~\cite{Henk_RISSLAM_VTM2020}.

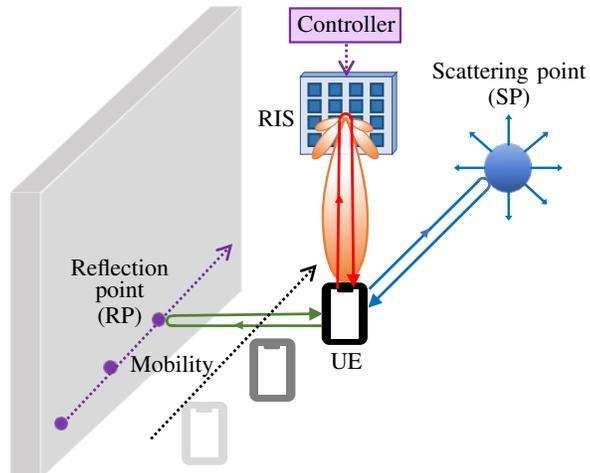
\begin{figure}
    \centering
    \begin{tikzpicture}
    \node (image) [anchor=south west]{\includegraphics[width=0.45\columnwidth]{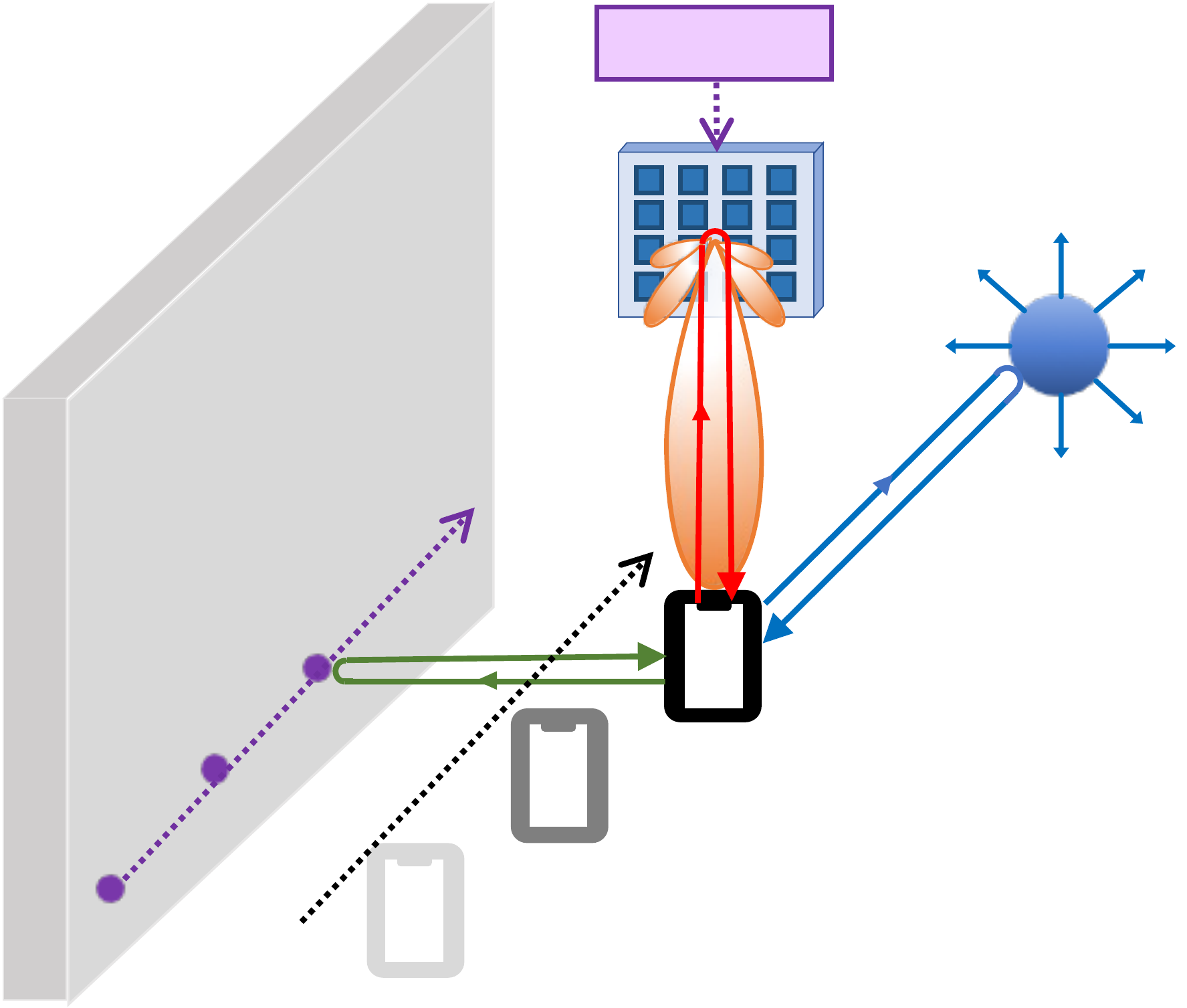}};
    \gettikzxy{(image.north east)}{\ix}{\iy};
    \node at (.6*\ix,.26*\iy){\footnotesize UE};
    \node at (.88*\ix,.84*\iy){\footnotesize Scattering point};
    \node at (.88*\ix,.79*\iy){\footnotesize (SP)};
    \node at (.3*\ix,.25*\iy){\footnotesize Mobility};
    \node at (.6*\ix,.94*\iy){\footnotesize Controller};
    \node at (.48*\ix,.75*\iy){\footnotesize RIS};
    \node at (.21*\ix,.45*\iy){\footnotesize Reflection};
    \node at (.21*\ix,.4*\iy){\footnotesize point};
    \node at (.21*\ix,.35*\iy){\footnotesize (RP)};
    \end{tikzpicture}
    \vspace{-.3cm}
    \caption{An example illustration of the considered \ac{ris}-enabled radio \ac{slam} setup that does not require the presence of a \ac{bs} or access point. The full-duplex \ac{ue} uses the \ac{ris} to localize itself and to place detected landmarks in the global frame of reference.}
    \vspace{-.6cm}
    \label{Fig:scenario}
\end{figure}

In this paper, we extend the \ac{ris}-enabled localization concept of~\cite{Kamran_RISloc_ICC2022}, intended for cases of absence of \acp{bs}, to \ac{slam}, considering that the full-duplex \ac{ue}~\cite{George_MIMO_VTM2022} transmits \ac{ofdm} signals and receives their back-scattered versions from the wireless propagation environment, as shown in Fig.~\ref{Fig:scenario}. To the best of our knowledge, this is the first study of \ac{ris}-enabled \ac{slam} with zero \acp{bs}.
In contrast to our recent work~\cite{Kamran_RISloc_ICC2022}, we map the landmarks in the environment and use this mapping to improve \ac{ue} localization. 
The contributions of this paper are summarized as follows.
\begin{itemize}
    \item 
    \textbf{RIS-aided SLAM modeling:} We propose a novel SLAM problem, where a full-duplex UE localizes itself with aid of an RIS and simultaneously maps objects in the environment, all based on single-bounce paths. In contrast to \cite{Yang_MetaSLAM_TWC2022}, a channel model for \ac{ofdm}-\ac{mimo} systems is considered which accounts for the \ac{ris} via the \ac{aoa}/\ac{aod} to/from it, the Doppler shift induced by mobility, and the random phase offset due to signal reflections.
    \item \textbf{Efficient RIS phase profile design:}
    To optimize SLAM performance, we present adaptive \ac{ris} phase profiles that take into account prior information on the \ac{ue} location and enable uniform illumination of the corresponding angular uncertainty region, while being flexible with the number of transmission slots and targeted coverage probability.
    In contrast to \cite{George_NFBeamManange_2022}, the angular uncertainty region is computed using \ac{ue} prior information over consecutive time steps. Specifically, the region is divided into multiple grids depending on the number of transmissions, and each grid is covered by the proposed beam per transmission slot.
    \item \textbf{RFS-based SLAM filter extension:} We present an extension of the \ac{mpmb}-\ac{slam} filter from the recent work \cite{Hyowon_MPMB_TVT2022}, which is capable to separate the reflected signal induced by the \ac{ris} from multipath signals, thus, facilitating the determination of landmarks in a global coordinate system. 
    \item \textbf{Performance evaluation:} We verify via numerical simulations that 
    the proposed \ac{ris}-enabled \ac{slam} approach, together with the designed adaptive \ac{ris} phase profiles, provides satisfactory performance even when small numbers of transmissions are used (due to, for example, channel coherence time limitations), outperforming the conventional random and directional phase profiles. We also incorporate the Doppler shift into our channel model to address \ac{ue} mobility scenarios, and it is shown that this  significantly improves the \ac{slam} performance through providing accurate UE speed estimation.
\end{itemize}

The remainder of this paper is organized as follows. Section~\ref{sec:Models} describes the various models for our \ac{ris}-enabled \ac{slam}. The proposed \ac{ris} phase profile design is presented in Section~\ref{sec:RISPhaseProf}, while, in Section~\ref{sec:FIM}, we derive the \acp{fim} of the channel parameters and \ac{ue} state. The proposed \ac{mpmb}-\ac{slam} filter for sequential Bayesian estimation is presented in Section~\ref{sec:MPMB}. Our framework's extensive numerical investigations are discussed in Section~\ref{sec:Results}, while Section~\ref{sec:Conclusions} concludes the paper. 

\subsubsection*{Notation}
The scalars, vectors, matrices, and sets are respectively represented by italic, bold lowercase, bold uppercase, and calligraphic fonts (e.g., $x$, $\mathbf{x}$, $\mathbf{X}$, and $\mathcal{X}$). The transpose and Hermitian of vectors and matrices are respectively expressed by the superscripts $\top$ and $\mathrm{H}$ (e.g., $\mathbf{x}^\top$, $\mathbf{X}^\top$,  $\mathbf{X}^\mathrm{H}$). In addition, $\odot$ denotes the Hadamard product, $\mathrm{Re}(\cdot)$ is the real part of a complex value, $\lVert \cdot \rVert$ returns the Euclidean distance, and $\lvert \cdot \rvert$ provides the amplitude of a complex number. We use the following indices: antenna element $n$; discrete time instant $k$; transmission interval $t$; subcarrier number $s$; landmark $i$; landmark type $m$; measurement $j$; and signal path $l$.

\vspace{-1mm}
\section{Considered Models and Design}\label{sec:Models}
\vspace{-1mm}
This section introduces the focused wireless signal propagation environment and presents the considered models for the received signal and channel, as well as our measurements' models.

\vspace{-1mm}
\subsection{Signal and Channel Models}
\vspace{-1mm}
    As shown in Fig.~\ref{Fig:scenario}, we consider wireless signal propagation environments, where there is no \ac{bs}.\footnote{It is noted that a \ac{bs} may still be present in the scenario to coordinate the transmissions, but it is not directly involved in the SLAM signals or processing thereof.} 
    The full-duplex \ac{ue} moves and transmits mmWave signals, and multipath components are generated depending on the considered propagation environment as follows:
    \textit{i}) reflected by the \ac{ris} elements, resulting in phase shifted versions  of the impinging signals at the RIS (\ac{ue}-\ac{ris}-\ac{ue} path); \textit{ii}) reflected by the \acp{rp} on the \ac{ls} (such as a wall), indicating specular reflection~\cite{HenkGlobecom2018,HyowonAsilomar2018,Erik_BPSLAM_TWC2019,Rico_BPSLAM_JSTSP2019}
    (\ac{ue}-\ac{rp}-\ac{ue} path); and \textit{iii}) scattered by the \acp{sp} (\ac{ue}-\ac{sp}-\ac{ue} path).
    During the same time-frequency resources, the \ac{ue} receives the back-propagated signals.
    Note that we consider only the one-bounce signal propagation~\cite{Erik_BPSLAM_TWC2019,Hyowon_TWC2020,Hyowon_MPMB_TVT2022}, consisting of $L+1$ paths ($L$ landmarks and 1 \ac{ris}); signals resulting from multiple bounces are considered as highly attenuated.
    The \ac{ue} and \ac{ris} are respectively equipped with $N_\mathrm{UE} =N_\mathrm{UE}^\mathrm{az} \times N_\mathrm{UE}^\mathrm{el}$ and $N_\mathrm{RIS} = N_\mathrm{RIS}^\mathrm{az} \times N_\mathrm{RIS}^\mathrm{el}$ radiating elements (in the azimuth and elevation) forming \acp{upa}.
    The \ac{ue} transmits at  discrete time intervals $k$, with period  $\Delta$, $T$ \ac{ofdm} pilot symbols spanning $N_\mathrm{SC}$ subcarriers.
    We model the received signal on the $s$-th subcarrier for each $t$-th transmission as follows:
\begin{align}\label{eq:signal_model}
    & y_{t,k,s} \triangleq \sqrt{E_s} \mathbf{w}_{t,k}^\mathrm{H} \sum_{l =0}^L
    \gamma_{l,k}\beta_{l,k} e^{j2\pi (f_{l,k}^{\mathrm{D}} T_\mathrm{o} t-\tau_{l,k} (s-1) \Delta_f)}
    \mathbf{a}_\mathrm{UE}(\bm{\theta}_{l,k})\mathbf{a}_\mathrm{UE}^\top (\bm{\theta}_{l,k})\mathbf{f}_{t,k} + n_{t,k,s}, 
\end{align}
    where $\gamma_{0,k} = \mathbf{a}_\mathrm{RIS}^\top(\bm{\phi}_{k})\bm{\Omega}_{\mathrm{RIS},t,k}\mathbf{a}_\mathrm{RIS}(\bm{\phi}_{k})$ and $\gamma_{l,k} = 1$ for $l>0$,
    in which we denote by $l$ the path index and define the \ac{ris} path for $l=0$ (\ac{ue}-\ac{ris}-\ac{ue}) and the \ac{nris} paths with $l \neq 0$ (\ac{ue}-\ac{rp}-\ac{ue} or \ac{ue}-\ac{sp}-\ac{ue}).
    Moreover,  $E_s$ denotes the energy per each pilot symbol, 
    $\mathbf{a}_\mathrm{RIS}(\bm{\phi}_{k})$ and $\mathbf{a}_\mathrm{UE}(\bm{\theta}_{k})$ are the response vectors for both \ac{aod} and \ac{aoa} at the \ac{ris} and \ac{ue} (defined in both azimuth and elevation, e.g., 
    $\bm{\phi}_{k} \triangleq [\phi_{k}^\mathrm{az}, \phi_{k}^\mathrm{el}]^\top$ for path $l=0$ and $\bm{\theta}_{l,k} \triangleq [\theta_{l,k}^\mathrm{az}, \theta_{l,k}^\mathrm{el}]^\top$ for any path $l$), $f_{l,k}^{\mathrm{D}} $ is the Doppler, $\tau_{l,k}$ is the \ac{toa}, and $\beta_{l,k}$ is the complex path gain, all of the $l$-th path. 
    In addition, $\mathbf{w}_{t,k}\in \mathbb{C}^{N_\mathrm{BS}\times 1}$ denotes the analog combining vector at the \ac{ue} receiver, $\mathbf{f}_{t,k}\in \mathbb{C}^{N_\mathrm{BS}\times 1}$ is the precoding vector at the \ac{ue} transmitter, and $n_{t,k,s}\sim \mathcal{CN}(0,\sigma_\mathrm{N}^2)$ is the \ac{awgn} with noise variance $\sigma_\mathrm{N}^2$.
    Finally,  $T_\mathrm{o} = T_\mathrm{S} + T_\mathrm{CP}$ denotes the signal duration,
    $\Delta_f$ is the subcarrier spacing,  $T_\mathrm{S} = 1/\Delta_f$ is the symbol duration, and $T_\mathrm{CP} = N_\mathrm{CP}/B$ is the \ac{cp} duration with $N_\mathrm{CP} = N_\mathrm{SC}\epsilon_\mathrm{CP}$ representing the number of \ac{cp} subcarriers, $B$ is the signal bandwidth with $N_\mathrm{SC}$ being the number of subcarriers, and $\epsilon_\mathrm{CP}$ is the \ac{cp} overhead. Finally, $\lambda = c/f_c$ is the wavelength where $c$ is the speed of light and $f_c$ is the carrier frequency.
    
    The array response vectors $\mathbf{a}_\mathrm{RIS}(\cdot)$ and $\mathbf{a}_\mathrm{UE}(\cdot)$ 
    are defined generically as follows:
\begin{align}
    \mathbf{a}_\mathrm{\star}(\bm{\vartheta}) = \exp(j\breve{\mathbf{X}}_\star^\top \mathbf{g}(\bm{\vartheta})),
\end{align}
    where $\star$ denotes the device type where antennas are mounted~(e.g., $\star$ is \ac{ue} or \ac{ris}), notation $\bm{\vartheta} \triangleq [\vartheta^\mathrm{az},\vartheta^\mathrm{el}]^\top$ represents the angular parameter vector corresponding to the array vector, $\breve{\mathbf{X}}_\star$ is the antenna element locations of device $\star$ in its local coordinate system (defined next in Section~\ref{sec:GeoModel}), and $\mathbf{g}(\cdot)$ is the wavenumber vector given by $\mathbf{g}(\bm{\vartheta}) \triangleq \frac{2 \pi}{ \lambda}[\cos (\vartheta^\mathrm{az}) \sin (\vartheta^\mathrm{el}),
    \sin (\vartheta^\mathrm{az}) \sin (\vartheta^\mathrm{el}),
    \cos (\vartheta^\mathrm{el})]^\top$. 
    For the RIS, $\bm{\Omega}_{\mathrm{RIS},t,k} \triangleq \operatorname{diag}(\bm{\omega}_{t,k})$, where $\bm{\omega}_{t,k} \in \mathbb{C}^{N_\mathrm{RIS}\times 1}$ is the \ac{ris} phase profile with elements constrained to lie on the unit circle. 
    We further define the following vector:
\begin{align} 
    \mathbf{b}(\bm{\phi}_{k}) 
    &\triangleq \mathbf{a}_\mathrm{RIS}(\bm{\phi}_{k}) \odot \mathbf{a}_\mathrm{RIS}(\bm{\phi}_{k})  
    = 
    \exp(2j 
    \breve{\mathbf{X}}_{\mathrm{RIS}}^\top \mathbf{g}( \bm{\phi}_{k}) )\label{eq:RIS_NAV},
\end{align}
    where $\odot$ denotes the Hadamard product, and hence, the \ac{ris} part in \eqref{eq:signal_model} can be simplified as:
\begin{align}
    \mathbf{a}_\mathrm{RIS}^\top(\bm{\phi}_{k})\bm{\Omega}_{\mathrm{RIS},t,k}\mathbf{a}_\mathrm{RIS}(\bm{\phi}_{k})=\bm{\omega}_{t,k}^\top\mathbf{b}(\bm{\phi}_{k}).    
\end{align}
    Note that the spacing between the nearest radiating elements at the \ac{ue} and \ac{ris} are respectively $\lambda/2$ and $\lambda/4$ for addressing the grating lobes, due to the combined response $\mathbf{b}(\bm{\phi}_{k})$~\cite{Kamran_RISloc_ICC2022}. 
    

\vspace{-1mm}
\subsection{Geometrical Representations}\label{sec:GeoModel}
\vspace{-1mm}

    Apart from the \ac{ue}, we  regard the \ac{ris}, \acp{sp}, and \acp{rp} as landmarks in this paper.
    The \ac{ue} state, \ac{sp} positions, and \ac{rp} positions are assumed unknown, whereas we consider that the \ac{ris} position and orientation are known. The numbers of \acp{sp} and \acp{rp} are a priori unknown. 
    We represent the \ac{ue} state with the polar velocity and known constant turn~\cite{Li_Dynamic_TAES2003}, denoted by $\mathbf{s}_k\triangleq[\mathbf{x}_{\mathrm{UE},k}^\top,\alpha_{\mathrm{UE},k}, v_{\mathrm{UE},k}]^\top$, where $\mathbf{x}_{\mathrm{UE},k}\triangleq[x_{\mathrm{UE},k},y_{\mathrm{UE},k},z_{\mathrm{UE},k}]^\top$, $\alpha_{\mathrm{UE},k}$, and $v_{\mathrm{UE},k}$
    are respectively the UE's \ac{3d} location, heading, and longitudinal speed.
    We represent the landmark types as $m\in \mathcal{M}\triangleq\{\mathrm{RIS},\mathrm{RP},\mathrm{SP}\}$ and
    their respective locations by $\mathbf{x}_\mathrm{RIS} = [x_\mathrm{RIS},y_\mathrm{RIS},z_\mathrm{RIS}]^\top$,
    $\mathbf{x}_{\mathrm{RP}} = [x_{\mathrm{RP}},y_{\mathrm{RP}},z_{\mathrm{RP}}]^\top$,
    and
    $\mathbf{x}_\mathrm{SP} = [x_\mathrm{SP},y_\mathrm{SP},z_\mathrm{SP}]^\top$. 

    %
    In the global coordinate system, the geometric center of antennas at the UE is located at $\mathbf{x}_{\mathrm{UE},k}$ and the location of the $n$-th antenna element is denoted by $\mathbf{x}_{\mathrm{UE},k}^n \triangleq [x_{\mathrm{UE},k}^n,y_{\mathrm{UE},k}^n,z_{\mathrm{UE},k}^n]^\top$.
    In the local coordinate system, the UE antenna lies in the $xz$ plane and the antenna element locations are denoted by $\breve{\mathbf{X}}_{\mathrm{UE},k} \triangleq [\breve{\mathbf{x}}_{\mathrm{UE},k}^1,\dots \breve{\mathbf{x}}_{\mathrm{UE},k}^{N_\mathrm{UE}}]$, where $\breve{\mathbf{x}}_\mathrm{UE}^n \triangleq [\breve{x}_{\mathrm{UE},k}^n,\breve{y}_{\mathrm{UE},k}^n,\breve{z}_{\mathrm{UE},k}^n]^\top$ is the location of the $n$-th antenna element.
    The relation between the \ac{3d} global and local locations of the \ac{ue} antennas are expressed as 
    $\breve{\mathbf{x}}_{\mathrm{UE},k}^n = \mathbf{O}_{\mathrm{UE},k}^\top (\mathbf{x}_{\mathrm{UE},k}^n - \mathbf{x}_{\mathrm{UE},k}), $
    where $\mathbf{O}_{\mathrm{UE},k}$ is the rotation matrix which rotates the local vector to the global one, and is defined as follows: $\mathbf{O}_{\mathrm{UE},k}=[\mathbf{o}_{\mathrm{UE},k,1}, \mathbf{o}_{\mathrm{UE},k,2}, \mathbf{o}_{\mathrm{UE},k,3}]$ with $\mathbf{o}_{\mathrm{UE},k,1}=[\cos \alpha_{\mathrm{UE},k}, \sin \alpha_{\mathrm{UE},k}, 0]^\top$, $\mathbf{o}_{\mathrm{UE},k,2}=[-\sin \alpha_{\mathrm{UE},k}, \cos \alpha_{\mathrm{UE},k}, 0]^\top$, and $\mathbf{o}_{\mathrm{UE},k,3}=[0,0,1]^\top$,
    indicating the counter-clockwise rotation of the local \ac{upa} configuration around the $z$-axis.
    Similarly, we denote by $\mathbf{x}_\mathrm{RIS}\triangleq[x_\mathrm{RIS},y_\mathrm{RIS},z_\mathrm{RIS}]^\top$ the center of \ac{ris} in the global coordinate system, where $\breve{\mathbf{x}}_\mathrm{RIS}^n \triangleq [\breve{x}_\mathrm{RIS}^n,\breve{y}_\mathrm{RIS}^n,\breve{z}_\mathrm{RIS}^n]^\top$ is the location of its $n$-th reflective unit element, and the matrix $\breve{\mathbf{X}}_\mathrm{RIS}\triangleq[\breve{\mathbf{x}}_\mathrm{RIS}^1,\dots,\breve{\mathbf{x}}_\mathrm{RIS}^n]$ collects the locations of all RIS elements. Finally,  $\mathbf{O}_{\mathrm{RIS}}$ represents the rotation matrix at the \ac{ris}, and we define $\breve{\mathbf{x}}_{\mathrm{RIS}}^n$ as
    $\breve{\mathbf{x}}_{\mathrm{RIS}}^n \triangleq \mathbf{O}_{\mathrm{RIS}}^\top (\mathbf{x}_{\mathrm{RIS}}^n - \mathbf{x}_{\mathrm{RIS}})$.
      
      Based on the latter definitions, the channel parameters (i.e., channel gains, Doppler shifts, \acp{toa}, \acp{aoa}, and \ac{aod}) are obtained by the geometry relations corresponding to the above three types of paths (\ac{ris}, \ac{rp}, and \ac{sp}), as detailed in Appendix~\ref{App:ChannelParameters}.

\vspace{-2mm}
\subsection{Measurements' Modeling}
\vspace{-1mm}

    The signal for each path in the received signal model~\eqref{eq:signal_model} comes from the landmark $(\mathbf{x},m)$.    
    The number of landmarks as well as each landmark's location and type are unknown, and thus, we model the landmarks by an \ac{rfs} $\mathcal{X} \triangleq\{(\mathbf{x}^1,m^1),\dots,(\mathbf{x}^{I},m^{I}) \}$, where $I$ denotes the random number of landmarks, having the set \ac{pdf} $f(\mathcal{X})$~\cite{mahler_book_2007}.
    
    Based on the received signals $y_{t,k,s}$'s for different $t$ and $s$, the \ac{ue} separates the \ac{ris} path~($l=0$) from the uncontrolled multipath~($l\neq 0$) (see the following Section~\ref{sec:RISPhaseProf}).   
    To this end, we consider that a channel parameter estimation technique\footnote{The channel parameters are assumed to be obtained by relevant channel estimation methods, such as the MD-ESPRIT~\cite{Florian_MDESPRIT_TSP2014,Fan_MDESPRIT_2021}) or maximum likelihood \cite{Kamran_RISMobility_JSTSP2022} that perform close to the \ac{crb}.} is adopted operating with $J_k+1$ measurements (one measurement for $l=0$ and a set of $J_k$ ones for $l\neq 0$). 
    Note that $J_k$ is not necessarily equal to $L$, due to missed detections, false alarm, or clutter. The clutter is modeled as random channel parameters within the \ac{fov} of the \ac{ue} in its signal propagation environment, where the number of clutter measurements follows the Poisson distribution with intensity $c(\mathbf{z})$. Missed detections occur with adaptive detection probability $\mathsf{p}_{\text{D},k}(\mathbf{s}_k,\mathbf{x},m) \in [0,1]$~\cite{wymeersch2020adaptive}, when the landmark is within the \ac{fov}. In particular, the measurement for $l=0$ is modeled as follows:
\begin{align}\label{eq:UEMea_model}
    \mathbf{z}_{k}^0 = [\bm{\phi}_{k}^\top,\tau_{0,k}, v_{0,k}, \bm{\theta}_{0,k}^\top]^\top
    + \mathbf{r}_{k}^0,
\end{align}
    with $\mathbf{r}_{k}^0\sim \mathcal{N}(\bm{0},\mathbf{R}_k^0)$ being the measurement noise, whereas the measurement for $l \neq 0$ is indexed by $j$ and modeled as:
\begin{align}\label{eq:ChParaMea_model}
    \mathbf{z}_{k}^j =\mathsf{h}(\mathbf{s}_k, \mathbf{x}, m) + \mathbf{r}_{k}^j,
\end{align}
    where $\mathsf{h}(\mathbf{s}_k, \mathbf{x}, m) \triangleq[\tau_{l,k},v_{l,k}, \bm{\theta}_{l,k}^\top]^\top$ with $\mathbf{r}_{k}^j\sim \mathcal{N}(\bm{0},\mathbf{R}_k^j)$ being the measurement noise. Here, $\mathbf{R}_k^0 \in \mathbb{R}^{6 \times 6}$ and $\mathbf{R}_k^j \in \mathbb{R}^{4 \times 4}$ are the measurement noise covariance matrices, 
    which are different for each path and depend on the precoder $\mathbf{f}_{t,k}$, combiner $\mathbf{w}_{t,k}$, and the \ac{ris} phase profile $\bm{\omega}_{t,k}$.
    Putting all above together, we represent the measurements as an \ac{rfs} $\mathcal{Z}_k\triangleq\{\mathbf{z}_{k}^0,...,\mathbf{z}_{k}^{J_k}\}$. It is important to stress here that the associations between the $L$ landmarks and the $J_k$ \ac{nris} signal paths is considered unknown in this paper.


    

\begin{figure}
    \centering
    \begin{tikzpicture}
    \node (image) [anchor=south west]{\includegraphics[width=0.55\columnwidth]{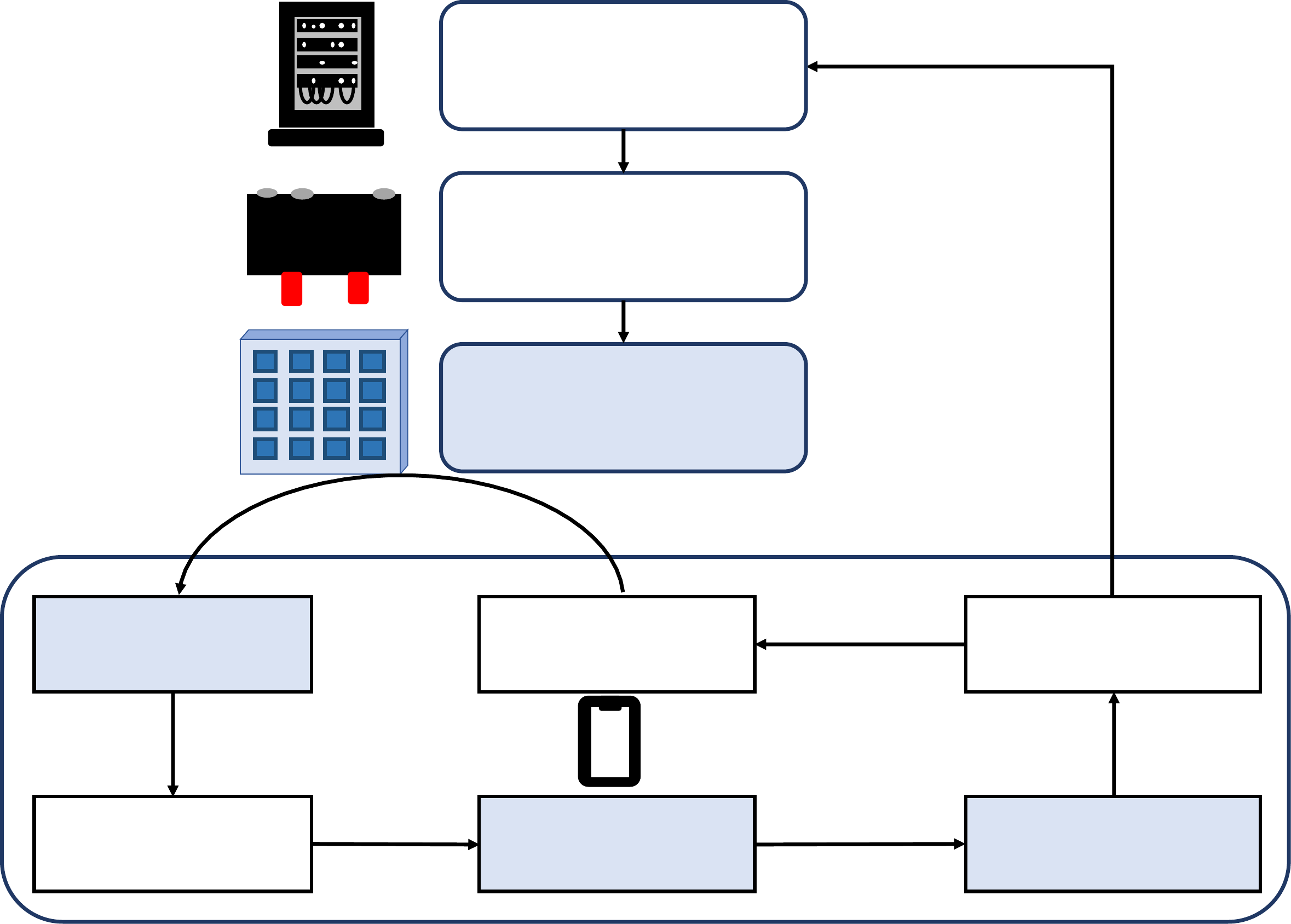}};
    \gettikzxy{(image.north east)}{\ix}{\iy};
    \node at (.41*\ix,.21*\iy){\footnotesize{UE}};
    \node at (.13*\ix,.91*\iy){\footnotesize{Server}};
    \node at (.1*\ix,.72*\iy){\footnotesize{Controller}};
    \node at (.15*\ix,.56*\iy){\footnotesize{RIS}};
    \node at (.485*\ix,.93*\iy){\footnotesize{Collect RIS profiles}};
    \node at (.485*\ix,.88*\iy){\footnotesize{Operate RISs}};
    \node at (.485*\ix,.76*\iy){\footnotesize{Control the RIS}};
    \node at (.485*\ix,.71*\iy){\footnotesize{phase profile}};
    \node at (.82*\ix,.67*\iy){\footnotesize \numcircledmod{2}};
    \node at (.485*\ix,.58*\iy){\footnotesize{Phase shift}};
    \node at (.485*\ix,.53*\iy){\footnotesize{(impinging signals)}};
    \node at
    (.73*\ix,.86*\iy){\footnotesize{$\{ \bm{\omega}_{t,k+1}\}_{t=1}^T$}};
    \node at
    (.73*\ix,.80*\iy){\footnotesize{Required RIS}};
    \node at
    (.73*\ix,.76*\iy){\footnotesize{phase profile}};
    \node at
    (.11*\ix,.38*\iy){\footnotesize{$\mathbf{y}_{t,k}$}};
    \node at 
    (.18*\ix,.38*\iy){\footnotesize \numcircledmod{3}};
    \node at
    (.15*\ix,.33*\iy){\footnotesize{Orthogonal}};
    \node at
    (.15*\ix,.29*\iy){\footnotesize{RIS phase}};
    \node at 
    (.18*\ix,.20*\iy){\footnotesize \numcircledmod{4}};
    \node at
    (.15*\ix,.12*\iy){\footnotesize{Channel}};
    \node at
    (.15*\ix,.085*\iy){\footnotesize{estimation}};
    \node at
    (.48*\ix,.33*\iy){\footnotesize{Pilot}};
    \node at
    (.48*\ix,.29*\iy){\footnotesize{transmission}};
    \node at
    (.48*\ix,.1025*\iy){\footnotesize{SLAM filter}};
    \node at
    (.85*\ix,.33*\iy){\footnotesize{Transmit control}};
    \node at
    (.85*\ix,.29*\iy){\footnotesize{information}};
    \node at
    (.85*\ix,.11*\iy){\footnotesize{RIS phase}};
    \node at
    (.85*\ix,.075*\iy){\footnotesize{profile design}};
    \node at (.76*\ix,.22*\iy){{$\scriptstyle\{\bm{\omega}_{t,k+1}\}_{t=1}^T$}};
    \node at (.665*\ix,.17*\iy){\footnotesize{UE prior}};
    \node at (.665*\ix,.13*\iy){{$\scriptstyle f_\mathrm{p}(\mathbf{s}_{k+1})$}};
    \node at (.66*\ix,.06*\iy){\footnotesize \numcircledmod{1}};
    \node at (.32*\ix,.17*\iy){\footnotesize{Measurements}};
    \node at (.32*\ix,.13*\iy){{$\scriptstyle \mathcal{Z}_{k}$}};
    \node at (.32*\ix,.06*\iy){\footnotesize \numcircledmod{5}};
    \end{tikzpicture}
    \vspace{-0.3cm}
    \caption{The block diagram of the proposed \ac{ris}-enabled \ac{slam} approach functioning without the intervention of a \acp{bs}. \hfill~}
    \vspace{-0.6cm}
    \label{Fig:Diagram}
\end{figure}

\vspace{-1mm}
\subsection{Objective and Process of RIS-enabled SLAM with Zero BSs}
\vspace{-1mm}
    Our objective in this paper is to perform \ac{slam} over time, i.e., localize the \ac{ue} while simultaneously performing mapping of its signal propagation environment. As a secondary objective, we 
    aim to design \ac{ris} phase profiles based on the \ac{ue} prior/predicted position, in order to cover its angular uncertainty at each time instant $k$. 
    The block diagram  depicted in Fig.~\ref{Fig:Diagram} introduces the key features of the proposed \ac{ris}-enabled \ac{slam} approach, as follows:
    \numcircledmod{1} Given the prior \ac{ue} density at instant $k$, as obtained from the prediction step realized by the \ac{slam} filter, we design the \ac{ris} phase profile for efficiently transmitting a limited number of \ac{ofdm} symbols; \numcircledmod{2} The designed phase profile is sent to the  \ac{ris} controller that collects all of them, and the \ac{ris} elements are configured; \numcircledmod{3} The full-duplex \ac{ue} receives its back-scattered transmitted signals; the \ac{ris} path can be separated from the \ac{nris} ones thanks to the designed time-domain \ac{ris} phase profiles; \numcircledmod{4} Then, channel estimation is initiated and the measurements for the channel parameters are collected; and \numcircledmod{5} Given the latter channel parameters' estimation, the \ac{slam} filter estimates the \ac{ue} state~(i.e., its location, heading, and speed) while simultaneously offering mapping of the UE's environment of signal coverage~(i.e., the location and type of landmarks). 

\vspace{-1mm}
\section{RIS Phase Profile Design}\label{sec:RISPhaseProf}
\vspace{-1mm}
    In this section, we present a spatio-temporal \ac{ris} phase profile design with time-balanced sequences, where at each time instant, a random or directional spatial profile is used, i.e., $\bm{\omega}_{t,k}$ for $t=1,\dots,T$, to adaptively cover the \ac{ue} position uncertainty.
    We first introduce the temporal \ac{ris} phase profiles intended for removing the \ac{nris} paths, that are actually uncontrolled multipath~\cite[Sec.~III-B]{Kamran_RISloc_ICC2022},\cite{Davide_RISLoc_TWC2021}, and the spatial \ac{ris} phase profile for which capitalizes on the availability of the \ac{ue} state information~\cite[Sec.~III-C]{Kamran_RISloc_ICC2022}.
    We then describe the proposed adaptive \ac{ris} phase profile design that enables the generation of multiple uniformly-spaced beams.
\vspace{-1mm}
\subsection{Overview of RIS Phase Profiles}
\vspace{-1mm}

\subsubsection{Time-Domain Design}\label{sec:OrthogonalPhase}
    In order to separate the RIS-induced path from the other ones, the \ac{ris} phase profile is designed to be balanced in the time domain.
    For removing all \ac{nris} paths with $l \neq 0$ in~\eqref{eq:signal_model}~(i.e., the \ac{ue}-\acp{sp}-\ac{ue} and \ac{ue}-\acp{rp}-\ac{ue} paths), we adopt the codebook design of~\cite[Sec.~III-B]{Kamran_RISloc_ICC2022}: $\tilde{\bm{\omega}}_{\tilde{t},k} \in \mathbb{C}^{N_\mathrm{RIS}}$ for $\tilde{t}=1,\dots,T/2$, and $\bm{\omega}_{2\tilde{t}-1,k} = \tilde{\bm{\omega}}_{\tilde{t},k}$,  $\bm{\omega}_{2\tilde{t},k} = -\tilde{\bm{\omega}}_{\tilde{t},k}$. In addition, we set the UE beamformers as $\mathbf{w}_{2\tilde{t}-1,k}=\mathbf{w}_{2\tilde{t},k} = \tilde{\mathbf{w}}_{\tilde{t},k}$ and $\mathbf{f}_{2\tilde{t}-1,k}=\mathbf{f}_{2\tilde{t},k} = \tilde{\mathbf{f}}_{\tilde{t},k}$.
    Following the latter definitions, we can remove the \ac{nris} paths via computing $\tilde{y}_{\tilde{t},k,s}^\mathrm{RIS} =0.5({y}_{2\tilde{t}-1,k,s}-{y}_{2\tilde{t},k,s})$, which yields the expression:
\begin{align}\label{eq:orth}
    &\tilde{y}_{\tilde{t},k,s}^\mathrm{RIS}= \dfrac{\sqrt{E_s} }{2}\tilde{\mathbf{w}}_{\tilde{t},k}^\mathrm{H} \Big( \sum_{l =1}^L 
    \beta_{l,k}e^{j2\pi f_{l,k}^{\mathrm{D}} T_\mathrm{o} (2\tilde{t}-1)} \mathbf{a}_\mathrm{UE}(\bm{\theta}_{l,k})\mathbf{a}_\mathrm{UE}^\top (\bm{\theta}_{l,k}) 
    e^{-j2\pi \tau_{l,k} (s-1) \Delta_f } (1-e^{j2\pi f_{l,k}^{\mathrm{D}} T_\mathrm{o}}) 
    \\ 
    &
    ~~~
    + 
    \beta_{0,k} e^{j2\pi f_{0,k}^{\mathrm{D}} T_\mathrm{o} (2\tilde{t}-1)} \mathbf{a}_\mathrm{UE}(\bm{\theta}_{0,k})
    \tilde{\bm{\omega}}_{\tilde{t},k}^\top\mathbf{b}(\bm{\phi}_{k})\mathbf{a}_\mathrm{UE}^\top (\bm{\theta}_{0,k}) 
    e^{-j2\pi \tau_{0,k} (s-1) \Delta_f } (1+e^{j2\pi f_{0,k}^{\mathrm{D}} T_\mathrm{o}}) \Big) \tilde{\mathbf{f}}_{\tilde{t},k} + \tilde{n}_{\tilde{t},k,s},\notag 
\end{align}
    where $\tilde{n}_{\tilde{t}}\sim \mathcal{CN}(0,\sigma_\mathrm{N}^2/2)$ is the \ac{awgn}. Similar to the approximation of \cite[Sec.~IV-B]{Kamran_RISMobility_JSTSP2022}, using the fact that $\lvert f_{0,k}^\mathrm{D}T_\mathrm{o} \rvert \ll 1$ is a negligible small value\footnote{In the considered mmWave setups with parameters as described in Table~\ref{tab:SimulPara}, the value for $f_{0,k}^\mathrm{D}T_\mathrm{o}$ is about $1.5\times 10^{-11}$.} for the \ac{ofdm} symbol duration, it holds for the residual terms caused by mobility that: $1- e^{j2\pi f_{l,k}^{\mathrm{D}} T_\mathrm{o}} \approx 0$ and $1+ e^{j2\pi f_{0,k}^{\mathrm{D}}T_\mathrm{o}} \approx 2$.
     Therefore, expression \eqref{eq:orth} can be simplified as follows: 
\begin{align}
    \tilde{y}_{\tilde{t},k,s}^\mathrm{RIS} &
     = \sqrt{E_s} 
    \beta_{0,k} e^{j2\pi f_{0,k}^{\mathrm{D}} T_\mathrm{o} (2\tilde{t}-1)} e^{-j2\pi \tau_{0,k} (s-1) \Delta_f } 
    \tilde{\mathbf{w}}_{\tilde{t},k}^\mathrm{H} \mathbf{a}_\mathrm{UE}(\bm{\theta}_{0,k})
    \tilde{\bm{\omega}}_{\tilde{t},k}^\top\mathbf{b}(\bm{\phi}_{k})\mathbf{a}_\mathrm{UE}^\top (\bm{\theta}_{0,k})   \tilde{\mathbf{f}}_{\tilde{t},k} + \tilde{n}_{\tilde{t},k,s},
\end{align}
    which implies that the \ac{ris}-reflected path can be processed separately.
    Similarly, we can obtain the \ac{nris} paths using the simple mathematical operation: $\tilde{y}_{\tilde{t},k,s}^\mathrm{NRIS} \approx 0.5({y}_{2\tilde{t}-1,k,s}+{y}_{2\tilde{t},k,s})$.
\subsubsection{Spatial-Domain Design}
    \label{sec:Spatial_Prof}
    The conventional spatial designs are the random or directional \ac{ris} phase profiles, which are described in the following.
\begin{itemize}
    \item \emph{Random Profile:}
    When the prior \ac{pdf} of the \ac{ue} location at each time step $k$, $f(\mathbf{x}_{\mathrm{UE},k})$, is uninformative or does not exist, random \ac{ris} phase profiles are commonly used, i.e., $\tilde{{\omega}}_{\tilde{t},k}^n = e^{j\phi_{\tilde{t},k}^n}$, where $\tilde{{\omega}}_{\tilde{t},k}^n$ denotes $\tilde{\bm{\omega}}_{\tilde{t},k}$'s $n$-th element and $\phi_{\tilde{t},k}^n \sim [0,2\pi)$ is the \ac{ris} phase which is assumed independent and identically distributed for $n$ and $\tilde{t}$ \cite{Davide_RISLoc_TWC2021,Kamran_RISMobility_JSTSP2022,lens_RIS_ICC_2021,RIS_beam_training_TWC_2021}.
\item \emph{Directional Profile:}
    When an informative prior $f(\mathbf{x}_{\mathrm{UE},k})$ is available, we may use directional \ac{ris} phase profiles $\tilde{\bm{\omega}}_{\tilde{t},k}$ \cite{Kamran_RISMobility_JSTSP2022,Ahmed_RISLoc_TSP2021,lens_RIS_ICC_2021}.
    From $f(\mathbf{x}_{\mathrm{UE},k})$, one can generate $T/2$ azimuth and elevation samples as $\{\bm{\phi}_{\tilde{t},k}\}_{\tilde{t}=1}^{T/2}$ with $\bm{\phi}_{\tilde{t},k} \triangleq [\phi_{\tilde{t},k}^\mathrm{az}, \phi_{\tilde{t},k}^\mathrm{el}]^\top$.
    Then, using \eqref{eq:RIS_NAV}, we yields:
\begin{align}
    \tilde{\bm{\omega}}_{\tilde{t},k} = \exp(-2j(\breve{\mathbf{X}}_\mathrm{RIS})^\top \mathbf{g}(\bm{\phi}_{\tilde{t},k})),
    \label{eq:DRP_w}
\end{align}
which leads to so-called fine pencil beams~\cite{AntenTheory_Book}. 
\end{itemize}

    The main drawback with the \ac{ris} directional phase profiles is that, when the \ac{ris} is large, each reflective beam is narrow, indicating that the $T/2$ \ac{ris} phase profiles may not adequately cover the angular uncertainty region determined by the prior $f(\mathbf{x}_{\mathrm{UE},k})$. In that case, there is a risk that the \ac{ue} location is missed by the finite number of directional beams. 

\vspace{-1mm}
\subsection{Proposed Adaptive RIS Phase Profiles}
\label{sec:ProposedRISProf}
\vspace{-1mm}
    In order to generate spatially-uniform beams, we herein design the \ac{ris} phase profile $\tilde{\bm{\omega}}_{\tilde{t},k}$ for $\tilde{t}=1,\dots,T/2$.
    Using the given \ac{ue} position prior $f(\mathbf{x}_{\mathrm{UE},k})$, we select the boundary of its angular uncertainty with the goal to uniformly illuminate it with some probability $P_\mathrm{E} \in [0,1)$.
   
\paragraph{Definition of the` Angular Uncertainty Region}
    Our objective is to first determine the angular uncertainty in the azimuth and elevation angles, denoted respectively by $[\phi_{\mathrm{min},k}^\mathrm{az},\phi_{\mathrm{max},k}^\mathrm{az}]$ and $[\phi_{\mathrm{min},k}^\mathrm{el},\phi_{\mathrm{max},k}^\mathrm{el}]$.
    We adopt the assumption that the density of the \ac{ue} position $\mathbf{x}_{\mathrm{UE},k}$ is Gaussian distributed,
    hence, the predicted \ac{ue} position \ac{pdf} is represented as $f(\mathbf{x}_{\mathrm{UE},k}|\mathcal{Z}_{1:k-1})=\mathcal{N}(\mathbf{x}_{\mathrm{UE},k|k-1},\mathbf{P}_{\mathrm{UE},k|k-1})$,
    where $\mathbf{x}_{\mathrm{UE},k|k-1}$
    and $\mathbf{P}_{\mathrm{UE},k|k-1}$
    are the mean and covariance, which can be extracted from the predicted density in the Bayesian recursion.
    Then, the position samples are drawn as $\mathbf{x}_{\mathrm{UE},k}^p \sim \mathcal{N}( \mathbf{x}_{\mathrm{UE},k|k-1},\mathbf{P}_{\mathrm{UE},k|k-1})$ and
    $N_\mathrm{pos}$ samples within the confidence region with a particular probability $P_\mathrm{E}$ are selected. 
    For the considered multivariate Gaussian distribution, the condition for satisfying the confidence region is~\cite{slotani1964tolerance}: $(\bm{\mu}^p)^\top\mathbf{P}_{\mathrm{UE},k|k-1}^{-1}\bm{\mu}^p \leq T_\mathrm{E}$,
    where $\bm{\mu}^p \triangleq \mathbf{x}_{\mathrm{UE},k|k-1}- \mathbf{x}_{\mathrm{UE},k}^p$ and $T_\mathrm{E}$ is a threshold computed as $T_\mathrm{E}=Q_{\chi_{D_\mathrm{F}}^2}(P_\mathrm{E})$, with $Q_{\chi_{D_\mathrm{F}}^2}(P_\mathrm{E})$ being the chi-square-distributed quantile function for $P_\mathrm{E}$ with $D_\mathrm{F}$ degrees of freedom for $\mathbf{x}_{\mathrm{UE},k}$.
    Using the latter considerations, we design the angular uncertainty ranges as follows:
\begin{align}
    \phi_{\mathrm{min},k}^\mathrm{az} & = \operatorname*{min}_{p} \operatorname{atan2}(\breve{y}_{\mathrm{UR},k}^p,\breve{x}_{\mathrm{UR},k}^p),\label{eq:az_min} &&
    \phi_{\mathrm{max},k}^\mathrm{az}  = \operatorname*{max}_{p} \operatorname{atan2}(\breve{y}_{\mathrm{UR},k}^p,\breve{x}_{\mathrm{UR},k}^p),\\
    \phi_{\mathrm{min},k}^\mathrm{el} & = \operatorname*{min}_{p} \operatorname{acos}(\breve{z}_{\mathrm{UR},k}^p/d_{\mathrm{UR},k}^p),
    &&
    \phi_{\mathrm{max},k}^\mathrm{el} = \operatorname*{max}_{p} \operatorname{acos}(\breve{z}_{\mathrm{UR},k}^p/d_{\mathrm{UR},k}^p),\label{eq:el_max}
\end{align}
    where $\breve{\mathbf{x}}_{\mathrm{UR},k}^p \triangleq [\breve{x}_{\mathrm{UR},k}^p,\breve{y}_{\mathrm{UR},k}^p,0]^\top$ and $d_{\mathrm{UR},k}^p = \lVert \breve{\mathbf{x}}_{\mathrm{UR},k}^p \rVert$, which can be obtained via the geometric relations in Appendix~\ref{App:ChannelParameters}.
    In addition, $\breve{\mathbf{x}}_{\mathrm{UR},k}^p =  \mathbf{O}_\mathrm{RIS}^\top \mathbf{x}_{\mathrm{UR},k}^p$ 
    with $\mathbf{x}_{\mathrm{UR},k}^p = \mathbf{x}_{\mathrm{UE},k}^p - \mathbf{x}_{\mathrm{RIS}}$.
    In order to represent the latter angular uncertainty ranges in the azimuth and elevation domains, we generate a rectangle (grid) with the four points
    $\bm{\phi}_\mathrm{sw} \triangleq [\phi_{\mathrm{min},k}^\mathrm{az},\phi_{\mathrm{min},k}^\mathrm{el}]^\top$,
    $\bm{\phi}_\mathrm{se} \triangleq [\phi_{\mathrm{max},k}^\mathrm{az},\phi_{\mathrm{min},k}^\mathrm{el}]^\top$,
    $\bm{\phi}_\mathrm{nw} \triangleq [\phi_{\mathrm{min},k}^\mathrm{az},\phi_{\mathrm{max},k}^\mathrm{el}]^\top$,
    $\bm{\phi}_\mathrm{ne} \triangleq [\phi_{\mathrm{max},k}^\mathrm{az},\phi_{\mathrm{max},k}^\mathrm{el}]^\top$, which is from now on represented by
    $B_{k}$.

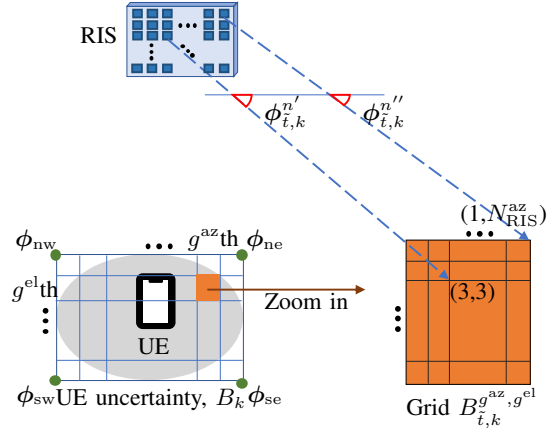
\begin{figure}
    \centering
    \begin{tikzpicture}
    \node (image) [anchor=south west]{\includegraphics[width=0.4\columnwidth]{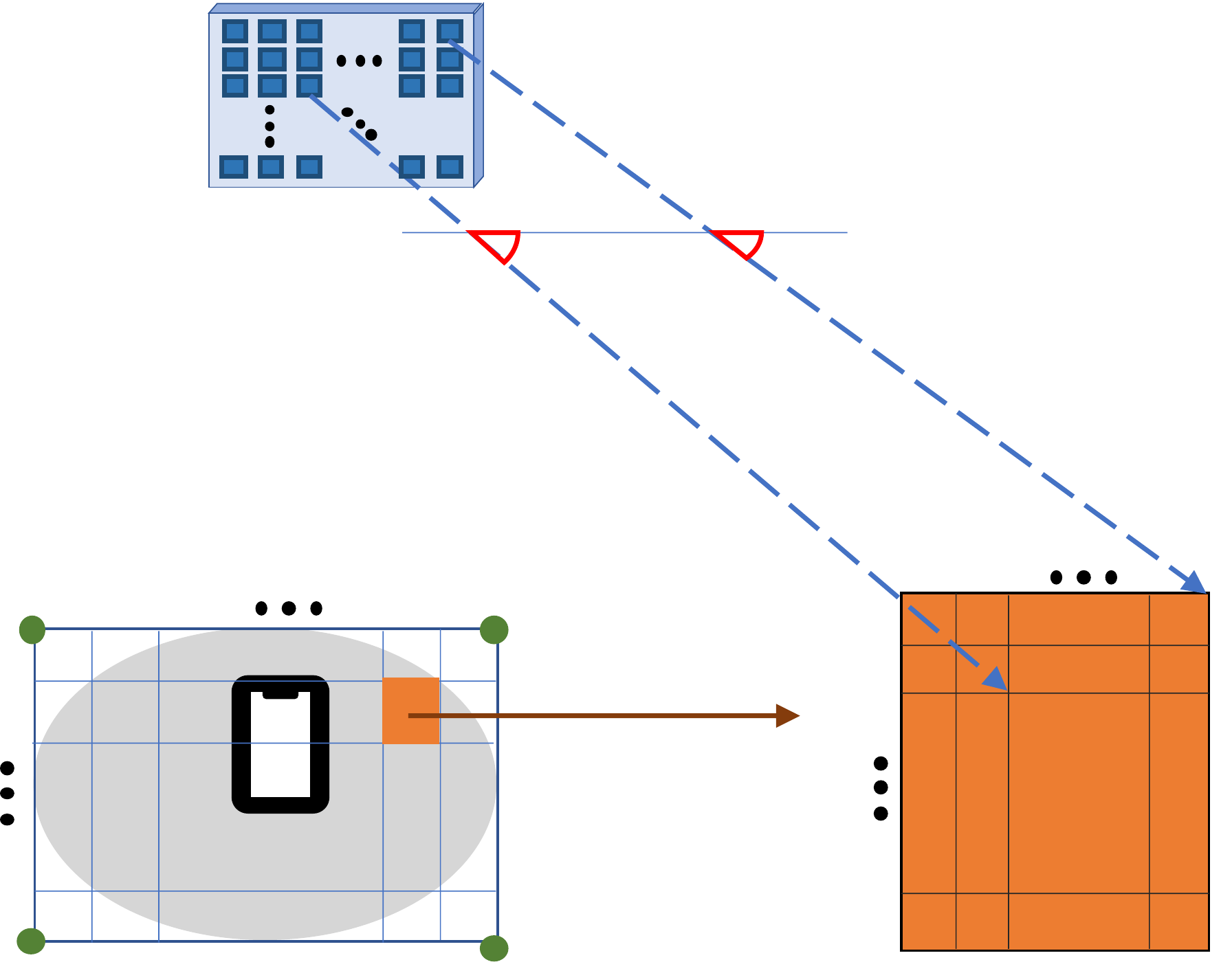}};
    \gettikzxy{(image.north east)}{\ix}{\iy};
    \node at (.235*\ix,.13*\iy){\footnotesize UE};
    \node at (.23*\ix,.0*\iy){\footnotesize UE uncertainty, $B_k$};
    \node at (.85*\ix,-0.02*\iy){\footnotesize Grid $B_{\tilde{t},k}^{g^\mathrm{az},g^\mathrm{el}}$};
    \node at (.13*\ix,.89*\iy){\footnotesize RIS};
    \node at (.49*\ix,.7*\iy){\footnotesize $\boldsymbol{\phi}_{\tilde{t},k}^{n'}$};
    \node at (.68*\ix,.7*\iy){\footnotesize $\boldsymbol{\phi}_{\tilde{t},k}^{n{''}}$};
    \node at (.007*\ix,.27*\iy){\footnotesize{$g^\mathrm{el}$th}};
    \node at (.35*\ix,.39*\iy){\footnotesize{$g^\mathrm{az}$th}};
    \node at (.53*\ix,.24*\iy){\footnotesize{Zoom in}};
    \node at (.85*\ix,.26*\iy){\footnotesize{(3,3)}};
    \node at (.91*\ix,.45*\iy){\footnotesize{(1,$N_\mathrm{RIS}^\mathrm{az}$)}};
    \node at (.01*\ix,.01*\iy){\footnotesize $\boldsymbol{\phi}_{\mathrm{sw}}$};
    \node at (.45*\ix,.01*\iy){\footnotesize $\boldsymbol{\phi}_{\mathrm{se}}$};
    \node at (.45*\ix,.39*\iy){\footnotesize $\boldsymbol{\phi}_{\mathrm{ne}}$};
    \node at (.01*\ix,.39*\iy){\footnotesize $\boldsymbol{\phi}_{\mathrm{nw}}$};
    \vspace{-0.6cm}
    \end{tikzpicture}
    \vspace{-0.6cm}
    \caption{An example illustration of the RIS phase profiles for multiple spatially-uniform beams. The \ac{ue} uncertainty is represented by a rectangle $B_k$, which is divided into $T/2$ grids, inspired by
    our proposed time-domain RIS profile design. This indicates that two consecutive transmissions suffice to cover the whole grid. During each transmission, each \ac{ris} element's phase is configured to direct the induced reflection in a slightly different direction than any other one, in order for all to cover all points of the grid. For example, the distinct elements $n'$ and $n''$ steer at the angles $\boldsymbol{\phi}_{\tilde{t},k}^{n'}$ and $\boldsymbol{\phi}_{\tilde{t},k}^{n''}$ with $\boldsymbol{\phi}_{\tilde{t},k}^{n'}\neq \boldsymbol{\phi}_{\tilde{t},k}^{n''}$.}
    \label{Fig:MUB}
    \vspace{-0.6cm}
\end{figure}     
\paragraph{Uniform Coverage of the Uncertainty Region}
    We now devise the \ac{ris} phase profiles $\tilde{\bm{\omega}}_{\tilde{t},k} \triangleq [\tilde{\omega}_{\tilde{t},k}^1,\dots,\tilde{\omega}_{\tilde{t},k}^{N_\mathrm{RIS}}]^\top$ in~\eqref{eq:DRP_w} by designing $\bm{\phi}_{\tilde{t},k}\triangleq[(\bm{\phi}_{\tilde{t},k}^1)^\top,\dots,(\bm{\phi}_{\tilde{t},k}^{N_\mathrm{RIS}})^\top]^\top$, as illustrated in Fig.~\ref{Fig:MUB}.
    We first note that, due to the half of transmission diversity offered by the time-domain RIS profiles, two consecutive pilot transmissions suffice to cover the angular uncertainty of the grid during each time interval $k$.
    We thus divide the rectangle $B_{k}$ into $T^\mathrm{az}$ horizontal and $T^\mathrm{el}$ vertical slices such that $T^\mathrm{az} \times T^\mathrm{el} = T/2$, where each $\tilde{t}$-th box, with $\tilde{t} = 1,\dots,T/2$, lies in between the $g^\mathrm{az}$-th column and $g^\mathrm{el}$-th row of the grid $B_{k}$. This defined the box $B_{\tilde{t},k}^{g^\mathrm{az}, g^\mathrm{el}}$ with $g^\mathrm{az}=\tilde{t}-T^\mathrm{az}(\mathrm{ceil}(\tilde{t}/T^\mathrm{az})-1)$ and $g^\mathrm{el}=\mathrm{ceil}(\tilde{t}/T^\mathrm{az})$, where $\mathrm{ceil}(x)$ returns the smallest integer that is greater or equal to $x$.  
    To illuminate all grid points $B_{\tilde{t},k}^{g^\mathrm{az}, g^\mathrm{el}}$ in each transmission $\tilde{t}$, each \ac{ris} element's phase is configured to a slightly different direction to point at a points of the grid. The phase configuration of each $n$-th RIS element is thus designed as $\bm{\phi}_{\tilde{t},k}^n = [\phi_{\mathrm{min},k}^\mathrm{az} + \zeta_k^\mathrm{az}(g^\mathrm{az}-1)/T^\mathrm{az} + \zeta_k^\mathrm{az}(n^\mathrm{az}-1)/(T^\mathrm{az}(N_\mathrm{RIS}^\mathrm{az}-1)),
    \phi_{\mathrm{min},k}^\mathrm{el} + \zeta_k^\mathrm{el}(g^\mathrm{el}-1)/T^\mathrm{az} + \zeta_k^\mathrm{el}(n^\mathrm{el}-1)/(T^\mathrm{az}(N_\mathrm{RIS}^\mathrm{el}-1))]^\top$, where $\zeta_k^\mathrm{az}= \phi_{\mathrm{max},k}^\mathrm{az}-\phi_{\mathrm{min},k}^\mathrm{az}$, $\zeta_k^\mathrm{el}= \phi_{\mathrm{max},k}^\mathrm{el}-\phi_{\mathrm{min},k}^\mathrm{el}$,
    $n^\mathrm{az}=n-N_\mathrm{az}^\mathrm{az}
    (\mathrm{ceil}(\tilde{n}/N_\mathrm{RIS}^\mathrm{az})-1)$, and $n^\mathrm{el}=\mathrm{ceil}(\tilde{n}/N_\mathrm{RIS}^\mathrm{az})$.
    By selecting the intersection points of evenly separated $N_\mathrm{RIS}^\mathrm{az}$ vertical and $N_\mathrm{RIS}^\mathrm{el}$ horizontal line segments on the grid, we can generate $N_\mathrm{RIS} = N_\mathrm{RIS}^\mathrm{az} \times N_\mathrm{RIS}^\mathrm{el}$ points at the grid.
    Finally, we design the phase profile for each $n$-th RIS element as follows:
\begin{align}
    \tilde{\omega}_{\tilde{t},k}^n = \exp(-2j(\breve{\mathbf{x}}_\mathrm{RIS}^n)^\top \mathbf{g}(\bm{\phi}_{\tilde{t},k}^n)).
    \label{eq:proposed_w}
\end{align}

\begin{rem}[Interpretation of \eqref{eq:proposed_w}]
    Conceptually, the phase of each $n$-th \ac{ris} element is configured with respect to the \ac{ris} phase center to  
    direct a beam in a slightly different direction than this. Hence, $\bm{\phi}_{\tilde{t},k}^n \neq  \bm{\phi}_{\tilde{t},k}^{n'}$ when $n\neq n'$, unless the uncertainty region is a single point, in which case \eqref{eq:proposed_w} reverts to a conventional pencil beam as in \eqref{eq:DRP_w}. 
    \end{rem}
    

\vspace{-1mm}
\section{Proposed RIS-Aided SLAM Estimation}
\label{sec:MPMB}
\vspace{-1mm}
    In this section, we propose an extension of the \ac{mpmb}-\ac{slam} filter\footnote{The \ac{mpmb}-\ac{slam} filter is efficient with respect to the performance trade-off between complexity and accuracy for \ac{rfs}-based sequential Bayesian estimation. Very recently, it has been applied for parameter estimation in mmWave vehicular systems~\cite{Hyowon_MPMB_TVT2022}.}~\cite{Hyowon_MPMB_TVT2022}, with the intention to account for the fact that the \ac{ris}-induced path can be separated using the time-domain RIS phase profiles, as presented in the previous Section~\ref{sec:OrthogonalPhase}.

    
    
    
\vspace{-1mm}
\subsection{Preliminaries}
\vspace{-1mm}
    We introduce the notation $I_{k-1}$ to denote the number of detected landmarks, with each of them indexed by $i$.    $f(\mathbf{s}_{k}|\mathcal{Z}_{1:k-1}) \triangleq f_{\mathsf{p},k}(\mathbf{s}_{k})$ and $f(\mathbf{s}_{k}|\mathcal{Z}_{1:k}) \triangleq f_{\mathsf{u},k}(\mathbf{s}_{k})$ represent the prior and updated \acp{pdf} of the \ac{ue} state. Undetected landmarks that have been never detected are modeled by a \ac{ppp}, which can be characterized by its intensity. To this end,  
    $\lambda(\mathbf{x},m|\mathcal{Z}_{1:k-1}) \triangleq \lambda_{\mathsf{p},k}(\mathbf{x},m)$ and $\lambda(\mathbf{x},m|\mathcal{Z}_{1:k}) \triangleq\lambda_{\mathsf{u},k}(\mathbf{x},m)$ are respectively the predicted and updated intensities for undetected landmarks of type $m$.
    Detected landmarks that have been previously detected at least once follow a \ac{mb} process, and each landmark is characterized by a Bernoulli density. $f^i(\mathbf{x},m|\mathcal{Z}_{1:k-1}) \triangleq f_{\mathsf{p},k}^i(\mathbf{x},m)$ ($f^i(\mathbf{x},m|\mathcal{Z}_{1:k}) \triangleq f_{\mathsf{u},k}^i(\mathbf{x},m)$) and $r_{\mathsf{p},k}^i$ ($r_{\mathsf{u},k}^i$) represent the predicted (updated) density of detected landmark $i$ of type $m$ and its existence probability, respectively. 
    $g(\mathbf{z}_k^0|\mathbf{s}_k)$ and
    $g(\mathbf{z}_k^j|\mathbf{s}_k,\mathbf{x},m)$ denote the measurement likelihood functions of~\eqref{eq:UEMea_model} and~\eqref{eq:ChParaMea_model}, respectively. 

    

\vspace{-1mm}
\subsection{Update Step}
\label{sec:MPMB_Up}
\vspace{-1mm}
    At each discrete time $k$, the following prior components are given: the intensities of the undetected landmarks  $\{\lambda_{\mathsf{p},k}(\mathbf{x},m)\}_{m\in\{\mathrm{RP},\mathrm{SP}\}}$, 
    the predicted \ac{ue} state density  $f_{\mathsf{p},{k}}(\mathbf{s}_{k})$, the \ac{ris} state density $f^{1}(\mathbf{x},\mathrm{RIS})$, the \ac{ris} existence probability $r^{1}$, as well as the densities and existence probabilities of the detected landmarks  $\{\{f_{\mathsf{p},k}^{i}(\mathbf{x},m)\}_{m\in \{\mathrm{RP},\mathrm{SP}\}},r_{\mathsf{p},k}^{i}\}_{i=2}^{I_{k-1}}$.

\subsubsection{Statistics and Map Updates}
    Due to the distinguishable \ac{ris} path, there is no update for the \ac{ris} Bernoulli components $f^{1}(\mathbf{x}_\mathrm{RIS})$ and $r^{1}$, whereas the \ac{pmb} components of the other landmarks are updated with the measurements $\widetilde{\mathcal{Z}}_k=\mathcal{Z}_k \setminus \{\mathbf{z}_k^0\}$, where `$\setminus$' indicates the set difference.
    In the \ac{pmb} density, a landmark set is divided into two independent subsets: a set of undetected landmarks (landmarks that exist but have not yet been detected); and a set of detected landmarks (landmarks that have been associated to at least one previous measurement). In the update step, each set goes through two cases: missed detections (when a landmark is not detected) and detections (when a landmark is detected by any measurement in $\widetilde{\mathcal{Z}}_k$), leading to an exponentially increasing number of cases. 
    We will next use the definitions $h_{\mathsf{p},k}^\mathrm{U}(\mathbf{s}_k,\mathbf{x},m) \triangleq f_{\mathsf{p},k}(\mathbf{s}_k)\lambda_{\mathsf{p},k}(\mathbf{x},m)$ and 
    $h_{\mathsf{p},k}^{\mathrm{D},i}(\mathbf{s}_k,\mathbf{x},m) \triangleq f_{\mathsf{p},k}(\mathbf{s}_k)f_{\mathsf{p},k}^i(\mathbf{x},m)$.
 \begin{enumerate}[i),left=0pt,noitemsep,topsep=0.5pt]
     \item \emph{Intensity of undetected landmarks that remained undetected:} The updated intensity 
     corresponds to the missed detection  $1- 
     \mathsf{p}_{\mathrm{D},k}(\mathbf{s}_k,\mathbf{x},m)$ of undetected landmarks $\lambda_{\mathsf{p},k}(\mathbf{x},m)$, integrated over the UE state density $f_{\mathsf{p},k}(\mathbf{s}_k)$. Mathematically, this is given by:
 \begin{align}
     \lambda_{\mathsf{u},k}(\mathbf{x},m) = \int \left( 1- 
     \mathsf{p}_{\mathrm{D},k}(\mathbf{s}_k,\mathbf{x},m)\right)
     h_{\mathsf{p},k}^\mathrm{U}(\mathbf{s}_k,\mathbf{x},m)
     \mathrm{d}\mathbf{s}_k. 
 \end{align}
    
     \item \emph{Density of a new landmark that was detected for the first time with a measurement $\mathbf{z}_k^j$:} With $\mathbf{z}_k^j \in \widetilde{\mathcal{Z}}_k$, new Bernoulli components are generated by the prior $h_{\mathsf{p},k}^\text{U}(\mathbf{s}_k,\mathbf{x},m)$ using the likelihood function $g(\mathbf{z}_k^j|\mathbf{s}_k,\mathbf{x},m) $, which is being weighted by the detection probability $\mathsf{p}_{\text{D},k}(\mathbf{s}_k,\mathbf{x},m)$, yielding the density: 
 \begin{align} \label{eq:MPMB-S2f}
     f_{\mathsf{u},k}^{j}(\mathbf{x},m) 
     \propto \int  \mathsf{p}_{\text{D},k}(\mathbf{s}_k,\mathbf{x},m)h_{\mathsf{p},k}^\text{U}(\mathbf{s}_k,\mathbf{x},m)g(\mathbf{z}_k^j|\mathbf{s}_k,\mathbf{x},m) \mathrm{d}\mathbf{s}_k ,
 \end{align}
 where $\rho_k^{j} = \sum_m \iint \mathsf{p}_{\text{D},k}(\mathbf{s}_k,\mathbf{x},m) h_{\mathsf{p},k}^\text{U}(\mathbf{s}_k,\mathbf{x},m) g(\mathbf{z}_k^j|\mathbf{s}_k,\mathbf{x},m) \mathrm{d}\mathbf{s}_k \mathrm{d}\mathbf{x}$ is the normalization constant.
     The updated existence probability is $r_{\mathsf{u},k}^{j} = \rho_k^{j}/(\rho_k^{j}+c(\mathbf{z}_k^j))$, accounting for the probability that the hypothesized landmark was actually clutter. 
      
     \item \emph{Density of a previously detected landmark that was miss-detected:}
     In this case, the prior $h_{\mathsf{p},k}^{\text{D},i}(\mathbf{s}_k,\mathbf{x},m)$  is multiplied by the missed detection probability $(1-\mathsf{p}_{\text{D},k}(\mathbf{s}_k,\mathbf{x},m))$ to obtain the density: 
\begin{align}\label{eq:MPMB-S3f}
     f_{\mathsf{u},k}^{0,i}(\mathbf{x},m) 
      \propto \int(1-\mathsf{p}_{\text{D},k}(\mathbf{s}_k,\mathbf{x},m))h_{\mathsf{p},k}^{\text{D},i}(\mathbf{s}_k,\mathbf{x},m)\mathrm{d}\mathbf{s}_k,
\end{align}
where $\rho_k^{0,i}=\sum_m  \iint(1-\mathsf{p}_{\text{D},k}(\mathbf{s}_k,\mathbf{x},m)) h_{\mathsf{p},k}^{\text{D},i}(\mathbf{s}_k,\mathbf{x},m) \mathrm{d} \mathbf{s}_k\mathrm{d}\mathbf{x}$ is the normalization constant. 
The updated existence probability in this case is given by  $r_{\mathsf{u},k}^{0,i} = {r_{\mathsf{p},k}^{i}\rho_k^{0,i}}/(1-r_{\mathsf{p},k}^{i} + r_{\mathsf{p},k}^{i}\rho_k^{0,i})$. 
     \item \emph{Density of a previously detected landmark that was detected again with a measurement $\mathbf{z}_k^j$:}
     With $\mathbf{z}_k^j \in \widetilde{\mathcal{Z}}_k$, the Bernoulli components of landmark $i$, for $i=\{2,\dots,I_{k-1}\}$, are updated by the product of the prior $ h_{\mathsf{p},k}^{\text{D},i}(\mathbf{s}_k,\mathbf{x},m)$ times the likelihood $g(\mathbf{z}_k^j|\mathbf{s}_k,\mathbf{x},m)$, while being corrected with the detection probability $\mathsf{p}_\text{D,k}(\mathbf{s}_k,\mathbf{x},m)$; the following expression is deduced:
\begin{align}
     f_{\mathsf{u},k}^{j,i}(\mathbf{x},m) 
      \propto \int \mathsf{p}_\text{D,k}(\mathbf{s}_k,\mathbf{x},m)
     h_{\mathsf{p},k}^{\text{D},i}(\mathbf{s}_k,\mathbf{x},m)
     g(\mathbf{z}_k^j|\mathbf{s}_k,\mathbf{x},m)
     \mathrm{d}\mathbf{s}_k, 
\end{align}
where $\rho_k^{j,i} = \sum_m\iint  \mathsf{p}_\text{D,k}(\mathbf{s}_k,\mathbf{x},m)
     h_{\mathsf{p},k}^{\text{D},i}(\mathbf{s}_k,\mathbf{x},m)
     g(\mathbf{z}_k^j|\mathbf{s}_k,\mathbf{x},m) \mathrm{d}\mathbf{s}_k\mathrm{d}\mathbf{x} $ is the normalization constant.
     The updated existence probability is now $r_{\mathsf{u},k}^{j,i} = 1$.
 \end{enumerate}
In conclusion, the landmark update procedure consists of the four steps i)--iv), and the updated densities and likelihood functions can be computed by a non-linear Kalman filter~\cite{Garcia-Fernandez2018}. 
We thus have $|\widetilde{\mathcal{Z}}_k|$ new landmarks (step ii) and for each of the $I_{k-1}$ previously detected landmarks, we have $|\widetilde{\mathcal{Z}}_k|+1$ possible Bernoulli densities (steps iii and iv). Since each non-RIS measurement\footnote{Due to the distinguishable \ac{ris} path, the associations corresponding to landmark $i=1$ and measurement $\mathbf{z}^0$ are known.} can only be associated to at most one landmark, the so-called marginal association probabilities should be computed \cite[Fig.~9]{williams2015marginal} based on the normalization constants $\rho_{k}^{i,0}$ and $\rho_{k}^{i,j}$. From these, each previously detected landmark gives rise to a mixture density with weights given by the marginal association probabilities, and each new landmark's existence probability is reduced by the corresponding marginal association probability \cite[Fig.~10]{williams2015marginal}.

\subsubsection{UE Update}
    We compute the global association probabilities through the product of the normalization constants of valid  track- and measurement-oriented associations.
    We calculate the most likely association using  Murty's algorithm \cite{murty1968algorithm} (a generalization of the Hungarian algorithm for solving linear assignment problems). Then, with each previously detected landmark $i$, we have an associated measurement set, denoted by $\mathcal{Z}_k^{{a}_k^i}$, which may be empty or contain one element from $\widetilde{\mathcal{Z}}_k$.  
    Note that the \ac{ris} measurement $\mathbf{z}_k^0$ is utilized in the \ac{ue} density update (as in \cite{Kamran_RISloc_ICC2022}) and never generates a newly detected landmark.
    The update of the \ac{ue} state density~\cite[Proposition~3]{Hyowon_MPMB_TVT2022} is thus simplified 
    as follows:
\begin{align}
    f_{\mathsf{u},k}(\mathbf{s}_k)  \propto f_{\mathsf{p},k}(\mathbf{s}_k) 
    g(\mathbf{z}_k^0|\mathbf{s}_k)
    \prod_{i=2}^{I_{k-1}}  q_i(\mathcal{Z}_k^{{a}_k^i}|\mathbf{s}_k),\label{eq:MarVehPosApp2}  
\end{align}
    where 
    $q_i(\mathcal{Z}_k^{a_k^i}|\mathbf{s}_k)$ for $i=2,\dots,I_{k-1}$ is given by
\begin{align}
    q_i(\mathcal{Z}_k^{a_k^i}|\mathbf{s}_k) 
    = 
\begin{cases}
    r_{\mathsf{p},k}^i \sum_{m} \int\mathsf{p}_\text{D}(\mathbf{s}_k,\mathbf{x},m) f_{\mathsf{p},k}^i(\mathbf{x},m)
    g(\mathbf{z}_k^j|\mathbf{s}_k,\mathbf{x},m) \mathrm{d}\mathbf{x},
    & \mathcal{Z}_k^{a_k^i}=\{\mathbf{z}_k^j \} 
     \\
    r_{\mathsf{p},k}^i \sum_{m} \int(1-\mathsf{p}_\text{D}(\mathbf{s}_k,\mathbf{x},m))
    f_{\mathsf{p},k}^i(\mathbf{x},m)\mathrm{d}\mathbf{x}+ 1-r_{\mathsf{p},k}^i,
    & \mathcal{Z}_k^{a_k^i}=\emptyset
\end{cases}.
     \label{eq:app-qZi}
\end{align}


   
\vspace{-1mm}
\subsection{Prediction Step}
\label{sec:MPMB_Pred}
\vspace{-1mm}

\subsubsection{Landmark}
    Since the undetected landmarks are fixed, the intensity for $m=\{\mathrm{RP},\mathrm{SP}\}$ is predicted as $\lambda_{\mathsf{p},k+1}(\mathbf{x}_{k+1},m) = \int P_\mathrm{S}  \lambda_{\mathsf{u},k}(\mathbf{x}_k,m) \mathrm{d} \mathbf{x} 
    + \lambda_{B,k+1}(\mathbf{x}_{k+1},m)$,
    where $P_\mathrm{S}$ is the survival probability 
    and $\lambda_{B,k+1}(\mathbf{x}_{k+1},m)$ is the birth intensity.
    The Bernoulli position density for the RP 
    depends on \ac{ue}'s heading and speed. For \ac{rp} prediction, we introduce the notation $\tilde{\mathbf{s}}_k \triangleq [\alpha_{\mathrm{UE},k},v_{\mathrm{UE},k}]^\top$ with $f_{\mathsf{u},k}(\tilde{\mathbf{s}}_k)$ extracted from $f_{\mathsf{u},k}({\mathbf{s}}_k)$ in~\eqref{eq:MarVehPosApp2}, and $f(\mathbf{x}_{k+1},\tilde{\mathbf{s}}_{k+1}|\mathbf{x}_{k},\tilde{\mathbf{s}}_k)$ being the known transition density.
    Then, the predicted density is obtained as $f_{\mathsf{p},k+1}^i(\mathbf{x},\mathrm{RP}) = \int f(\mathbf{x}_{k+1},\tilde{\mathbf{s}}_{k+1}|\mathbf{x}_{k},\tilde{\mathbf{s}}_k)
    f_{\mathsf{u},k}(\tilde{\mathbf{s}}_k)
    P_\mathrm{S}
    f_{\mathsf{u},k}^i(\mathbf{x}_k,\mathrm{RP}) \mathrm{d}\mathbf{x}_{k} \mathrm{d} 
    \tilde{\mathbf{s}}_k
    \mathrm{d} 
    \tilde{\mathbf{s}}_{k+1}$.
    For the fixed \ac{sp}, the Bernoulli components are predicted as $f_{\mathsf{p},k+1}^i(\mathbf{x}_{k+1},\mathrm{SP}) = f_{\mathsf{u},k}^i(\mathbf{x},\mathrm{SP})$ and
    $r_{\mathsf{p},k+1}^i = P_\mathrm{S} r_{\mathsf{u},k}^i$.

\subsubsection{UE}
    The \ac{ue} density is finally predicted as $f_{\mathsf{p},k+1}(\mathbf{s}_{k+1}) = \int f(\mathbf{s}_{k+1}|\mathbf{s}_{k}) f_{\mathsf{u},k}(\mathbf{s}_{k}) \mathrm{d} \mathbf{s}_{k}$,
    in which $f(\mathbf{s}_{k+1}|\mathbf{s}_{k})$ is the known transition density of the dynamic model for the \ac{ue} state.
    We finally extract the prior density of the \ac{ue} position $f_{\mathsf{p},k+1}(\mathbf{x}_{\mathrm{UE},k+1})$, using $f_{\mathsf{p},k+1}(\mathbf{s}_{k+1})$, which is utilized for the \ac{ris} phase profiles of Section~\ref{sec:RISPhaseProf} at each next time instant $k+1$.

\vspace{-1mm}
\section{Numerical Results and Discussion}\label{sec:Results}
\vspace{-1mm}
In this section, we present simulation results demonstrating the efficiency of the proposed SLAM method in conjunction with the new RIS phase profiles. We first detail the simulation setup and then investigate the performance of the proposed RIS-enabled SLAM scheme over various operation parameters.

\vspace{-1mm}
\subsection{Simulation Setup}
\vspace{-1mm}
    The simulation parameters used in our performance evaluations 
    are summarized in Table~\ref{tab:SimulPara}. The size of the propagation environment was set to [40, 100]~m $\times$ [-30, 70]~m $\times$ [0, 40]~m, and the size of the road, where the \ac{ue} moves, was [40, 60]~m $\times$ [-20, 60]~m with $z=0$.
    In the propagation environment, we have considered a single \ac{ris}, a single \ac{ls}, four \acp{sp}, and a single \ac{ue}.
    During $K=20$ time intervals, the \ac{ue} dynamics followed the constant turn model~\cite{Li_Dynamic_TAES2003} with polar velocity and known turn rate. Hence, the \ac{ue} state evolution was given by $\mathbf{s}_{\mathrm{UE},k+1} = \mathsf{m}(\mathbf{s}_{\mathrm{UE},k}) + \mathbf{q}_k$,
    where $\mathsf{m}(\cdot)$ is the transition function and $\mathbf{q}_k\sim \mathcal(\bm{0},\mathbf{Q}_k)$ is the process noise.
    The function $\mathsf{m}(\cdot)$, considering the known constant turn rate $\rho_{\mathrm{UE},k}$, can be derived as~\cite{Li_Dynamic_TAES2003}:
    $\mathsf{m}(\mathbf{s}_{\mathrm{UE},k+1})= \mathbf{s}_{\mathrm{UE},k} + [\Delta_{x,\mathrm{UE}}, \Delta_{y,\mathrm{UE}},0,\rho_{\mathrm{UE},k}\Delta,v_{\mathrm{UE},k}]^\top$, where
    $\Delta_{x,\mathrm{UE}}=\frac{v_{\mathrm{UE},k}}{\rho_{\mathrm{UE},k}}(\sin(\alpha_{\mathrm{UE},k}+\rho_{\mathrm{UE},k}\Delta)-\sin\alpha_{\mathrm{UE},k} )$ and 
    $\Delta_{y,\mathrm{UE}}=\frac{v_{\mathrm{UE},k}}{\rho_{\mathrm{UE},k}}\left(-\cos(\alpha_{\mathrm{UE},k}+\rho_{\mathrm{UE},k}\Delta)+\cos\alpha_{\mathrm{UE},k}\right)$,
    and $\mathbf{Q}_k = 
    \mathrm{diag}(\sigma_x^2, \sigma_y^2, \sigma_z^2, \sigma_\alpha^2, \sigma_v^2)$ with $\sigma_x = 0.2$ m, $\sigma_y = 0.2$ m, $\sigma_z = 0$ m, $\sigma_\alpha = 10^{-3}$ rad, and $\sigma_v^2=0.2$ m/s; we have set 
    $\Delta = 0.5$ s and $\rho_{\mathrm{UE},k} = 2 \times 10^{-6}$ rad/s.
    The \ac{ris} lied on a vertical wall on the $yz$ plane, the \ac{ue} movement was parallel to the \ac{ls}, and one \ac{rp} was visible from the \ac{ls} at every time instant $k$. Thus, $K$ \acp{rp} were detected and the \ac{rp} locations were set to $\mathbf{x}_{\mathrm{RP},k} = [100,y_{\mathrm{UE},k},0]$ m for $k=1,\dots,20$.
    The \acp{sp} locations were set to $\mathbf{s}_\mathrm{SP}^1=[40, -20, z_\mathrm{SP}]^\top$ m, 
    $\mathbf{s}_\mathrm{SP}^2=[60, 5, z_\mathrm{SP}]^\top$ m, 
    $\mathbf{s}_\mathrm{SP}^3=[40, 30, z_\mathrm{SP}]^\top$ m, and $\mathbf{s}_\mathrm{SP}^4=[60, 55, z_\mathrm{SP}]^\top$ m with $z_\mathrm{SP} \sim \mathcal{U}(0,10)$.
    The \ac{ris} and \acp{rp} were always visible, while \acp{sp} were only visible when the  \acp{sp} exist within a \ac{fov} distance of $r_\mathrm{FOV} = 50$~m from the\footnote{The \ac{nris} paths were only visible if their \acp{snr} were sufficiently high.} \ac{ue}. Within the \ac{fov}, we set the detection probability as $P_\mathrm{D}=0.95$, \acp{rp} and \acp{sp} were detected with $P_\mathrm{D}$, and the \ac{ris} was always detected.
    The clutter intensity was set to $c(\mathbf{z}) = 2.1 \times 10^{-6}$, the number of clutter was Poisson distributed with mean $\mu_\mathrm{Poi}=1$, and the measurement for clutter $\mathbf{z}_k$ was randomly generated within the propagation environment.
    The probability for the confidence region $P_\mathrm{E}$ was set to 0.99 with $D_\mathrm{F}=2$ degrees of freedom, yielding $T_\mathrm{E}= -2\log P_\mathrm{E}$.
    

\begin{table}[t!]
\centering
\caption{The Simulation Parameters Used in the Performance Evaluations.}
\vspace{-2mm}
\label{tab:SimulPara}
\begin{tabular}{llll}
	\hlineB{3}
	Parameter & Value & Parameter & Value \\ \hline \hline
	\ac{ris} array size & $N_\mathrm{RIS} = 2500$~($50\times50$) 
&
	\ac{ue} array size & $N_\mathrm{UE} = 16$~($4\times4$) \\
	Number of transmissions & $T=20$ 
 &
        Carrier frequency & $f_c = 30$ GHz \\
	Speed of light & $c = 3 \times 10^8$ m/s 
&
	Wavelength & $\lambda = 1$ cm \\
        {Bandwidth} &  $B=200$ MHz 
&
        {Subcarrier spacing} & {$\Delta_f = 120$ kHz} \\
	Number of subcarriers & $N_\mathrm{SC}= 1600$ 
 &
	\ac{cp} overhead & $\epsilon_\mathrm{CP}=0.07$ \\
	Number of \ac{cp} subcarriers & $N_\mathrm{CP}= 112$ 
 &
	Transmission power & $E_sN_\mathrm{SC} \Delta_f = 20$ dBm \\
	Symbol duration & $T_\mathrm{S}=8$ $\mu$s 
 &
	\ac{cp} duration & $T_\mathrm{CP}=0.56$ $\mu$s \\
	Signal duration & $T_o = T_\mathrm{S} + T_\mathrm{CP}= 8.07$ $\mu$s 
 &
	Noise PSD & $N_0 = -174$ dBm/Hz \\
	Noise figure & $N_\mathrm{NF} = 8$ dB 
 &
	Signal noise variance & $\sigma_\mathrm{N}^2 = N_0N_\mathrm{NF}= -166$ dBm/Hz \\
	\ac{ris} 3D position & $\mathbf{x}_\mathrm{RIS}=[40,0,20]^\top$ m 
 &  
    \ac{ls} plane's point & $x=100$ m\\
	\hlineB{3}
\end{tabular}
\vspace{-0.6cm}
\end{table}    
    We adopted the complex path gain model $\beta_{l,k} = \lvert \beta_{l,k} \rvert e^{{-j(2\pi f_c \tau_{l,k}}+\nu_\mathrm{G})}$ for $l=0,\dots,L$, where the amplitudes $\lvert \beta_{l,k} \rvert$ for specular reflection, scatter one~\cite[eq.~(45)]{Zohair_5GFIM_TWC2018}, and for \ac{ris} reflection~\cite[eq.~(22)]{Steven_RISGain_2021} were set as:
    $\lvert \beta_{l,k} \rvert = \frac{\lambda }{4\pi}\frac{\lambda \cos(\varphi_0)^{2q_0}} {16d_{\mathrm{UR},k}^2}
    $ for the RIS; $\lvert \beta_{l,k} \rvert=\frac{\lambda }{4\pi}\frac{\sqrt{\Gamma_\mathrm{R}}}
    {2d_{\mathrm{IU},k}^{(\mathrm{RP})}}$ for the \ac{rp}; and $\lvert \beta_{l,k} \rvert=\frac{\lambda }{4\pi}\frac{\sqrt{S}_\mathrm{RCS}}
    {\sqrt{4\pi} (d_{\mathrm{IU},k}^{(\mathrm{SP})})^2}$ for the \ac{sp}.
    We have considered
    $\nu_\mathrm{G} \sim \mathcal{U}[0,2\pi)$ being the unknown phase offset, 
    $q_0 =0.285$, $\cos(\varphi_0) = (\mathbf{x}_\mathrm{UR}^\top \mathbf{n}_\mathrm{RIS}/d_{\mathrm{UR},k})$,  $\mathbf{n}_\mathrm{RIS}=[1,0,0]^\top$ (normal vector of the plane where the \ac{ris} elements lie), $\Gamma_\mathrm{R}=0.7$ was the reflection coefficient, and $S_\mathrm{RCS}= 50$~m$^2$ was the square of the radar cross section.
    
    
    To implement the \ac{mpmb}-\ac{slam} filter, we adopted the parameters of~\cite[Sec.~VI.A]{Hyowon_MPMB_TVT2022}. Specifically, in steps i) and ii) of Section~\ref{sec:MPMB_Up}, we set $\mathsf{p}(\mathbf{s}_k,\mathbf{x},m) = 0.95$, and in steps iii) and iv), we set $\mathsf{p}(\mathbf{s}_k,\mathbf{x},m)= 0.95$ if  $M_\text{d}<T_\text{G}$, otherwise, $\mathsf{p}(\mathbf{s}_k,\mathbf{x},m)~\approx 0$, where $M_\text{d}$ is the ellipsoidal gating distance with
    gate probability $P_\text{G}=0.99$. 
    We have set the survival probability as $P_\mathrm{S}=0.99$, and the approximated Gaussian densities were computed via the cubature Kalman filter approximation~\cite{HaykinCKF2009,Hyowon_TWC2020,Hyowon_MPMB_TVT2022} with the measurement noise covariance ${\mathbf{R}_{\mathsf{U},k}^j=4\times \mathbf{R}_k^j}$ for  numerical robustness.
    The weight thresholds for pruning a Bernoulli, Gaussian of the intensity, and global hypothesis were respectively $10^{-5}$, $5\times 10^{-10}$, and $10^{-4}$.
    The landmark $i$ was detected if $r_{\mathrm{u},k}^i > T_\mathrm{EP}=0.4$, and its landmark type and position were respectively estimated as  $\hat{m} = \max_{m} \rho_k^i(m)$ and $\hat{\mathbf{x}} = \int \mathbf{x} f_{\mathrm{u},k}^i(\mathbf{x},\hat{m})\mathrm{d}\mathbf{x}$.
    The performance results that follows were obtained by averaging over $500$ Monte Carlo simulation runs. To evaluate the performance of localization and radio mapping, the \ac{mae} and the \ac{gospa} distance~\cite{RahmathullahGFS:2017} metrics were utilized.

\begin{figure}
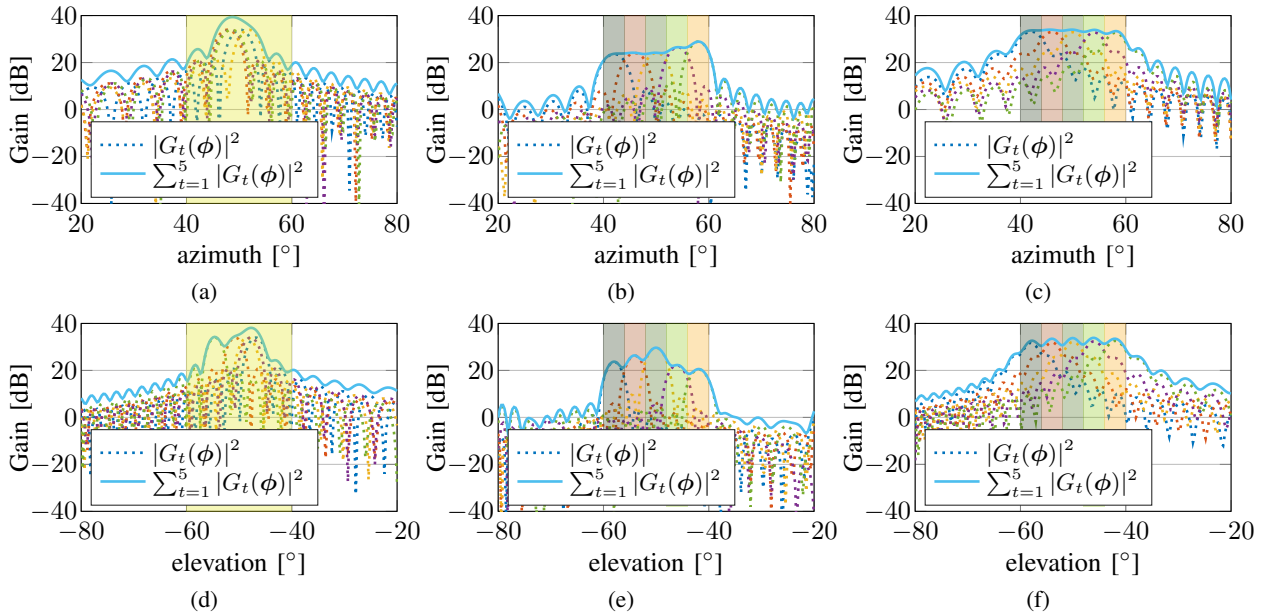

    \captionsetup[subfigure]{aboveskip=-1mm,belowskip=-1mm}
    \centering
\hspace{-3mm}
\begin{subfigure}{0.33\textwidth}
    \centering
    {\input{Figures/1D_NarAz}}
    \subcaption{}
    \label{Fig:1D_azSharp}
\end{subfigure}
\hspace{-2mm}
\begin{subfigure}{0.33\textwidth}
    \centering
    {\input{Figures/1D_OptAz.tex}}
    \subcaption{}
    \label{Fig:1D_azOpt}
\end{subfigure}
\hspace{-2mm}
\begin{subfigure}{0.33\textwidth}
    \centering
    {\input{Figures/1D_UniAz.tex}}
    \subcaption{}
    \label{Fig:1D_azUniform}
\end{subfigure}\\
\hspace{-3mm}
\begin{subfigure}{0.33\textwidth}
    \centering
    {\input{Figures/1D_NarEl}}
    \subcaption{}
    \label{Fig:1D_elSharp}
\end{subfigure}
\hspace{-2mm}
\begin{subfigure}{0.33\textwidth}
    \centering
    {\input{Figures/1D_OptEl.tex}}
    \subcaption{}
    \label{Fig:1D_elOpt}
\end{subfigure}
\hspace{-2mm}
\begin{subfigure}{0.33\textwidth}
    \centering
    {\input{Figures/1D_UniEl.tex}}
    \subcaption{}
    \label{Fig:1D_elUniform}
\end{subfigure}
    \caption{Beampatterns of the 2D \ac{ris} in the azimuth domain ((a)--(c)) with the fixed elevation $\phi_\mathrm{0}^\mathrm{el}= -50 ^\circ$, and in the elevation domain ((e)--(f)) with the fixed azimuth $\phi_\mathrm{0}^\mathrm{az}= 50 ^\circ$, aiming the highlighted UE angular uncertainty region spanning onto $[40^\circ, 60^\circ]$ in the azimuth and $[-60^\circ, 40^\circ]$ in the elevation. 
    The beampatterns in (a) and (d) were generated with directional phases; (b) and (e) with the optimized phase approach in~\cite{Mustafa_ArbitraryBeam_6G2022}; and in (c) and (f) via the proposed uniform phase in Section~\ref{sec:ProposedRISProf}.}
    \label{Fig:1D_Beampatterns}
 \vspace{-5mm}
\end{figure}
        

\begin{figure*}[t!]
\captionsetup[subfigure]{aboveskip=-1mm,belowskip=-1mm}
\centering
\begin{centering}
\begin{subfigure}{0.49\textwidth}
    \centering 
    {\includegraphics [width=.48\columnwidth]{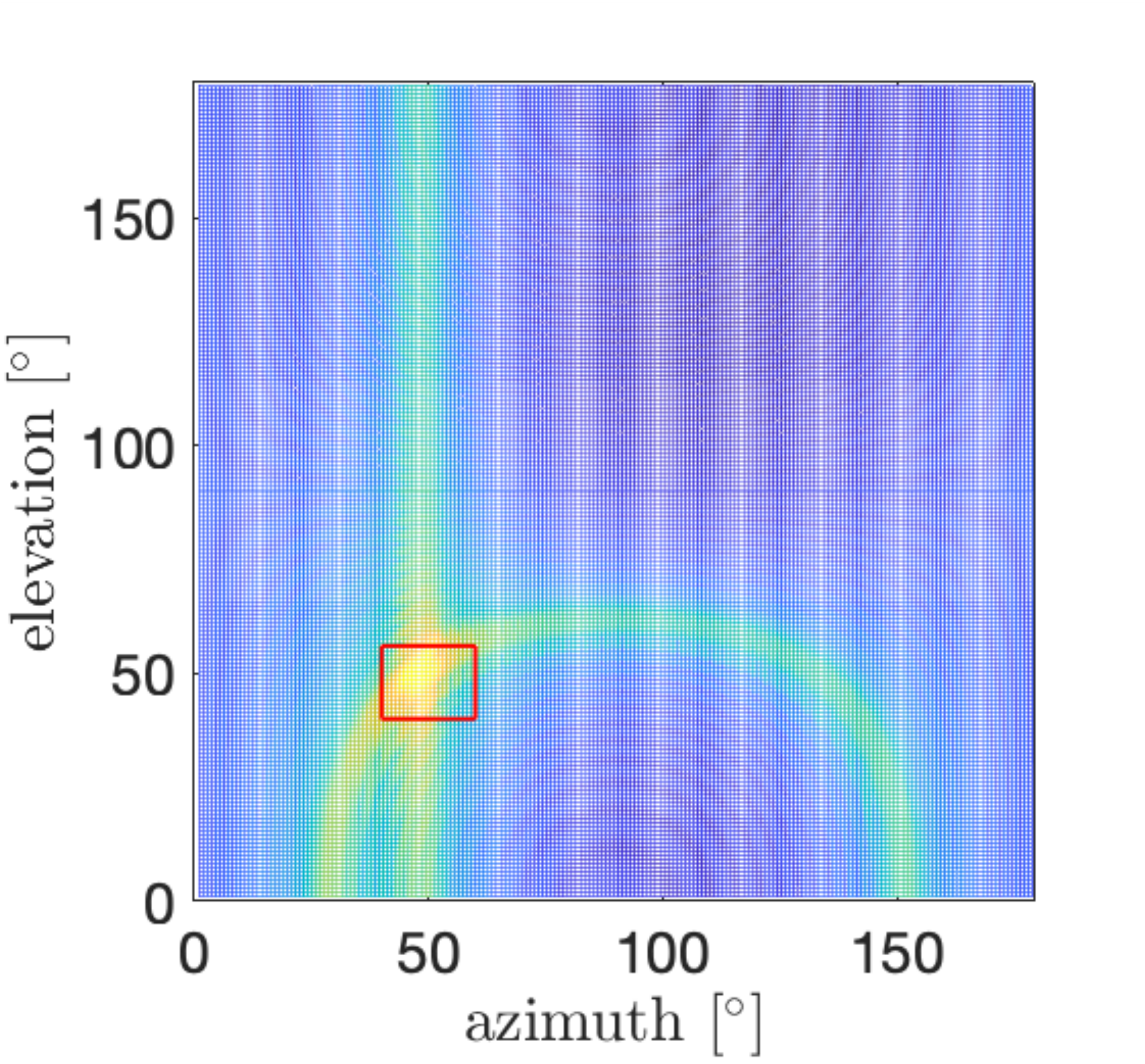}
    \includegraphics [width=.48\columnwidth]{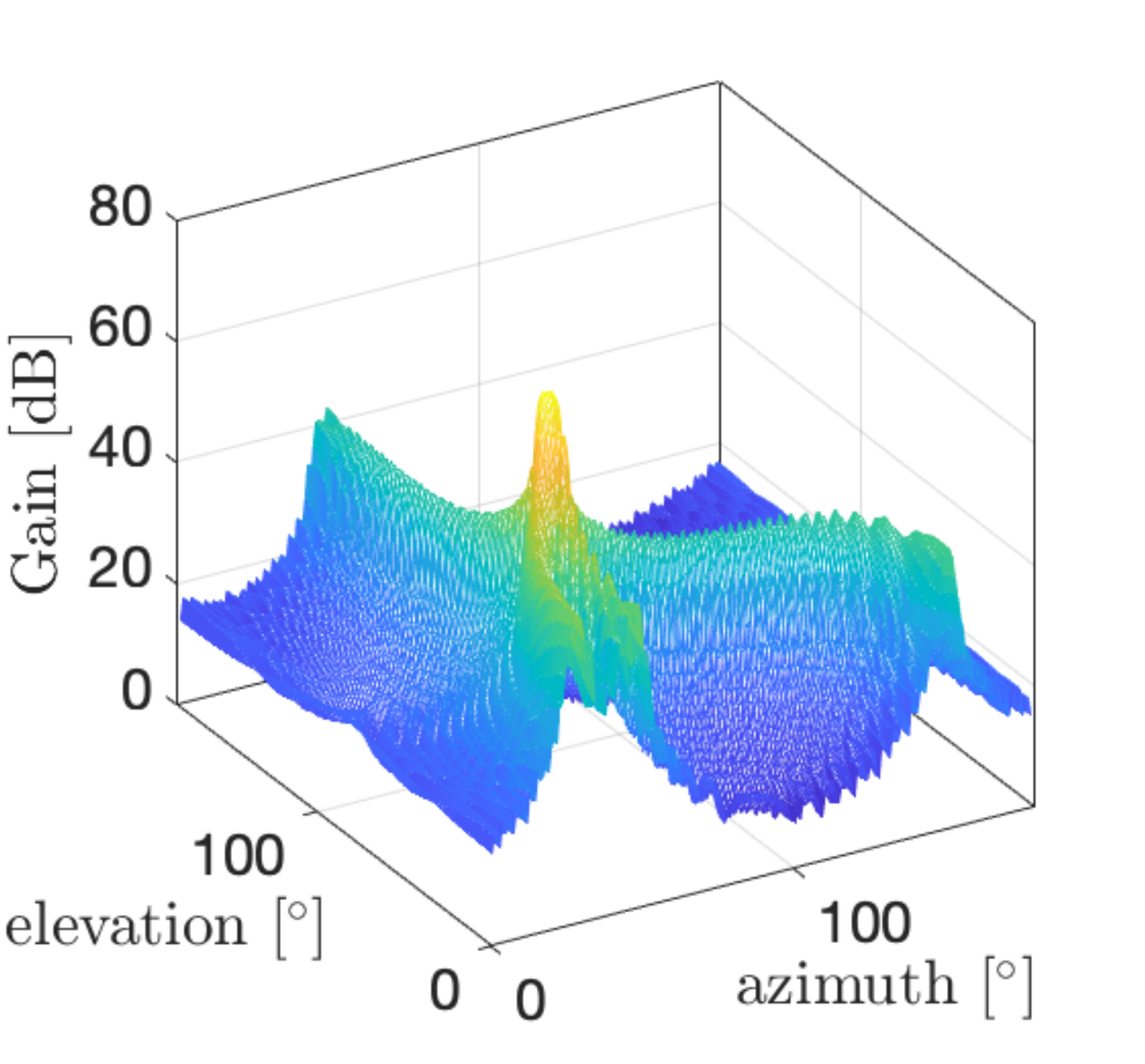}}
    \subcaption{}
    \label{Fig:2D_Narrow}
\end{subfigure}
\begin{subfigure}{0.49\textwidth}
    \centering 
    {\includegraphics [width=.48\columnwidth]{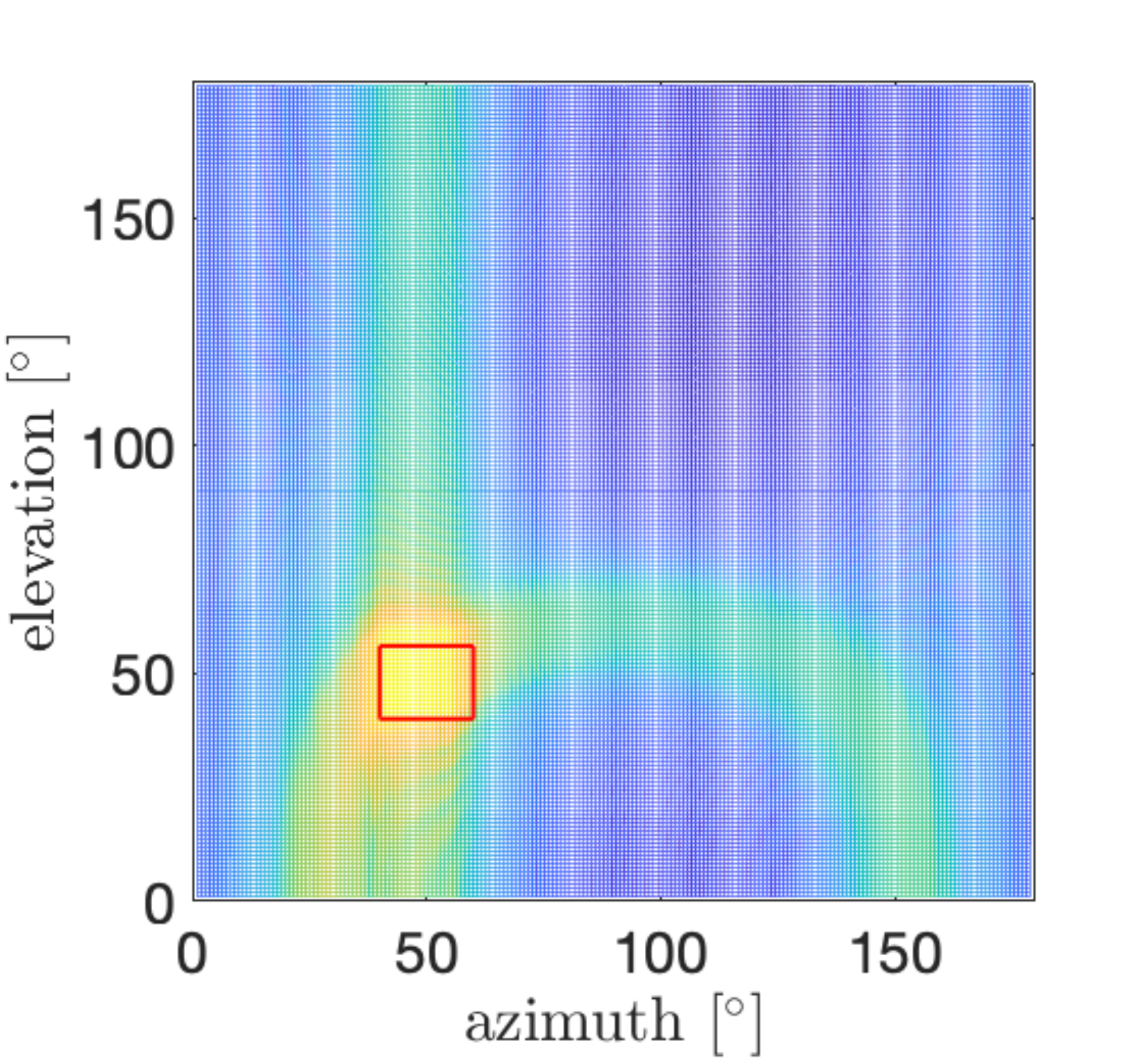}
    \includegraphics [width=.48\columnwidth]{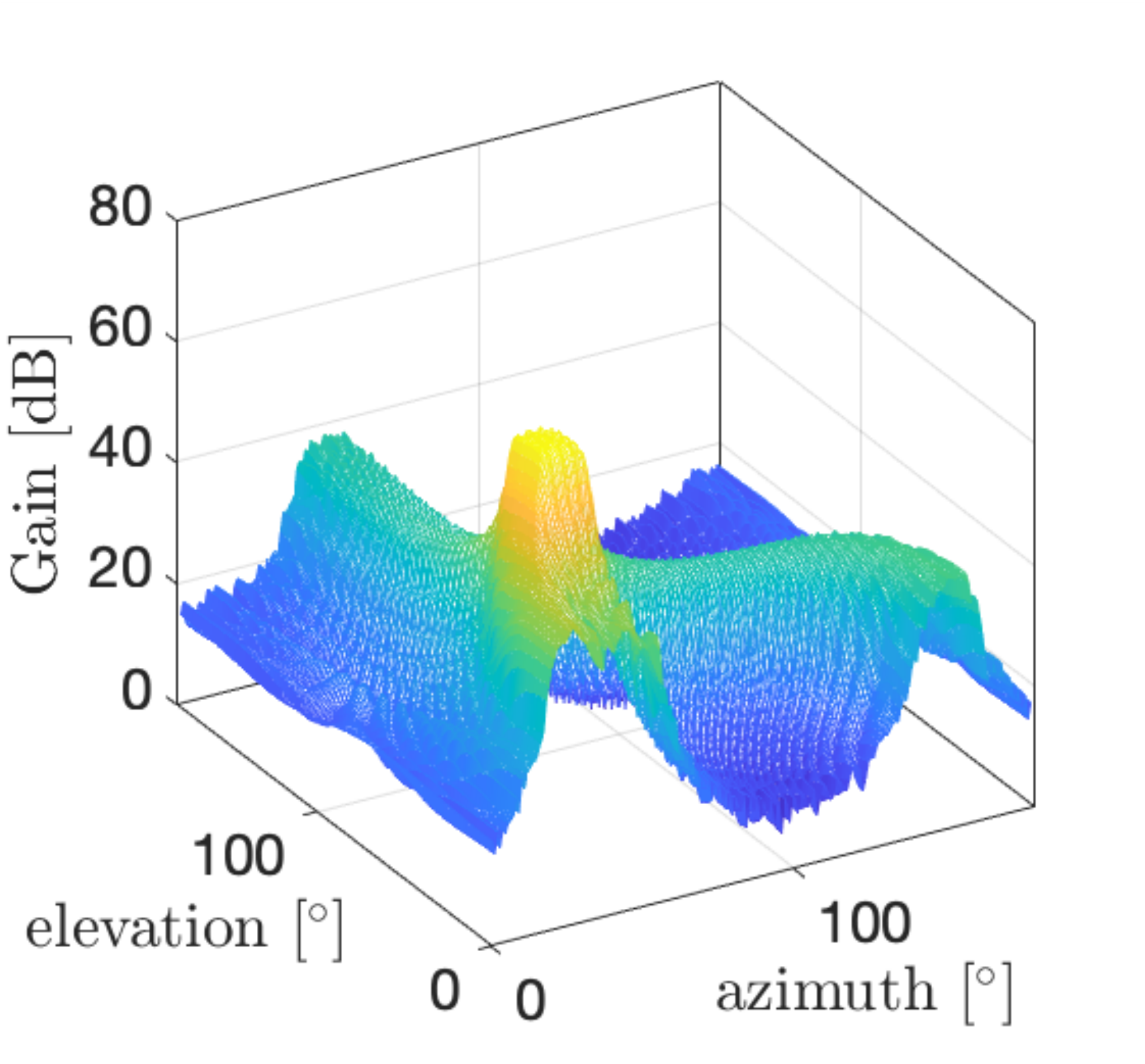}}
    \subcaption{}
    \label{Fig:2D_Uniform}
\end{subfigure}
\caption{The gain $\sum_{t=1}^{20} |G_t(\bm{\phi})|^2$ in dB for exemplary beampatterns of a 2D \ac{ris} using: (a) directional phase profiles; and (b) the proposed uniform phase profile in Section~\ref{sec:ProposedRISProf}. The red squares at left of each subfigure demonstrate the angular uncertainty that needs to be illuminated with sufficient beamforming gain.}
\vspace{-5mm}
\label{Fig:2Dbeam}
\par
\end{centering}
\end{figure*}


\vspace{-1mm}
\subsection{Positioning Performance}
\vspace{-1mm}
    As benchmark reflective beampatterns, we have used the following 
    \ac{ris} phase profiles: \textit{i}) random phases, where the individual element phases were drawn randomly from $[0,2\pi]$~(see, Section~\ref{sec:Spatial_Prof}); and \textit{ii}) the directional phases of~\cite{Kamran_RISloc_ICC2022,Kamran_RISMobility_JSTSP2022}, where the element phases were sampled from the Gaussian angular uncertainty.
    It is noted that the \acp{crlb} of the \ac{ue} state were obtained by only using the \ac{ris} path, due to the unknown data association between the measurements of the \ac{nris} paths~(i.e., the uncontrolled ones) and landmarks, loosening the gap between the \acp{crlb} and the performance of the proposed estimation approach.

\subsubsection{Beampattern Gain}\label{sec:Conv_DirProf}
    Even though the directional \ac{ris} phase profiles provide performance improvements over the random ones~\cite{Kamran_RISloc_ICC2022,Kamran_RISMobility_JSTSP2022}, the possibility that a directional beam cannot provide sufficient gain increases as the uncertainty of the \ac{ue} position prior increases. This happens due to the narrow beam width and the limited number of transmissions. To assess the beam quality, we define the following metric that quantifies the total received power at the \ac{ue}, due to the \ac{ris} reflection, with a given \ac{ris} phase profile
    over $T$ transmission intervals: $ G \triangleq \sum_{{t}=1}^{T} \lvert G_{{t}}(\bm{\phi}_{k}) \rvert^2$,
    where $G_{{t}}(\bm{\phi}_{k}) = {\bm{\omega}}_{{t},k}^\top \mathbf{b}(\bm{\phi}_k)$.
    
    Aiming to cover the highlighted UE angular uncertainty region spanning onto $[40^\circ, 60^\circ]$ in the azimuth and $[-60^\circ, 40^\circ]$ in the elevation, Fig.~\ref{Fig:1D_Beampatterns} considers a $T=5$ transmission interval and illustrates the beampatterns of a 2D \ac{ris} in the azimuth domain~(Fig.~\ref{Fig:1D_azSharp}--~\ref{Fig:1D_azUniform}) with the fixed elevation $\phi_\mathrm{0}^\mathrm{el}= -50 ^\circ$, and in the elevation domain~(Fig.~\ref{Fig:1D_elSharp}--~\ref{Fig:1D_elUniform}) with the fixed azimuth $\phi_\mathrm{0}^\mathrm{az}= 50 ^\circ$. 
    The beampatterns are visualized for the directional phases~(Fig.~\ref{Fig:1D_azSharp} and \ref{Fig:1D_elSharp}), the optimized beam design of~\cite{Mustafa_ArbitraryBeam_6G2022}~(Fig.~\ref{Fig:1D_azOpt} and \ref{Fig:1D_elOpt}), and the proposed adaptive beams in Section~~\ref{sec:ProposedRISProf}~(Fig.~\ref{Fig:1D_azUniform} and \ref{Fig:1D_elUniform}). As it can be seen from the resulting sharp beams, the directional phases cannot uniformly illuminate the desired angular sector, and thus, the \ac{ue} cannot be provided with a sufficient gain if it lies close to the edges of the uncertainty region. 
	The \ac{ris} phase optimization in~\cite{Mustafa_ArbitraryBeam_6G2022} enables generating beampatterns that are similar to rectangular pulse shapes, where the gain over the desired angular sector is maximized. In this case, a resulting gain of more than 20 dB is achievable within the desired sector.	However, it comes at the cost of increased computational complexity due to the required iterative optimization. On the other hand, the proposed \ac{ris} phase configurations enable uniform illumination of the uncertainty angular region in a more efficient way than the approach in~\cite{Mustafa_ArbitraryBeam_6G2022}, in terms of achievable power gain and computational complexity. 
    This happens because it is relatively harder to find the global optimum point as the UE prior is larger.
    Finally, Fig.~\ref{Fig:2Dbeam} includes the visualization of the sum of beampatterns over $T=20$ transmissions, as generated by the directional phases~(Fig.~\ref{Fig:2D_Narrow}) and the proposed ones~(Fig.~\ref{Fig:2D_Uniform}).
    The desired angular sector 
    is marked by red squares in the left subfigures in Figs.~\ref{Fig:2D_Narrow} and~\ref{Fig:2D_Uniform}.
    The results confirm that the entire angular sector can be illuminated by the proposed phase profiles with sufficient power; this does not happen, however, using the directional profiles.


\begin{figure*}[t!]
\captionsetup[subfigure]{aboveskip=-1mm,belowskip=-1mm}
\centering
\begin{centering}
\begin{subfigure}{0.49\textwidth}
    {
%
\definecolor{mycolor1}{rgb}{0.46667,0.67451,0.18824}%
\begin{tikzpicture}

\begin{axis}[%
width=67mm,
height=30mm,
at={(0mm,0mm)},
scale only axis,
xmin=0,
xmax=10,
ymin=0.1,
ymax=100,
ymode=log,
yticklabel style = {font=\small,xshift=0.5ex},
xticklabel style = {font=\small,yshift=0ex},
axis background/.style={fill=white},
xmajorgrids,
ymajorgrids,
legend style={legend pos=south east,legend cell align=left, align=left, draw=white!15!black,style={row sep=-0.1cm}}
]

\addplot [color=mycolor1, line width=1.0pt, mark=+, mark options={solid, mycolor1}]
  table[row sep=crcr]{%
0.01	8.65877159329929\\
0.1	8.65877159329929\\
0.2	8.65877159329929\\
0.3	8.65877159329929\\
0.4	8.65877159329929\\
0.5	8.65877159329929\\
0.6	8.65877159329929\\
0.7	8.65877159329929\\
0.8	8.65877159329929\\
0.9	8.65877159329929\\
1	8.65877159329929\\
2	8.65877159329929\\
3	8.65877159329929\\
4	8.65877159329929\\
5	8.65877159329929\\
6	8.65877159329929\\
7	8.65877159329929\\
8	8.65877159329929\\
9	8.65877159329929\\
10	8.65877159329929\\
};
\addlegendentry{\footnotesize{Random profile}}

\addplot [color=blue, line width=1.0pt, mark=o, mark options={solid, blue}]
  table[row sep=crcr]{%
0.01	38.1291485915582\\
0.1	3.83817300437474\\
0.2	1.95623600895186\\
0.3	1.34602295007486\\
0.4	1.05487735647951\\
0.5	0.892688458674215\\
0.6	0.796468179805877\\
0.7	0.739569093660476\\
0.8	0.708958941989795\\
0.9	0.697587462226162\\
1	0.70123022078779\\
2	0.933885725560279\\
3	1.40506403501427\\
4	2.57523978493561\\
5	3.41729525539841\\
6	4.5226780275594\\
7	7.55794201972935\\
8	9.35799551929238\\
9	12.9557113685996\\
10	15.3668815728465\\
};
\addlegendentry{\footnotesize Directional profile}

\addplot [color=red, dashed, line width=1.0pt, mark=x, mark options={solid, red}]
  table[row sep=crcr]{%
0.01	20.5525006524443\\
0.1	2.07428346391296\\
0.2	1.06005427858459\\
0.3	0.731045941539933\\
0.4	0.57377833331909\\
0.5	0.485603862492259\\
0.6	0.432307643032828\\
0.7	0.399204731442262\\
0.8	0.378952818858566\\
0.9	0.367520085013714\\
1	0.362576804618955\\
2	0.517474278738619\\
3	0.884428595977512\\
4	1.39128913922309\\
5	1.98592806433056\\
6	2.58726733523046\\
7	2.96151556131542\\
8	3.33970062513529\\
9	3.74158112155263\\
10	5.74652283463067\\
};
\addlegendentry{\footnotesize Uniform profile (proposed)}

\end{axis}

\node[rotate=0,fill=white] (BOC6) at (3.35cm,-.7cm){\small UE uncertainty [m]};
\node[rotate=90] at (-8mm,12.5mm){\small PEB [m]};

\end{tikzpicture}%
}
    \subcaption{}
    \label{Fig:Ini_Perform}
\end{subfigure}
\begin{subfigure}{0.49\textwidth}
    {
%
\definecolor{mycolor1}{rgb}{0.46667,0.67451,0.18824}%
\begin{tikzpicture}

\begin{axis}[%
width=67mm,
height=30mm,
at={(0mm,0mm)},
scale only axis,
xmin=20,
xmax=100,
ymin=1,
ymax=100,
ymode=log,
yticklabel style = {font=\small,xshift=0.5ex},
xticklabel style = {font=\small,yshift=0ex},
axis background/.style={fill=white},
xmajorgrids,
ymajorgrids,
legend style={legend cell align=left, align=left, draw=white!15!black,style={row sep=-0.1cm}}
]

\addplot [color=mycolor1, line width=1.0pt, mark=+, mark options={solid, mycolor1}]
  table[row sep=crcr]{%
20	8.65877159329929\\
40	5.45657546041957\\
64	4.7087013324276\\
90	4.20532198726261\\
100	3.97266070147064\\
};
\addlegendentry{\footnotesize{Random profile}}

\addplot [color=blue, line width=1.0pt, mark=o, mark options={solid, blue}]
  table[row sep=crcr]{%
20	15.3668815728465\\
40	4.23427724208612\\
64	1.90806115650633\\
90	1.45400642865742\\
100	1.35054938111432\\
};
\addlegendentry{\footnotesize Directional profile}

\addplot [color=red, dashed, line width=1.0pt, mark=x, mark options={solid, red}]
  table[row sep=crcr]{%
20	5.74652283463067\\
40	2.47776676616586\\
64	1.93821273542773\\
90	1.36946412244752\\
100	1.39888672236702\\
};
\addlegendentry{\footnotesize Uniform profile (proposed)}

\end{axis}


\node[rotate=0,fill=white] (BOC6) at (3.35cm,-.65cm){\small \# transmissions};
\node[rotate=90] at (-8mm,15mm){\small PEB [m]};

\end{tikzpicture}%
}
    \subcaption{}
    \label{Fig:OFDM_Perform}
\end{subfigure}
    \caption{PEBs on the \ac{ue} state as function of (a) the \ac{ue} uncertainties in meters for $T=20$ transmission intervals; and (b) the number of transmission intervals for the case where the \ac{ue} position uncertainty is 10~m.
	}
\vspace{-5mm}
\label{Fig:PEBs}
\par
\end{centering}
\end{figure*}
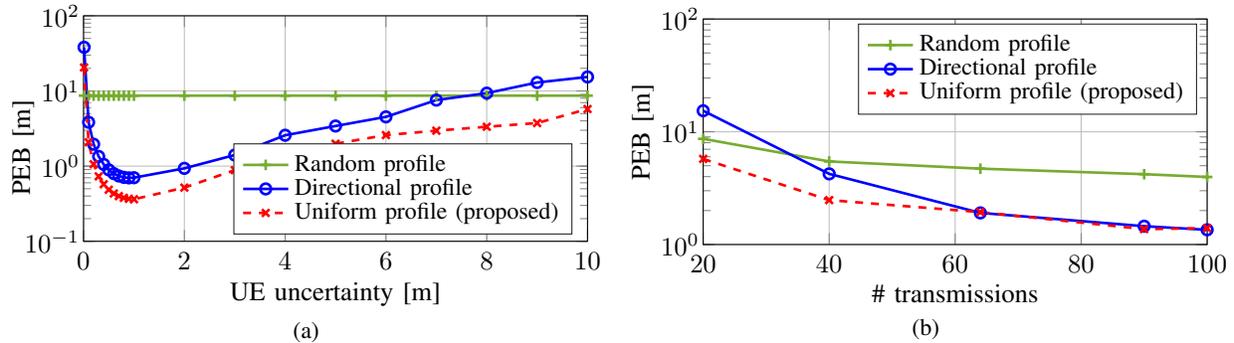


\begin{figure}[t!]
\captionsetup[subfigure]{aboveskip=-1mm,belowskip=-1mm}
\centering
\begin{centering}
\begin{subfigure}{0.49\textwidth}
\centering
    {\includegraphics [width=.95\columnwidth]{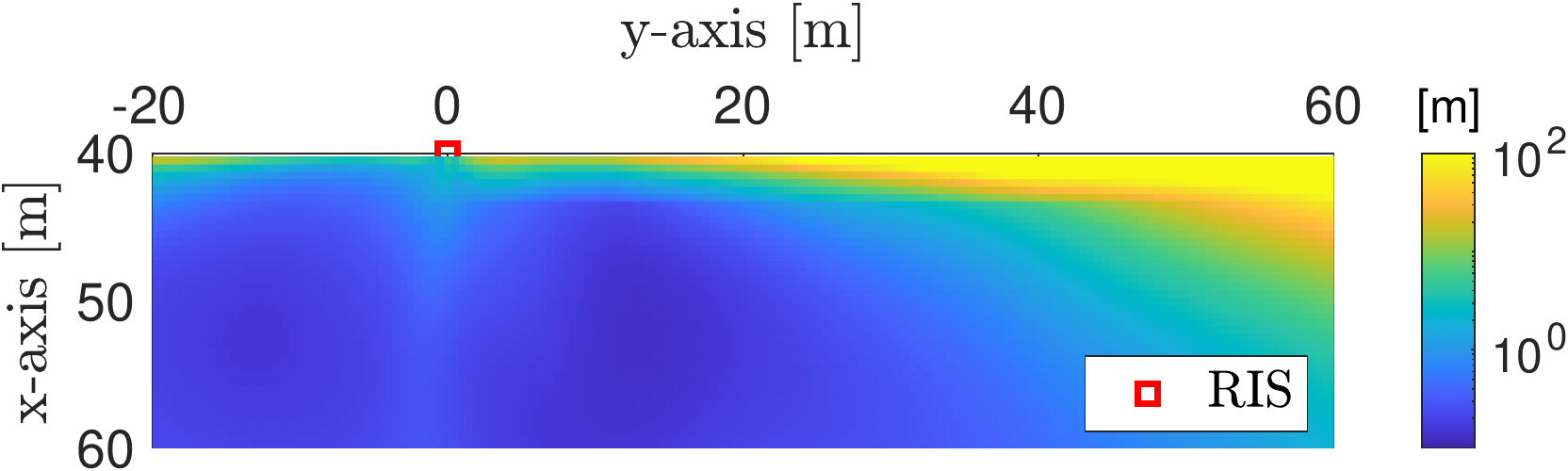}}
    \subcaption{}
    \label{Fig:visual_PEB}
\end{subfigure}
\begin{subfigure}{0.49\textwidth}
\centering
    {\includegraphics [width=.95\columnwidth]{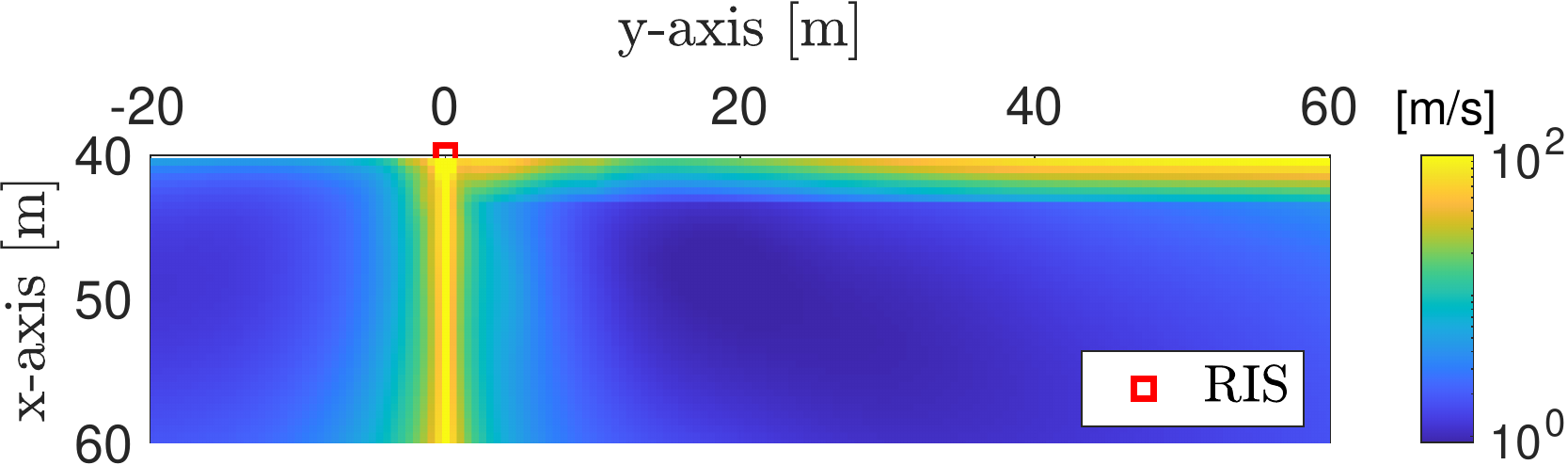}}
    \subcaption{}
    \label{Fig:visual_SEB}
\end{subfigure}
	\caption{Visualization of the (a) PEB and (b) SEB with different \ac{ue} positions, using only the \ac{ris} path and $T=20$ transmission intervals. The \ac{ue} position uncertainty was set to 1~m.}
\vspace{-5mm}
\label{Fig:visual_UE}
\par
\end{centering}
\end{figure}

\begin{figure}[t!]
\captionsetup[subfigure]{aboveskip=-1mm,belowskip=-1mm}
\hspace{-2mm}
\begin{subfigure}{0.49\textwidth}
    {
%
%
\definecolor{mycolor1}{rgb}{0.46667,0.67451,0.18824}%

\begin{tikzpicture}
\begin{axis}[%
width=67mm,
height=30mm,
at={(0mm,0mm)},
scale only axis,
xmin=0,
xmax=20,
ymin=1e-2,
ymax=12,
ymode=log,
yticklabel style = {font=\small,xshift=0.5ex},
xticklabel style = {font=\small,yshift=0ex},
axis background/.style={fill=white},
axis background/.style={fill=white},
xmajorgrids,
ymajorgrids,
legend style={legend pos=south east, legend cell align=left, align=left, draw=white!15!black,style={row sep=-0.1cm}}
]
\addplot [color=mycolor1, line width=1.0pt, mark=+, mark options={solid, mycolor1}]
  table[row sep=crcr]{%
0	10.17398890293\\
1	5.63319531337099\\
2	3.50578834188848\\
3	2.43708054048907\\
4	1.50838897578263\\
5	1.60081690407719\\
6	1.26521723672566\\
7	1.24346592596176\\
8	1.2130120751119\\
9	1.24115734754726\\
10	1.24428478709886\\
11	1.25088978509903\\
12	1.28108857230273\\
13	1.28512058557187\\
14	1.29129821764236\\
15	1.30973749447731\\
16	1.32404495351723\\
17	1.32825277284891\\
18	1.34132936005669\\
19	1.35037496313692\\
20	1.35758196970716\\
};
\addlegendentry{\footnotesize{Random profile}}

\addplot [color=blue, line width=1.0pt, mark=o, mark options={solid, blue}]
  table[row sep=crcr]{%
0	10.17398890293\\
1	5.23740362681919\\
2	2.76855240216738\\
3	1.41875728570019\\
4	0.669461700115266\\
5	0.554681908265174\\
6	0.381797048993645\\
7	0.304162228514755\\
8	0.301498257268999\\
9	0.312570163991554\\
10	0.344780259281012\\
11	0.366392445143578\\
12	0.388520581615144\\
13	0.409763408610985\\
14	0.425348898250867\\
15	0.457348756575948\\
16	0.48770557186908\\
17	0.506906466303236\\
18	0.531094115710629\\
19	0.551366287945949\\
20	0.57906011794559\\
};
\addlegendentry{\footnotesize Directional profile}

\addplot [color=red, dashed, line width=1.0pt, mark=x, mark options={solid, red}]
  table[row sep=crcr]{%
0	10.17398890293\\
1	3.66978655525223\\
2	1.19910071019329\\
3	0.3925466902136\\
4	0.119918136785063\\
5	0.0616038874825896\\
6	0.0598018219094055\\
7	0.0470645712073228\\
8	0.0473390982572155\\
9	0.0680525057955013\\
10	0.103415656283854\\
11	0.1297185592882\\
12	0.164775329839123\\
13	0.190708807128822\\
14	0.204524830908851\\
15	0.234531186627942\\
16	0.259563621826579\\
17	0.275456515568049\\
18	0.29922107281187\\
19	0.317647521548285\\
20	0.344373660070324\\
};
\addlegendentry{\footnotesize Uniform profile (proposed)}
\end{axis}

 \node[rotate=0,fill=white] (BOC6) at (3.35cm,-.65cm){\small time $k$};
\node[rotate=90] at (-10mm,15mm){\small 
 Position MAE [m]};
\end{tikzpicture}%
}
    \subcaption{}
    \label{Fig:time_Pos}
\end{subfigure}
\begin{subfigure}{0.49\textwidth}
    {
%
%
\definecolor{mycolor1}{rgb}{0.46667,0.67451,0.18824}%

\begin{tikzpicture}
\begin{axis}[%
width=67mm,
height=30mm,
at={(0mm,0mm)},
scale only axis,
xmin=0,
xmax=20,
ymin=1e-3,
ymax=1,
ymode=log,
yticklabel style = {font=\small,xshift=0.5ex},
xticklabel style = {font=\small,yshift=0ex},
axis background/.style={fill=white},
axis background/.style={fill=white},
xmajorgrids,
ymajorgrids,
legend style={legend cell align=left, align=left, draw=white!15!black,style={row sep=-0.1cm}}
]
\addplot [color=mycolor1, line width=1.0pt, mark=+, mark options={solid, mycolor1}]
  table[row sep=crcr]{%
0	0.174720400779609\\
1	0.174603936542137\\
2	0.171942658093947\\
3	0.156137468083029\\
4	0.127478375572262\\
5	0.0231774519335173\\
6	0.0044685989061608\\
7	0.00197591388189541\\
8	0.00155657172051934\\
9	0.00148718847005513\\
10	0.00143508199294681\\
11	0.00136585587114801\\
12	0.00114805982578441\\
13	0.00120919625138427\\
14	0.00126421376182882\\
15	0.00126626872781077\\
16	0.0011998148332707\\
17	0.00112432412219915\\
18	0.00119667381368147\\
19	0.00125907874935939\\
20	0.00131774156691152\\
};
\addlegendentry{\footnotesize{Random profile}}

\addplot [color=blue, line width=1.0pt, mark=o, mark options={solid, blue}]
  table[row sep=crcr]{%
0	0.174720400779609\\
1	0.174584128176331\\
2	0.167950233884614\\
3	0.122908905501972\\
4	0.072444450688885\\
5	0.0151579812382084\\
6	0.00447594257891766\\
7	0.00208043005082134\\
8	0.0013892894560622\\
9	0.00133169580077849\\
10	0.0012726286144023\\
11	0.00126124095346162\\
12	0.00108312996327991\\
13	0.00112943961199466\\
14	0.0012038398638132\\
15	0.00119597967689671\\
16	0.00114869469461144\\
17	0.00107838090522691\\
18	0.00114477951525266\\
19	0.00120018063189694\\
20	0.00125123289865971\\
};
\addlegendentry{\footnotesize Directional profile}

\addplot [color=red, dashed, line width=1.0pt, mark=x, mark options={solid, red}]
  table[row sep=crcr]{%
0	0.174720400779609\\
1	0.174683722647439\\
2	0.161494059533308\\
3	0.0992956452444846\\
4	0.051391478050147\\
5	0.00494532812301097\\
6	0.00194946910636306\\
7	0.00135048020332199\\
8	0.00113187132643139\\
9	0.00119840508496019\\
10	0.00117208029963697\\
11	0.00122111047925345\\
12	0.00110236933211636\\
13	0.00111759601280564\\
14	0.00123948106521611\\
15	0.00116880878058015\\
16	0.00116143900325434\\
17	0.00107605441292477\\
18	0.00111966552446532\\
19	0.00117767096294886\\
20	0.00121856593210923\\
};
\addlegendentry{\footnotesize Uniform profile (proposed)}
\end{axis}

\node[rotate=0,fill=white] (BOC6) at (3.35cm,-.65cm){\small time $k$};
\node[rotate=90] at (-10mm,15mm){\small Heading MAE [rad]};
\end{tikzpicture}%
}
    \subcaption{}
    \label{Fig:time_Head}
\end{subfigure}
\caption{Localization performance versus the time steps for $T=20$ transmission intervals: \acp{mae} of the (a) position; and (b) heading estimates.}
\vspace{-6mm}
\label{Fig:time_5DLoc}
\par
\end{figure}
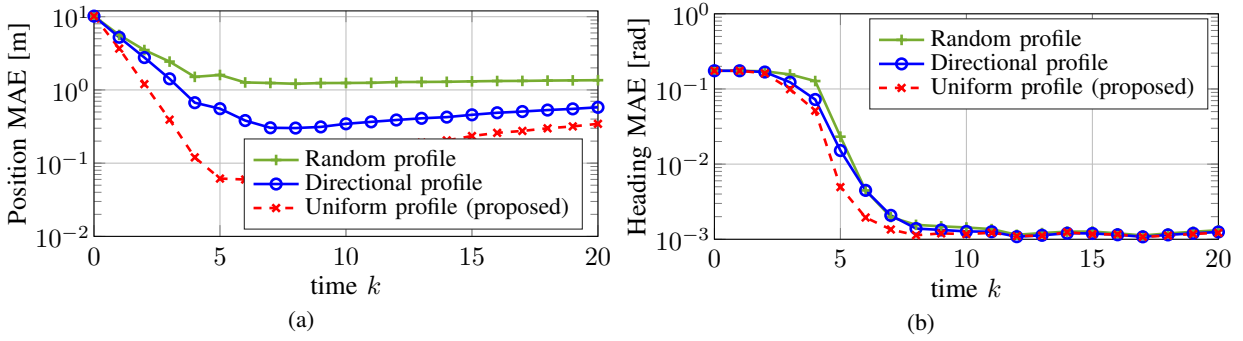

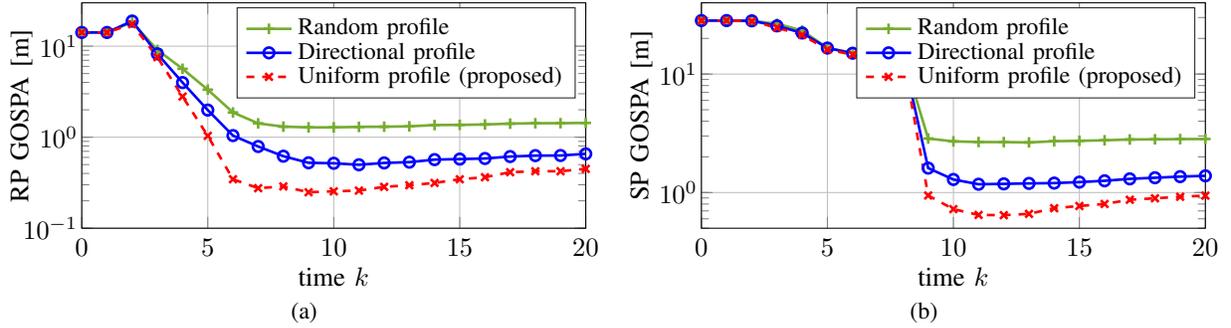
\begin{figure}[t!]
\captionsetup[subfigure]{aboveskip=-1mm,belowskip=-1mm}
\centering
\hspace{-2mm}
\begin{subfigure}{0.49\textwidth}
    \centering
    {
%
%
\definecolor{mycolor1}{rgb}{0.46667,0.67451,0.18824}%

\begin{tikzpicture}

\begin{axis}[%
width=67mm,
height=30mm,
at={(0mm,0mm)},
scale only axis,
xmin=0,
xmax=20,
ymin=0.1,
ymax=30,
ymode=log,
yticklabel style = {font=\small,xshift=0.5ex},
xticklabel style = {font=\small,yshift=0ex},
axis background/.style={fill=white},
axis background/.style={fill=white},
xmajorgrids,
ymajorgrids,
legend style={legend cell align=left, align=left, draw=white!15!black,style={row sep=-0.1cm}}
]

\addplot [color=mycolor1, line width=1.0pt, mark=+, mark options={solid, mycolor1}]
  table[row sep=crcr]{%
0	14.1421356237309\\
1	14.1421356237309\\
2	18.3391161601545\\
3	9.0955531581174\\
4	5.63732034081161\\
5	3.32894705204925\\
6	1.87008900422584\\
7	1.41851757922041\\
8	1.30548393631564\\
9	1.28217284178726\\
10	1.28294711279296\\
11	1.2978334545216\\
12	1.30414122820148\\
13	1.31628313120439\\
14	1.35916007203526\\
15	1.36583640339286\\
16	1.37883236187327\\
17	1.40923014236524\\
18	1.42346729119116\\
19	1.42515362366728\\
20	1.43453460472136\\
};
\addlegendentry{\footnotesize{Random profile}}

\addplot [color=blue, line width=1.0pt, mark=o, mark options={solid, blue}]
  table[row sep=crcr]{%
0	14.1421356237309\\
1	14.1421356237309\\
2	18.8806202737549\\
3	8.2386973363075\\
4	3.98282992274254\\
5	1.98586911328517\\
6	1.0413555669352\\
7	0.790407319218241\\
8	0.618426611686808\\
9	0.523328215374523\\
10	0.516955676517564\\
11	0.497844927546767\\
12	0.521439089588655\\
13	0.531852409154976\\
14	0.56721151861425\\
15	0.574820700195456\\
16	0.583335102485883\\
17	0.612865265423939\\
18	0.627459822763889\\
19	0.629219107614974\\
20	0.656145446708414\\
};
\addlegendentry{\footnotesize Directional profile}

\addplot [color=red, dashed, line width=1.0pt, mark=x, mark options={solid, red}]
  table[row sep=crcr]{%
0	14.1421356237309\\
1	14.1421356237309\\
2	17.4464453864661\\
3	7.57430369422146\\
4	2.79517230725079\\
5	1.03510512786881\\
6	0.345241202594345\\
7	0.275429567059698\\
8	0.287892620392698\\
9	0.248390369565978\\
10	0.254248020093312\\
11	0.259185182855801\\
12	0.284770445566436\\
13	0.296453497798654\\
14	0.314340544157455\\
15	0.344962072141179\\
16	0.363359400272258\\
17	0.412518949184862\\
18	0.422711124593967\\
19	0.422635748738772\\
20	0.449014808540448\\
};
\addlegendentry{\footnotesize Uniform profile (proposed)}
\end{axis}

\node[rotate=0,fill=white] (BOC6) at (3.35cm,-.65cm){\small time $k$};
\node[rotate=90] at (-8mm,15mm){\small RP GOSPA [m]};
\end{tikzpicture}%
}
    \subcaption{}
    \label{Fig:time_RP}
\end{subfigure}
\begin{subfigure}{0.49\textwidth}
    \centering
    {
%
%
\definecolor{mycolor1}{rgb}{0.46667,0.67451,0.18824}%

\begin{tikzpicture}

\begin{axis}[%
width=67mm,
height=30mm,
at={(0mm,0mm)},
scale only axis,
xmin=0,
xmax=20,
ymin=0.5,
ymax=40,
ymode=log,
yticklabel style = {font=\small,xshift=0.5ex},
xticklabel style = {font=\small,yshift=0ex},
axis background/.style={fill=white},
axis background/.style={fill=white},
xmajorgrids,
ymajorgrids,
legend style={legend cell align=left, align=left, draw=white!15!black,style={row sep=-0.1cm}}
]

\addplot [color=mycolor1, line width=1.0pt, mark=+, mark options={solid, mycolor1}]
  table[row sep=crcr]{%
0	28.2842712474617\\
1	28.2842712474617\\
2	28.1589635632483\\
3	26.6056434605984\\
4	23.2671133787047\\
5	16.6429906882995\\
6	15.0849535823066\\
7	14.5788469715943\\
8	14.3990306478028\\
9	2.85679009016383\\
10	2.70331126820888\\
11	2.67004478107174\\
12	2.66570687379182\\
13	2.64846239277442\\
14	2.7124463693882\\
15	2.7338255950043\\
16	2.76149689770404\\
17	2.80478530027318\\
18	2.81362274455022\\
19	2.82558625953664\\
20	2.83179324695124\\
};
\addlegendentry{\footnotesize{Random profile}}

\addplot [color=blue, line width=1.0pt, mark=o, mark options={solid, blue}]
  table[row sep=crcr]{%
0	28.2842712474617\\
1	28.2842712474617\\
2	28.1392661332627\\
3	25.5209941481803\\
4	22.2120044343188\\
5	16.5739664942646\\
6	14.9958176040194\\
7	14.6528301027096\\
8	14.4538426873092\\
9	1.60808648233716\\
10	1.28901718209757\\
11	1.17516100441585\\
12	1.18402124405917\\
13	1.19436427390061\\
14	1.20163652147993\\
15	1.22477485930104\\
16	1.25666867102599\\
17	1.30492672435358\\
18	1.33330919274044\\
19	1.36268359672463\\
20	1.3841652364264\\
};
\addlegendentry{\footnotesize Directional profile}

\addplot [color=red, dashed, line width=1.0pt, mark=x, mark options={solid, red}]
  table[row sep=crcr]{%
0	28.2842712474617\\
1	28.2842712474617\\
2	27.8514023561685\\
3	24.6338046465948\\
4	21.3726951892167\\
5	16.0834111212272\\
6	14.5886765418306\\
7	14.2428285244408\\
8	14.1774886233974\\
9	0.944440803773932\\
10	0.72470264758502\\
11	0.647343411445377\\
12	0.644099214146357\\
13	0.660919952898282\\
14	0.737610905338954\\
15	0.770321170650626\\
16	0.800738875170354\\
17	0.869525414145898\\
18	0.894468391174574\\
19	0.921648777997215\\
20	0.941900036653275\\
};
\addlegendentry{\footnotesize Uniform profile (proposed)}
\end{axis}

\node[rotate=0,fill=white] (BOC6) at (3.35cm,-.65cm){\small time $k$};
\node[rotate=90] at (-8mm,15mm){\small SP GOSPA [m]};
\end{tikzpicture}%
}
    \subcaption{}
    \label{Fig:time_SP}
\end{subfigure}
\caption{Mapping performance 
for $T=20$ transmission intervals: \ac{gospa} distances of the (a) \ac{rp} and (b) \ac{sp}.\hfill~}
\vspace{-5mm}
\label{Fig:Mapping_time}
\par
\end{figure}

\begin{figure}[t!]
\captionsetup[subfigure]{aboveskip=-1mm,belowskip=-1mm}
\centering
\hspace{-2mm}
\begin{subfigure}{0.49\textwidth}
    \centering
    {
%
%
\definecolor{mycolor1}{rgb}{0.46667,0.67451,0.18824}%

\begin{tikzpicture}

\begin{axis}[%
width=67mm,
height=30mm,
at={(0mm,0mm)},
scale only axis,
xmin=0,
xmax=20,
ymin=1e-2,
ymax=16,
ymode=log,
yticklabel style = {font=\small,xshift=0.5ex},
xticklabel style = {font=\small,yshift=0ex},
axis background/.style={fill=white},
axis background/.style={fill=white},
xmajorgrids,
ymajorgrids,
legend style={legend cell align=left, align=left, draw=white!15!black,style={row sep=-0.1cm}}
]

\addplot [color=blue, line width=1.0pt, mark=o, mark options={solid, blue}]
  table[row sep=crcr]{%
0	5.04814464312805\\
1	5.04814464312805\\
2	2.7663083556085\\
3	1.04771618277245\\
4	0.523985770063641\\
5	0.380614395584846\\
6	0.311180113503674\\
7	0.29312763235861\\
8	0.289503845204544\\
9	0.282017071442709\\
10	0.275417881554131\\
11	0.280949550070587\\
12	0.279541236194602\\
13	0.280281738270354\\
14	0.291217803303001\\
15	0.27937354527315\\
16	0.279911183890345\\
17	0.286477154275133\\
18	0.299556867690522\\
19	0.297862334895851\\
20	0.289851365859702\\
};
\addlegendentry{\footnotesize w/o Doppler shift}

\addplot [color=red, dashed, line width=1.0pt, mark=x, mark options={solid, red}]
  table[row sep=crcr]{%
0	5.04814464312805\\
1	4.80108784245706\\
2	2.45050954417892\\
3	0.85228517759386\\
4	0.440129181333366\\
5	0.079384667378811\\
6	0.0461106893551595\\
7	0.0624776241364016\\
8	0.0502310329560575\\
9	0.043790606517369\\
10	0.0393284518744248\\
11	0.0480914099719975\\
12	0.0638732310762264\\
13	0.0486331279462773\\
14	0.0403051040827605\\
15	0.0466799645289435\\
16	0.0701065239157468\\
17	0.0624079495267382\\
18	0.0559663713577612\\
19	0.0507040865486245\\
20	0.0647568508475907\\
};
\addlegendentry{\footnotesize with Doppler shift}
\end{axis}

\node[rotate=0,fill=white] (BOC6) at (3.35cm,-.65cm){\small time $k$};
\node[rotate=90] at (-10mm,15mm){\small Speed MAE [m/s]};
\end{tikzpicture}%
}
    \subcaption{}
    \label{Fig:Doppler_5DLoc}
\end{subfigure}
\hspace{-2mm}
\begin{subfigure}{0.49\textwidth}
    \centering
    {
%
%
\definecolor{mycolor1}{rgb}{0.46667,0.67451,0.18824}%

\begin{tikzpicture}

\begin{axis}[%
width=67mm,
height=30mm,
at={(0mm,0mm)},
scale only axis,
xmin=0,
xmax=20,
ymin=0.1,
ymax=30,
ymode=log,
yticklabel style = {font=\small,xshift=0.5ex},
xticklabel style = {font=\small,yshift=0ex},
axis background/.style={fill=white},
axis background/.style={fill=white},
xmajorgrids,
ymajorgrids,
legend style={legend cell align=left, align=left, draw=white!15!black,style={row sep=-0.1cm}}
]

\addplot [color=blue, line width=1.0pt, mark=o, mark options={solid, blue}]
  table[row sep=crcr]{%
0	14.1421356237309\\
1	14.1421356237309\\
2	17.486269019598\\
3	7.85184674174991\\
4	3.3091957309127\\
5	1.40801504219962\\
6	0.730917743004414\\
7	0.592271252638197\\
8	0.508379310722194\\
9	0.455015324476051\\
10	0.450865409272845\\
11	0.457651998978801\\
12	0.498708213900761\\
13	0.523621326443391\\
14	0.553806538476066\\
15	0.5928093736436\\
16	0.647493571306817\\
17	0.674782870545179\\
18	0.718032888445657\\
19	0.771686735089253\\
20	0.8247729642405\\
};
\addlegendentry{\footnotesize w/o Doppler shift}

\addplot [color=red, dashed, line width=1.0pt, mark=x, mark options={solid, red}]
  table[row sep=crcr]{%
0	14.1421356237309\\
1	14.1421356237309\\
2	17.4464453864661\\
3	7.57430369422146\\
4	2.79517230725079\\
5	1.03510512786881\\
6	0.345241202594345\\
7	0.275429567059698\\
8	0.287892620392698\\
9	0.248390369565978\\
10	0.254248020093312\\
11	0.259185182855801\\
12	0.284770445566436\\
13	0.296453497798654\\
14	0.314340544157455\\
15	0.344962072141179\\
16	0.363359400272258\\
17	0.412518949184862\\
18	0.422711124593967\\
19	0.422635748738772\\
20	0.449014808540448\\
};
\addlegendentry{\footnotesize with Doppler shift}
\end{axis}

\node[rotate=0,fill=white] (BOC6) at (3.35cm,-.65cm){\small time $k$};
\node[rotate=90] at (-8mm,15mm){\small RP GOSPA [m]};
\end{tikzpicture}%
}
    \subcaption{}
    \label{Fig:Doppler_Mapping}
\end{subfigure}
\caption{
SLAM performance using the proposed RIS phase profile with and without Doppler shift: (a) \Ac{mae} of the speed estimates;
and 
(b) GOSPA distances of the RP using the proposed RIS phase profile for
$T=20$ transmission intervals.
}
\vspace{-6mm}
\label{Fig:Dopplers}
\par
\end{figure}
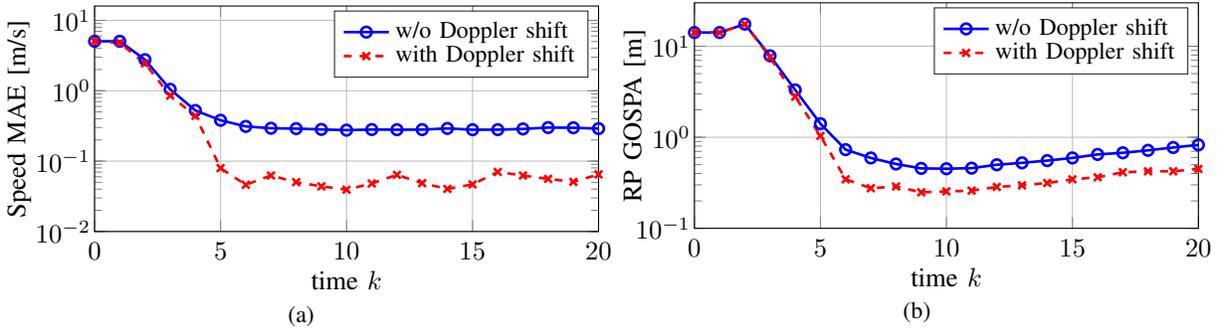

\subsubsection{UE State Error Bounds}
    The \ac{peb}, \ac{heb}, and \ac{seb} are henceforth evaluated over prior uncertainties, different \ac{ue} positions, and numbers of transmissions.
    
\paragraph{Impact of Prior Uncertainties}
    The \ac{peb} performance versus the \ac{ue} position prior uncertainties 
    is plotted in Fig.~\ref{Fig:Ini_Perform}, using the \acp{crlb} on the \ac{ue} state, as derived in Appendix~\ref{sec:FIM}. We have omitted the \ac{heb} and \ac{seb}, which tended to be similar to the \ac{peb}. It is obvious from the figure that, as the prior uncertainty increases, the \acp{peb} for the proposed uniform phases and the directional ones increase, while the random phases do not depend on the uncertainty. In addition, the gain of the proposed method gradually increases. This happens because when the UE uncertainty becomes large, the probability that the \ac{ue} gets illuminated using \ac{ris}-controlled signals, and thus provided with sufficient received signal power, decreases. However, the \ac{peb} also increases when the \ac{ue} uncertainty becomes smaller than $1$ m, due to the lack of direction diversity. It can be also seen in the figure that the \ac{peb} offered by the directional phases is larger than that when random phases are used, when the \ac{ue} uncertainty is larger than $8$ m.

\paragraph{Impact of the Number of Transmissions}

    The \ac{peb} performance as a function of the number of transmissions $T$ is depicted in Fig.~\ref{Fig:OFDM_Perform} for the case where the \ac{ue} uncertainty is $10$~m    
    ($\bar{\mathbf{P}}_0=\mathrm{diag}(50,50,0)$).
    We have omitted the \ac{heb} and \ac{seb} performances, which were shown to have the same trend with \ac{peb}.
    It can be seen that, as $T$ increases during the coherence time, the \acp{peb}  for all considered methods decreases. It is shown that the gain of the proposed method gradually decreases compared to the benchmarks. In all transmissions, the \acp{eb} of the proposed phase profiles are better than the random ones, and similar to, or slightly better than, the directional ones.
    This happens because, as $T$ increases, the angular uncertainty, that is not illuminated, gradually decreases.
    When $T=20$, the \acp{eb} of the directional phases are worse than that with random phases, but this trend gets reversed for $T>40$.
     

\paragraph{Impact of the UE Positions}
    We visualize the \ac{peb} and \ac{seb} performances for different \ac{ue} positions in the $xy$ plane in Fig.~\ref{Fig:visual_UE}.
    We have omitted the \ac{heb}, which was found to be similar to the \ac{peb}.
    We can see that the \acp{eb} become usually larger when the \ac{ue} position is further away from the \ac{ris}.
    The \ac{peb} and \ac{seb} at about $x=40$~m are relatively larger than other positions, because the \ac{ris} elements lie in the $yz$ plane heading to $\mathbf{n}_\mathrm{RIS} = [1,0,0]^\top$.
    The \ac{seb} at about $y=0$~m is extremely larger than the other positions, since in this case, the UE velocity vector is orthogonal to the radial direction from the RIS to the UE, meaning that Doppler measurements provide no information on the longitudinal speed. 
    

\vspace{-1mm}
\subsection{SLAM Performance}
\vspace{-1mm}
    We next investigate the \ac{slam} performance over time for the proposed modified \ac{mpmb}-\ac{slam} filter presented in Section~\ref{sec:MPMB}, using the proposed \ac{ris} phase profile design in Section~\ref{sec:ProposedRISProf}.
    
\subsubsection{The Role of the Phase Profile}
The \acp{mae} of the \ac{ue} position, heading, and speed estimates over the discrete time index $k$ are illustrated in Fig.~\ref{Fig:time_5DLoc}. As shown, the \acp{mae} of the estimates increase after $k=8$, while the gain of the proposed uniform phase profile decreases when compared with the benchmarked ones. This happens because the \ac{ue} moves, getting close to the \ac{ris} until $k=8$, and far away from it after this time instant. It is also depicted that the localization performance with the directional phase profiles (i.e., the conventional and proposed ones) is always better than that with random phases. 
We omit the \ac{seb}, which was found to be similar to the \ac{heb}.
The averaged mapping \ac{gospa} distances for the \ac{rp} and \acp{sp} are depicted in Fig.~\ref{Fig:Mapping_time}. It is confirmed that the proposed framework with the \ac{nris} paths can well estimate the landmark positions and classify the landmark types, while being robust to the missed detections and false alarms.  Due to the correlation between the \ac{ue} and landmarks in the geometric parameters, the mapping performance gain tends to be similar to the \ac{peb}, even though the \ac{ris} phase profile design never affects the \ac{fim} of the \ac{nris} paths, which have been set to be uncontrollable.


\subsubsection{Impact of the Doppler Shift}
We next present results of Doppler-assisted \ac{slam}, where the Doppler shift is used as an additional measurement in the update step of the proposed \ac{slam} filter. To showcase the efficiency of the Doppler consideration in \ac{slam}, the performance without the Doppler measurement is used as benchmark, and thus, evaluated.

The \ac{mae} of the speed estimates is illustrated in Fig.~\ref{Fig:Doppler_5DLoc}, while the \acp{mae} of the heading and position estimates were similar and omitted.
It is depicted that, thanks to the Doppler shift measurement, the speed estimate is improving. Consequently, it is turned into a benefit in both position and heading estimates. 
The \ac{gospa} distances of the \ac{rp} are demonstrated in Fig.~\ref{Fig:Doppler_Mapping}; the \ac{gospa} distances of \acp{sp} were similar, and thus, omitted. As before, it is shown that there is a performance improvement in the mapping with the Doppler shift estimate. Due to the large error of the \ac{ue} speed estimates without the Doppler estimate, the potential landmarks (that have not been detected or have been detected with low existence probabilities) are located at errant positions. This causes frequent errors in data association, such that \acp{rp} measurements are associated with \acp{sp}, leading to false alarms in the \ac{rp} mapping. Inevitably, there are missed detections in the \ac{sp} mapping; the potentially true \acp{sp} cannot be associated with the \ac{sp} measurements, due to the latter data association error.
    
\vspace{-1mm}
\section{Conclusions}\label{sec:Conclusions}
\vspace{-1mm}
In this paper, we presented a novel \ac{ris}-enabled radio \ac{slam} framework, considering a multi-antenna \ac{ue}, a passive reflective \ac{ris}, a large surface, scattering points, and no \acp{bs}.
An efficient method for \ac{ris} phase profile design, which can uniformly illuminate any desired angular sector where the \ac{ue} is probabilistically located, was designed. We have also derived the \ac{crlb} of the channel parameters and the \ac{ue} state, using only the \ac{ris}-induced path. According to the proposed online framework, the \ac{ris} phase design and the \ac{slam} filter are recursively designed over time. We evaluated the proposed \ac{slam} approach in the considered propagation environment with respect to different \ac{ue} uncertainties and positions, number of transmissions, as well as with/without Doppler shift. Our results demonstrated that the proposed \ac{ris} phase profile outperforms directional and random reflective beams, in terms of the \ac{ris}-based power gain for wireless communications, as well as in terms of localization and mapping accuracy for \ac{slam}. It was also showcased that the proposed \ac{ris} profiles and modified \ac{mpmb}-\ac{slam} filter including Doppler shift estimation yield improved estimation performance. 

For future work, we intend to study sensing and \ac{slam} aided by \acp{ris} for integrated sensing, localization, and communications.
The availability of prior SLAM information will be exploited for beamforming and combining designs that can recursively improve the sensing and localization performance over time. In addition, we will study near-field propagation environments aided by extremely large \ac{ris} in THz frequency band. We will finally investigate practical hardware impairments in RIS-aided sensing, \ac{slam}, and communications.

\appendices
\vspace{-1mm}
\section{Geometry Relations of the Channel Parameters}\label{App:ChannelParameters}
\vspace{-1mm}
In this appendix, we define the geometry relations of the parameters for each channel path $l$. For notation simplicity, we omit the time index $k$ and the transmission index $t$. We denote the location of the incident point for $l$-th path by $\mathbf{x}_l$, which is defined as $\mathbf{x}_l= \mathbf{x}_\mathrm{RIS}$ for $l=0$ and $\mathbf{x}_l= \mathbf{x}_\mathrm{IP}^{(m)}$ for $l \neq 0$, where $\mathbf{x}_\mathrm{IP}^{(\mathrm{SP})}=\mathbf{x}^{(\mathrm{SP})}$ and
    $\mathbf{x}_{\mathrm{IP}}^{(\mathrm{RP})}=\mathbf{x}^{(\mathrm{RP})}$.
    
    The Doppler shift is defined as\footnote{The Doppler shift is generally given by $f^{\mathrm{D}} = f_c\Delta v/c$ with $\Delta v$ denoting the relative velocity of the receiver to the transmitter. For example, let us suppose that the \ac{ue} moves to a landmark at its speed $v$. Then, the \ac{ue} radial velocity and the landmark radial velocity viewed by the \ac{ue}, are respectively $v$ and $-v$, and it holds $\Delta v = 2v$.} $f_{l}^{\mathrm{D}} = 2f_c v_{l} /c$, where $v_{l}$ denotes the relative radial velocity to the landmark corresponding to\footnote{We define the radial velocity $v_{l}$ as the scalar projection of the velocity vector $\mathbf{v}_\mathrm{UE}$ onto $\mathbf{e}_{l}$, i.e., $v_{l} = \mathbf{v}_\mathrm{UE}^\top \mathbf{e}_{l}$, where $\mathbf{v}_\mathrm{UE} \triangleq [v_\mathrm{UE} \cos \alpha_\mathrm{UE}, v_\mathrm{UE} \sin \alpha_\mathrm{UE}, 0]^\top$ denotes the \ac{ue} velocity vector in the Cartesian coordinate system and $\mathbf{e}_{l} \triangleq (\mathbf{x}_{l} - \mathbf{x}_{\mathrm{UE}}) / \lVert \mathbf{x}_{l} - \mathbf{x}_{\mathrm{UE}} \rVert$ is the unit-vector from the \ac{ue} location $\mathbf{x}_{\mathrm{UE}}$ to the object $\mathbf{x}_{l}$. Note that $\mathbf{v}_\mathrm{UE}$ is represented by the \ac{ue} heading and speed in the polar coordinate system.} path $l$.
    The channel parameters are defined as:
\begin{align}
    v_{l} 
    &=
    \begin{cases}
        \mathbf{v}_\mathrm{UE}^\top \mathbf{x}_\mathrm{RU}/d_\mathrm{RU}, & l = 0\\
        \mathbf{v}_\mathrm{UE}^\top \mathbf{x}_{\mathrm{IU}}^{(m)}/d_{\mathrm{IU}}^{(m)}, & l \neq 0
    \end{cases},
    &
    {\phi}^\mathrm{az} & = 
        \operatorname{atan2}(\breve{y}_{\mathrm{UR}},\breve{x}_{\mathrm{UR}}),
        &
    {\phi}^\mathrm{el}  &=         
        \operatorname{acos}(\breve{z}_{\mathrm{UR}} / d_\mathrm{UR}),
    \notag
    \\
    {\theta}_{l}^\mathrm{az} & = 
        \begin{cases}
            \operatorname{atan2}(\breve{y}_{\mathrm{RU}},\breve{x}_{\mathrm{RU}}), 
            & l=0\\
            \operatorname{atan2}(\breve{y}_{\mathrm{IU}},\breve{x}_{\mathrm{IU}}), & l\neq 0
        \end{cases},
    &
    {\theta}_{l}^\mathrm{el}  &= 
        \begin{cases}
            \operatorname{acos}({\breve{z}_{\mathrm{RU}}} / 
            d_\mathrm{RU}),
            & l=0\\
            \operatorname{acos}({\breve{z}_{\mathrm{IU}}} / 
            d_\mathrm{IU}), & l\neq 0
        \end{cases},
        &
    \tau_{l} &= 
        \begin{cases}
            2 d_\mathrm{UR},
            & l=0\\
            2 d_\mathrm{IU}, & l\neq 0
        \end{cases},
        \notag
\end{align}
 where $\breve{\mathbf{x}}_{\mathrm{RU}} = [\breve{x}_{\mathrm{RU}},\breve{y}_{\mathrm{RU}},\breve{z}_{\mathrm{RU}}]^\top =\mathbf{O}_\mathrm{UE}^\top  \mathbf{x}_{\mathrm{RU}}$,
    $\breve{\mathbf{x}}_{\mathrm{UR}} = [\breve{x}_{\mathrm{UR}},\breve{y}_{\mathrm{UR}},\breve{z}_{\mathrm{UR}}]^\top =\mathbf{O}_\mathrm{RIS}^\top  \mathbf{x}_{\mathrm{UR}}$,
    $\mathbf{x}_{\mathrm{RU}} =\mathbf{x}_{\mathrm{RIS}} - \mathbf{x}_{\mathrm{UE}}$,
    $\mathbf{x}_{\mathrm{UR}} =\mathbf{x}_{\mathrm{UE}} - \mathbf{x}_{\mathrm{RIS}}$,
    $d_{\mathrm{RU}}=d_{\mathrm{UR}} =\lVert \mathbf{x}_{\mathrm{UR}} \rVert 
    = \lVert \breve{\mathbf{x}}_{\mathrm{UR}} \rVert 
    =\lVert \mathbf{x}_{\mathrm{RU}} \rVert 
    = \lVert \breve{\mathbf{x}}_{\mathrm{RU}} \rVert$,
    $\mathbf{x}_{\mathrm{IU}}^{(m)} = \mathbf{x}_\mathrm{IP}^{(m)} - \mathbf{x}_\mathrm{UE}$, and $d_{\mathrm{IU}}^{(m)} = \lVert \mathbf{x}_{\mathrm{IU}}^{(m)} \rVert$.
    In the latter definitions, $\mathbf{O}_\mathrm{RIS} = [\mathbf{o}_{\mathrm{RIS},1}, \mathbf{o}_{\mathrm{RIS},2}, \mathbf{o}_{\mathrm{RIS},3}]$ and $\mathbf{O}_\mathrm{UE} = [\mathbf{o}_{\mathrm{UE},1}, \mathbf{o}_{\mathrm{UE},2}, \mathbf{o}_{\mathrm{UE},3}]$. By denoting the real and imaginary parts as $\beta_{l}^\mathrm{Re}$ and $\beta_{l}^\mathrm{Im}$, respectively, the complex channel gain of each $l$-th path is defined as follows: $\beta_{l} = \beta_{l}^\mathrm{Re} + j \beta_{l}^\mathrm{Im}$.

\section{Fisher Information Matrix of the Channel Parameters and UE State}
\label{sec:FIM}
We describe the \ac{fim} derivation, providing the lower bounds on the estimation error of any unbiased estimator. We next present the \ac{fim} of the channel parameters for $l=0,\dots,L$, given the received signal of~\eqref{eq:signal_model}. We also derive the \ac{fim} of the \ac{ue} state from the \ac{fim} of the channel parameters corresponding to the \ac{ris} path, leveraging the variable transformation from the equivalent \ac{fim} (EFIM) of the channel parameters for the \ac{ris} path. This holds due to the considered time-domain \ac{ris} phase profiles, which enable separation of the \ac{ris} path ($l=0$) from the other \ac{nris} paths (i.e., $\forall$$l\neq 0$) without any data association process, as discussed in Section~\ref{sec:OrthogonalPhase}. For notational simplicity, we omit the time index $k$ in the following.
    

\vspace{-1mm}
\subsection{FIM of the Channel Parameters}
\label{sec:ChParaFIM}
\vspace{-1mm}
    We introduce the channel parameter vector notation $\bar{\bm{\eta}} \triangleq 
    [\tilde{{\bm{\eta}}}^\top,
    (\bm{\beta}^\mathrm{Re})^\top, (\bm{\beta}^\mathrm{Im})^\top]^\top$,
    where
    $\tilde{{\bm{\eta}}} = [\bm{\phi}_{0}^\top,\bm{\tau}^\top,
    \mathbf{v}^\top, \bm{\theta}^\top]^\top$ with
    $\bm{\tau} = [\tau_{0},\dots,\tau_{L}]^\top$, 
    $\mathbf{v} = [v_{0},\dots,v_{L}]^\top$, and
    $\bm{\theta} = [\bm{\theta}_{0}^\top,\dots,\bm{\theta}_{L}^\top]^\top$, 
    $\bm{\beta}^\mathrm{Re} =[\beta_{0}^\mathrm{Re},\dots,\beta_{L}^\mathrm{Re}]^\top$, and
    $\bm{\beta}^\mathrm{Im} = [\beta_{0}^\mathrm{Im},\dots,\beta_{L}^\mathrm{Im}]^\top$.
We also represent the noiseless part of the received signal in~\eqref{eq:signal_model} by $\bar{y}_{t,s}$ and we define $\bar{\mathbf{y}}_{t}\triangleq[\bar{y}_{t,1},\dots,\bar{y}_{t,N_\mathrm{SC}}]^\top$ including all $N_\mathrm{SC}$ subcarriers. Using the latter definitions, the \ac{fim} of the channel parameters can be computed as follows~\cite{Kay1993}:
\begin{align}
    \mathbf{J}(\bar{\bm{\eta}}) = \frac{2E_s}{N_0} \sum_{t=0}^T \mathrm{Re}
    \left\{ 
    \left( \frac{\partial \bar{\mathbf{y}}_{t}}{\partial {\bar{\bm{\eta}}_k}} \right)^\mathsf{H} 
     \frac{\partial \bar{\mathbf{y}}_{t}}{\partial {\bar{\bm{\eta}}}} \right\},
\end{align}
    where $\mathrm{Re}\{\cdot\}$ is the function for extracting the real part of a complex number, ${\partial}/{\partial \bm{\lambda}}$ indicates the partial derivative with respect to the channel parameters (such as \acp{toa}, radial velocities, \acp{aod}, and path gains), and $\mathbf{J}(\bar{\bm{\eta}}) \in \mathbb{R}^{(2+6(L+1))\times (2+6(L+1))}$ is of the following form:
\begin{align}
    \mathbf{J}(\bar{\bm{\eta}}) = 
    \left[
    \begin{array}{c : c}
        \left<\mathbf{J}(\bar{\bm{\eta}})\right>_{11} & \left<\mathbf{J}(\bar{\bm{\eta}})\right>_{12} \\ \hdashline
        \left<\mathbf{J}(\bar{\bm{\eta}})\right>_{12}^\top & \left<\mathbf{J}(\bar{\bm{\eta}})\right>_{22}
    \end{array}
    \right],
\end{align}
\begin{align}
    \left<\mathbf{J}(\bar{\bm{\eta}})\right>_{11} \in \mathbb{R}^{\left( 2+4(L+1) \right) \times 
    \left( 2+4(L+1)\right)}, 
    && \left<\mathbf{J}(\bar{\bm{\eta}})\right>_{12} \in \mathbb{R}^{ \left(2+4(L+1) \right) \times 
    2(L+1) },
    && \left<\mathbf{J}(\bar{\bm{\eta}})\right>_{22} \in \mathbb{R}^{2(L+1)\times
    2(L+1)}. \notag
\end{align}
    We also define $\tilde{{\bm{\eta}}}$ for the channel parameter vector $\bar{\bm{\eta}}$ when the channel coefficient $\bm{\beta}^\mathrm{Re}$ and $\bm{\beta}^\mathrm{Im}$ are excluded.
    To remove the effect of these nuisance parameters, we compute an EFIM as follows: $\mathbf{J}(\tilde{{\bm{\eta}}}) = \left<\mathbf{J}(\bar{\bm{\eta}})\right>_{11} - \left<\mathbf{J}(\bar{\bm{\eta}})\right>_{12} \left<\mathbf{J}(\bar{\bm{\eta}})\right>_{22}^{-1} \left<\mathbf{J}(\bar{\bm{\eta}})\right>_{12}^\top$.
To obtain the \ac{fim} of the channel parameters $\bm{\eta} = [\bm{\eta}_{0}^\top,\dots,\bm{\eta}_{L}^\top]^\top$ with
    $\bm{\eta}_{0} = [\bm{\phi}^\top,\tau_{0}, v_{0}, \bm{\theta}_{0}^\top]^\top$
    and $\bm{\eta}_{l} = [\tau_{l},v_{l}, \bm{\theta}_{l}^\top]^\top$ for $l=1,\dots,L$,
    we re-order the rows and columns of $\mathbf{J}(\tilde{{\bm{\eta}}})\in \mathbb{R}^{(2+4(L+1))\times (2+4(L+1))}$ and obtain $\mathbf{J}(\bm{\eta})$.
    The measurements noise covariance of~\eqref{eq:UEMea_model} and \eqref{eq:ChParaMea_model} are then determined by extracting the sub-matrix $\mathbf{R}^j=\mathbf{J}^{-1}(\bm{\eta}_j)$ (corresponding to the channel parameters $\bm{\eta}_{j}$) for the measurement indices $j=0,\dots,J$.



\vspace{-1mm}
\subsection{FIM of the UE State Using the RIS Path}
\vspace{-1mm}

    Given the FIM expression $\mathbf{J}(\bm{\eta})$, we can compute an EFIM corresponding to $\bm{\eta}_{0}$ as follows:
\begin{align}
    \mathbf{J}(\bm{\eta}_{0}) = \left<\mathbf{J}({\bm{\eta}})\right>_{11} - \left<\mathbf{J}({\bm{\eta}})\right>_{12} \left<\mathbf{J}({\bm{\eta}})\right>_{22}^{-1} \left<\mathbf{J}({\bm{\eta}})\right>_{12}^\top,
\end{align}
    where $\left<\mathbf{J}({\bm{\eta}})\right>_{11} \in \mathbb{R}^{6 \times 6}$, $\left<\mathbf{J}({\bm{\eta}})\right>_{12} \in \mathbb{R}^{6 \times 4L}$, and $\left<\mathbf{J}({\bm{\eta}})\right>_{22} \in \mathbb{R}^{4L \times 4L}$.
    Finally, we calculate the \ac{fim} referring to the \ac{ue} state $\mathbf{s}$ by $\mathbf{J}(\mathbf{s}) = \mathbf{T}^\top \mathbf{J}(\bm{\eta}_{0}) \mathbf{T}$,
    where $\mathbf{T} \in \mathbb{R}^{6 \times 5}$ denotes the Jacobian transformation matrix, which is defined as $\mathbf{T} = {\partial \bm{\eta}_{0}}/{\partial \mathbf{s}}$;
    the derivatives are detailed in the following Appendix~\ref{App:Derivatives}.
    The \ac{peb}, \ac{heb}, and \ac{seb} are finally given by
\begin{align}
    \mathrm{PEB} = \sqrt{\mathrm{tr}(\left[\mathbf{J}^{-1}(\mathbf{s})\right]_{1:3,1:3})}, &&
    \mathrm{HEB} = \sqrt{\left[\mathbf{J}^{-1}(\mathbf{s})\right]_{4,4}},&&
    \mathrm{SEB} = \sqrt{\left[\mathbf{J}^{-1}(\mathbf{s})\right]_{5,5}}.
\end{align}
    
\section{The Jacobian Transformation Matrix 
}\label{App:Derivatives}
    The derivatives for the Jacobian transformation matrix $\mathbf{T}$ 
    are derived as follows:
\begin{align}
    \frac{\partial \phi^\mathrm{az}}
    {\partial \mathbf{x}_{\mathrm{UE}}} 
    &= 
    \frac{\breve{x}_{\mathrm{UR}}\mathbf{o}_{\mathrm{RIS},2} 
    -\breve{y}_{\mathrm{UR}}\mathbf{o}_{\mathrm{RIS},1}}
    {\breve{x}_{\mathrm{UR}}^2 + \breve{y}_{\mathrm{UR}}^2},
    &
    \dfrac{\partial \phi^\mathrm{el}}
    {\partial \mathbf{x}_{\mathrm{UE}}} 
    &= 
    \frac{1}
    {\sqrt{\breve{x}_{\mathrm{UR}}^2 + \breve{y}_{\mathrm{UR}}^2}}
    (\frac{\breve{z}_{\mathrm{UR}}}{d_{\mathrm{UR}}^2}\mathbf{x}_{\mathrm{UR}} - \mathbf{o}_{\mathrm{RIS},3}), \notag
    \\
    \frac{\partial \theta_{0}^\mathrm{az}}
    {\partial \mathbf{x}_{\mathrm{UE}}} 
    &=
    \frac{ \breve{y}_{\mathrm{RU}}\mathbf{o}_{\mathrm{UE},1} - \breve{x}_{\mathrm{RU}}\mathbf{o}_{\mathrm{UE},2} }
    {\breve{x}_{\mathrm{RU}}^2 + \breve{y}_{\mathrm{UR}}^2},
    &
    \frac{\partial \theta_{0}^\mathrm{el}}
    {\partial \mathbf{x}_{\mathrm{UE}}}
    &=
    \frac{1}
    {\sqrt{\breve{x}_{\mathrm{RU}}^2 + \breve{y}_{\mathrm{RU}}^2}}
    (\mathbf{o}_{\mathrm{UE},3} - \dfrac{\breve{z}_{\mathrm{RU}}}{d_{\mathrm{RU}}^2}\mathbf{x}_{\mathrm{RU}}),
    \notag
    \\
    \frac{\partial \tau_{0}}
    {\partial \mathbf{x}_{\mathrm{UE}}} 
    &= 
    \frac{\mathbf{x}_{\mathrm{UR}}}{d_{\mathrm{UR}}},
    &
    \frac{\partial v_{0}}
    {\partial \mathbf{x}_{\mathrm{UE}}}
    &=
    \frac{v_{0}}{d_{\mathrm{RU}}} \mathbf{x}_{\mathrm{RU}} - \dfrac{1}{d_{\mathrm{RU}}} \mathbf{v}_{\mathrm{UE}},
    \notag
    \\
    \frac{\partial \theta_{0}^\mathrm{az}}
    {\partial \alpha_{\mathrm{UE}}} 
    &= 
    \frac{-( \breve{x}_{\mathrm{RU}}\mathbf{o}_{\mathrm{UE},1}^\top + \breve{y}_{\mathrm{RU}} \mathbf{o}_{\mathrm{UE},2}^\top)\mathbf{x}_{\mathrm{RU}}}
    {\breve{x}_{\mathrm{RU}}^2 + \breve{y}_{\mathrm{RU}}^2},
    &
    \frac{\partial \theta_{0}^\mathrm{el}}
    {\partial \alpha_{\mathrm{UE}}}
    &= 0,
    \notag
    \\
    \dfrac{\partial v_{0}}{\partial \alpha_{\mathrm{UE}}}
    &=\!\dfrac{-v_{\mathrm{UE}} \sin \alpha_{\mathrm{UE}} x_{\mathrm{RU}} \!+\! v_{\mathrm{UE}} \cos \alpha_{\mathrm{UE}} y_{\mathrm{RU}}}
    {d_{\mathrm{RU}}},
    &
    \dfrac{\partial v_{0}}{\partial v_{\mathrm{UE}}} 
    &=
    \dfrac{1}{d_{\mathrm{RU}}}
    (x_{\mathrm{RU}}\cos \alpha_{\mathrm{UE}}
    + 
    y_{\mathrm{RU}}\sin \alpha_{\mathrm{UE}}).
    \notag
\end{align}

\bibliographystyle{IEEEtran}
\bibliography{refs}

\begin{thebibliography}{10}
\providecommand{\url}[1]{#1}
\csname url@samestyle\endcsname
\providecommand{\newblock}{\relax}
\providecommand{\bibinfo}[2]{#2}
\providecommand{\BIBentrySTDinterwordspacing}{\spaceskip=0pt\relax}
\providecommand{\BIBentryALTinterwordstretchfactor}{4}
\providecommand{\BIBentryALTinterwordspacing}{\spaceskip=\fontdimen2\font plus
\BIBentryALTinterwordstretchfactor\fontdimen3\font minus
  \fontdimen4\font\relax}
\providecommand{\BIBforeignlanguage}[2]{{%
\expandafter\ifx\csname l@#1\endcsname\relax
\typeout{** WARNING: IEEEtran.bst: No hyphenation pattern has been}%
\typeout{** loaded for the language `#1'. Using the pattern for}%
\typeout{** the default language instead.}%
\else
\language=\csname l@#1\endcsname
\fi
#2}}
\providecommand{\BIBdecl}{\relax}
\BIBdecl

\bibitem{Hyowon_TWC2020}
H.~Kim, K.~Granstr\"{o}m, L.~Gao \emph{et~al.}, ``{5G mmWave cooperative
  positioning and mapping using multi-model PHD},'' \emph{IEEE Trans. Wireless
  Commun.}, vol.~19, no.~6, pp. 3782--3795, Mar. 2020.

\bibitem{Yu_EK-PMBM_JSAC2021}
Y.~Ge, O.~Kaltiokallio, H.~Kim \emph{et~al.}, ``A computationally efficient
  {EK-PMBM} filter for bistatic {mmWave} radio {SLAM},'' \emph{IEEE J. Sel.
  Areas Commun.}, vol.~40, no.~7, Jul. 2022.

\bibitem{Hyowon_MPMB_TVT2022}
H.~Kim, K.~Granstrom, L.~Svensson \emph{et~al.}, ``{PMBM-based SLAM filters in
  5G mmWave vehicular networks},'' \emph{IEEE Trans. Veh. Technol.}, vol.~71,
  no.~8, pp. 8646--8661, Aug. 2022.

\bibitem{HenkGlobecom2018}
H.~Wymeersch, N.~Garcia, H.~Kim \emph{et~al.}, ``{5G mmWave downlink vehicular
  positioning},'' in \emph{Proc. IEEE Global Commun. Conf. (GLOBECOM)}, Abu
  Dhabi, UAE, Dec. 2018, pp. 206--212.

\bibitem{HyowonAsilomar2018}
H.~Kim, H.~Wymeersch, N.~Garcia \emph{et~al.}, ``{5G mmWave vehicular
  tracking},'' in \emph{Proc. IEEE 52nd Asilomar Conf. Signals, Syst.,
  Comput.}, Pacific Grove, CA, USA, Oct. 2018, pp. 541--547.

\bibitem{Erik_BPSLAM_TWC2019}
E.~Leitinger, F.~Meyer, F.~Hlawatsch \emph{et~al.}, ``A belief propagation
  algorithm for multipath-based {SLAM},'' \emph{IEEE Trans. Wireless Commun.},
  vol.~18, no.~12, pp. 5613--5629, Sep. 2019.

\bibitem{Rico_BPSLAM_JSTSP2019}
R.~Mendrzik, F.~Meyer, G.~Bauch \emph{et~al.}, ``{Enabling situational
  awareness in millimeter wave massive MIMO systems},'' \emph{IEEE J. Sel.
  Topics Signal Process.}, vol.~13, no.~5, pp. 1196--1211, Aug. 2019.

\bibitem{yassin2018mosaic}
A.~Yassin, Y.~Nasser, A.~Y. Al-Dubai \emph{et~al.}, ``{MOSAIC}: Simultaneous
  localization and environment mapping using {mmWave} without a-priori
  knowledge,'' \emph{IEEE Access}, vol.~6, pp. 68\,932--68\,947, Nov. 2018.

\bibitem{palacios2018communication}
J.~Palacios, G.~Bielsa, P.~Casaril \emph{et~al.}, ``Communication-driven
  localization and mapping for millimeter wave networks,'' in \emph{Proc. IEEE
  Int. Conf. Comput. Commun. (INFOCOM)}, Honolulu, HI, USA, Apr. 2018, pp.
  2402--2410.

\bibitem{Durrant2006SLAM1}
H.~Durrant-Whyte and T.~Bailey, ``Simultaneous localization and mapping: {P}art
  {I},'' \emph{IEEE Robot. Autom. Mag.}, vol.~13, no.~2, pp. 99--110, Jun.
  2006.

\bibitem{Durrant2006SLAM2}
T.~Bailey and H.~Durrant-Whyte, ``Simultaneous localization and mapping
  {(SLAM)}: {P}art {II},'' \emph{IEEE Robot. Autom. Mag.}, vol.~13, no.~3, pp.
  108--117, Sep. 2006.

\bibitem{Taranto_LocAC_SPM2014}
R.~D. Taranto, S.~Muppirisetty, R.~Raulefs \emph{et~al.}, ``Location-aware
  communications for {5G} networks: How location information can improve
  scalability, latency, and robustness of {5G},'' \emph{IEEE Signal Process.
  Mag.}, vol.~31, no.~6, pp. 102--112, Nov. 2014.

\bibitem{Koivisto_LocAC_CM2017}
M.~Koivisto, A.~Hakkarainen, M.~Costa \emph{et~al.}, ``High-efficiency device
  positioning and location-aware communications in dense {5G} networks,''
  \emph{IEEE Commun. Mag.}, vol.~55, no.~8, pp. 188--195, Aug. 2017.

\bibitem{An_ISAC_Survey2022}
A.~Liu, Z.~Huang, M.~Li \emph{et~al.}, ``{A survey on fundamental limits of
  integrated sensing and communication},'' \emph{IEEE Commun. Surv. Tutor.},
  vol.~24, no.~2, pp. 994--1034, 2nd Quart. 2022.

\bibitem{Fan_ISAC_JSAC2022}
F.~Liu, Y.~Cui, C.~Masouros \emph{et~al.}, ``{Integrated sensing and
  communications: Towards dual-functional wireless networks for 6G and
  beyond},'' \emph{IEEE journal on selected areas in communications}, Jun.
  2022.

\bibitem{Sundeep_RISISAC_2022}
S.~P. Chepuri, N.~Shlezinger, F.~Liu \emph{et~al.}, ``Integrated sensing and
  communications with reconfigurable intelligent surfaces,'' \emph{arXiv
  preprint arXiv:2211.01003}, 2022.

\bibitem{TR2022}
G.~C. Alexandropoulos, A.~Mokh, R.~Khayatzadeh \emph{et~al.}, ``Time reversal
  for {6G} wireless communications: {N}ovel experiments, opportunities, and
  challenges,'' \emph{IEEE Veh. Technol. Mag.}, early access, 2022.

\bibitem{huang2019reconfigurable}
C.~Huang, A.~Zappone, G.~C. Alexandropoulos \emph{et~al.}, ``Reconfigurable
  intelligent surfaces for energy efficiency in wireless communication,''
  \emph{IEEE Trans. Wireless Commun.}, vol.~18, no.~8, pp. 4157--4170, Aug.
  2019.

\bibitem{Tsinghua_RIS_Tutorial}
M.~Jian, G.~C. Alexandropoulos, E.~Basar \emph{et~al.}, ``Reconfigurable
  intelligent surfaces for wireless communications: {O}verview of hardware
  designs, channel models, and estimation techniques,'' \emph{Intell. Converged
  Netw.}, vol.~3, no.~1, pp. 1--32, Mar. 2022.

\bibitem{Emil_RIS_SPM2022}
E.~Bj{\"o}rnson, H.~Wymeersch, B.~Matthiesen \emph{et~al.}, ``{Reconfigurable
  intelligent surfaces: A signal processing perspective with wireless
  applications},'' \emph{IEEE Signal Process. Mag.}, vol.~39, no.~2, pp.
  135--158, Mar. 2022.

\bibitem{RISE6G_COMMAG}
E.~Calvanese~Strinati, G.~C. Alexandropoulos, H.~Wymeersch \emph{et~al.},
  ``Reconfigurable, intelligent, and sustainable wireless environments for {6G}
  smart connectivity,'' \emph{IEEE Commun. Mag.}, vol.~59, no.~10, pp. 99--105,
  Oct. 2021.

\bibitem{Henk_RISSLAM_VTM2020}
H.~Wymeersch, J.~He, B.~Denis \emph{et~al.}, ``{Radio localization and mapping
  with reconfigurable intelligent surfaces: Challenges, opportunities, and
  research directions},'' \emph{IEEE Veh. Technol. Mag.}, vol.~15, no.~4, pp.
  52--61, 2020.

\bibitem{Zhang_RISSL_Proc2022}
H.~Zhang, B.~Di, K.~Bian \emph{et~al.}, ``Toward ubiquitous sensing and
  localization with reconfigurable intelligent surfaces,'' \emph{Proc. IEEE},
  vol. 110, no.~9, pp. 1401--1422, Sept. 2022.

\bibitem{Keykhosravi2022infeasible}
K.~Keykhosravi, B.~Denis, G.~C. Alexandropoulos \emph{et~al.}, ``Leveraging
  {RIS}-enabled smart signal propagation for solving infeasible localization
  problems,'' \emph{arXiv preprint arXiv:2204.11538}, 2022.

\bibitem{Kamran_RISMobility_JSTSP2022}
K.~Keykhosravi, M.~F. Keskin, G.~Seco-Granados \emph{et~al.}, ``{RIS-Enabled
  SISO} localization under user mobility and spatial-wideband effects,''
  \emph{IEEE J. Sel. Topics Signal Process.}, vol.~16, no.~5, pp. 1125--1140,
  Aug. 2022.

\bibitem{Ahmed_RISLoc_TSP2021}
A.~Elzanaty, A.~Guerra, F.~Guidi \emph{et~al.}, ``Reconfigurable intelligent
  surfaces for localization: Position and orientation error bounds,''
  \emph{IEEE Trans. Signal Process.}, vol.~69, pp. 5386--5402, Aug. 2021.

\bibitem{chen2022tutorial}
H.~Chen, H.~Sarieddeen, T.~Ballal \emph{et~al.}, ``A tutorial on terahertz-band
  localization for {6G} communication systems,'' \emph{IEEE Commun. Surv.
  Tutor.}, 3rd Quart. 2022.

\bibitem{Davide_RISLoc_TWC2021}
D.~Dardari, N.~Decarli, A.~Guerra \emph{et~al.}, ``{LOS/NLOS} near-field
  localization with a large reconfigurable intelligent surface,'' \emph{IEEE
  Trans, Wireless Commun.}, vol.~21, no.~6, pp. 4282--4294, Jun. 2022.

\bibitem{Boyu_RISLoc_JSTSP2022}
B.~Teng, X.~Yuan, R.~Wang \emph{et~al.}, ``Bayesian user localization and
  tracking for reconfigurable intelligent surface aided {MIMO} systems,''
  \emph{IEEE J. Sel. Topics Signal Process.}, vol.~16, no.~5, pp. 1040--1054,
  2022.

\bibitem{Kamran_RISloc_ICC2022}
K.~Keykhosravi, G.~Seco-Granados, G.~C. Alexandropoulos \emph{et~al.},
  ``{RIS-enabled self-localization: Leveraging controllable reflections with
  zero access points},'' in \emph{Proc. IEEE Int. Conf. Commun. Workshops
  (ICC)}, Seoul, South Korea, May 2022.

\bibitem{Aubry_RIS-Sensing_2022}
A.~Aubry, A.~De~Maio, and M.~Rosamilia, ``{RIS-aided radar sensing in NLOS
  environment},'' in \emph{Proc. IEEE 8th International Workshop on Metrology
  for AeroSpace (MetroAeroSpace)}, Jun. 2021, pp. 277--282.

\bibitem{Buzzi_RIS-Sensing_ICASSP2022}
S.~Buzzi, E.~Grossi, M.~Lops \emph{et~al.}, ``{RIS-aided monostatic MIMO radar
  with co-located antennas},'' in \emph{IEEE Int. Conf. Acoust., Speech, Signal
  Process. (ICASSP)}, May 2022, pp. 4998--5002.

\bibitem{Reza_SensingRIS_2022}
R.~Ghazalian, K.~Keikhosravi, H.~Chen \emph{et~al.}, ``Bi-static sensing for
  near-field {RIS} localization,'' \emph{arXiv preprint arXiv:2206.13915},
  2022.

\bibitem{Yang_MetaSLAM_TWC2022}
Z.~Yang, H.~Zhang, H.~Zhang \emph{et~al.}, ``{MetaSLAM}: Wireless simultaneous
  localization and mapping using reconfigurable intelligent surfaces,''
  \emph{IEEE Trans. Wireless Commun.}, 2022.

\bibitem{He_RIS-Sensing_JSAC20222}
Y.~He, Y.~Cai, H.~Mao \emph{et~al.}, ``{RIS}-assisted communication radar
  coexistence: Joint beamforming design and analysis,'' \emph{IEEE J. Sel.
  Areas Commun.}, Jul. 2022.

\bibitem{loc_awareness_JSAC_2022}
Z.~Wang, Z.~Liu, Y.~Shen \emph{et~al.}, ``Location awareness in beyond {5G}
  networks via reconfigurable intelligent surfaces,'' \emph{IEEE J. Sel. Areas
  Commun.}, vol.~40, no.~7, pp. 2011--2025, Jul. 2022.

\bibitem{RIS_beam_training_TWC_2021}
W.~Wang and W.~Zhang, ``Joint beam training and positioning for intelligent
  reflecting surfaces assisted millimeter wave communications,'' \emph{IEEE
  Transactions on Wireless Communications}, vol.~20, no.~10, pp. 6282--6297,
  Oct. 2021.

\bibitem{RIS_Loc_JSTSP_2022}
A.~Fascista, M.~F. Keskin, A.~Coluccia \emph{et~al.}, ``{RIS}-aided joint
  localization and synchronization with a single-antenna receiver: Beamforming
  design and low-complexity estimation,'' \emph{IEEE J. Sel. Topics Signal
  Process.}, vol.~16, no.~5, pp. 1141--1156, Aug. 2022.

\bibitem{Mustafa_ArbitraryBeam_6G2022}
M.~Rahal, B.~Denis, K.~Keykhosravi \emph{et~al.}, ``Arbitrary beam pattern
  approximation via riss with measured element responses,'' in \emph{Joint
  European Conference on Networks and Communications \& 6G Summit (EuCNC/6G
  Summit)}, Jun. 2022, pp. 506--511.

\bibitem{Kaitao_RISSensing_TC2022}
K.~Meng, Q.~Wu, R.~Schober \emph{et~al.}, ``Intelligent reflecting surface
  enabled multi-target sensing,'' \emph{IEEE Trans. Commun.}, 2022.

\bibitem{Vahid_BeamChannelEst_CL2022}
V.~Jamali, G.~C. Alexandropoulos, R.~Schober \emph{et~al.},
  ``Low-to-zero-overhead irs reconfiguration: {D}ecoupling illumination and
  channel estimation,'' \emph{IEEE Commun. Lett.}, vol.~26, no.~4, pp.
  932--936, Apr. 2022.

\bibitem{George_NFBeamManange_2022}
G.~C. Alexandropoulos, V.~Jamali, R.~Schober \emph{et~al.}, ``Near-field
  hierarchical beam management for {RIS}-enabled millimeter wave multi-antenna
  systems,'' in \emph{IEEE Sensor Array and Multichannel Signal Processing
  Workshop}, Trondheim, Norway, Jun. 2022, pp. 1--5.

\bibitem{Gosan_Highspeed_WC2020}
G.~Noh, B.~Hui, and I.~Kim, ``High speed train communications in {5G}: design
  elements to mitigate the impact of very high mobility,'' \emph{IEEE Wireless
  Commun.}, vol.~27, no.~6, pp. 98--106, Dec. 2020.

\bibitem{Han_Doppler_TIT2015}
Y.~Han, Y.~Shen, X.-P. Zhang \emph{et~al.}, ``Performance limits and geometric
  properties of array localization,'' \emph{IEEE Trans. Inf. Theory}, vol.~62,
  no.~2, pp. 1054--1075, Dec. 2015.

\bibitem{kakkavas2019performance}
A.~Kakkavas, M.~H.~C. Garc{\'\i}a, R.~A. Stirling-Gallacher \emph{et~al.},
  ``Performance limits of single-anchor millimeter-wave positioning,''
  \emph{IEEE Trans. Wireless Commun.}, vol.~18, no.~11, pp. 5196--5210, Aug.
  2019.

\bibitem{Hui_Doppler_Globecom2022}
H.~Chen, F.~Jiang, Y.~Ge \emph{et~al.}, ``Doppler-enabled single-antenna
  localization and mapping without synchronization,'' \emph{arXiv preprint
  arXiv:2205.15427}, 2022.

\bibitem{George_MIMO_VTM2022}
G.~C. Alexandropoulos, M.~A. Islam, and B.~Smida, ``Full-duplex massive
  multiple-input, multiple-output architectures: Recent advances, applications,
  and future directions,'' \emph{IEEE Vehicular Technology Magazine}, pp.
  2--10, 2022.

\bibitem{Li_Dynamic_TAES2003}
X.~R. Li and V.~P. Jilkov, ``{Survey of maneuvering target tracking. Part I.
  Dynamic models},'' \emph{IEEE Trans. Aero. Electron. Syst.}, vol.~39, no.~4,
  pp. 1333--1364, Oct. 2003.

\bibitem{mahler_book_2007}
R.~Mahler, \emph{Statistical Multisource-Multitarget Information Fusion}.\hskip
  1em plus 0.5em minus 0.4em\relax Norwood, MA, USA: Artech House, 2007.

\bibitem{Florian_MDESPRIT_TSP2014}
F.~Roemer, M.~Haardt, and G.~Del~Galdo, ``Analytical performance assessment of
  multi-dimensional matrix-and tensor-based {ESPRIT}-type algorithms,''
  \emph{IEEE Trans. Signal Process.}, vol.~62, no.~10, pp. 2611--2625, May
  2014.

\bibitem{Fan_MDESPRIT_2021}
F.~Jiang, F.~Wen, Y.~Ge \emph{et~al.}, ``Beamspace multidimensional {ESPRIT}
  approaches for simultaneous localization and communications,'' \emph{arXiv
  preprint arXiv:2111.07450}, 2021.

\bibitem{wymeersch2020adaptive}
H.~Wymeersch and G.~Seco-Granados, ``Adaptive detection probability for
  mm{W}ave 5{G} {SLAM},'' in \emph{6G Wireless Summit (6G SUMMIT)}, Mar. 2020.

\bibitem{lens_RIS_ICC_2021}
Z.~Abu-Shaban, K.~Keykhosravi, M.~F. Keskin \emph{et~al.}, ``Near-field
  localization with a reconfigurable intelligent surface acting as lens,'' in
  \emph{ICC 2021 - IEEE International Conference on Communications}, 2021, pp.
  1--6.

\bibitem{AntenTheory_Book}
C.~A. Balanis, \emph{Antenna theory: analysis and design}.\hskip 1em plus 0.5em
  minus 0.4em\relax John wiley \& sons, 2015.

\bibitem{slotani1964tolerance}
M.~Slotani, ``Tolerance regions for a multivariate normal population,''
  \emph{Annals of the Institute of Statistical Mathematics}, vol.~16, no.~1,
  pp. 135--153, 1964.

\bibitem{Garcia-Fernandez2018}
{\'{A}}.~F. Garc{\'{i}}a-Fern{\'{a}}ndez, J.~L. Williams, K.~Granstr{\"{o}}m
  \emph{et~al.}, ``Poisson multi-{Bernoulli} mixture filter: {Direct}
  derivation and implementation,'' \emph{IEEE Trans. Aerosp. Electron. Syst.},
  vol.~54, no.~4, pp. 1883--1901, Aug. 2018.

\bibitem{williams2015marginal}
J.~L. Williams, ``{Marginal multi-Bernoulli filters: RFS derivation of MHT,
  JIPDA, and association-based MeMBer},'' \emph{IEEE Trans. Aerosp. Electron.
  Syst.}, vol.~51, no.~3, pp. 1664--1687, Jul. 2015.

\bibitem{murty1968algorithm}
K.~G. Murty, ``An algorithm for ranking all the assignments in order of
  increasing costs,'' \emph{Oper. Res.}, vol.~16, no.~3, pp. 682--687, 1968.

\bibitem{Zohair_5GFIM_TWC2018}
Z.~Abu-Shaban, X.~Zhou, T.~Abhayapala \emph{et~al.}, ``{Error bounds for uplink
  and downlink 3D localization in 5G millimeter wave systems},'' \emph{IEEE
  Trans. Wireless Commun.}, vol.~17, no.~8, pp. 4939--4954, Aug. 2018.

\bibitem{Steven_RISGain_2021}
S.~W. Ellingson, ``Path loss in reconfigurable intelligent surface-enabled
  channels,'' in \emph{Proc. IEEE Int. Symp. on Personal, Indoor and Mob. Radio
  Commun., (PIMRC)}, Sep. 2021, pp. 829--835.

\bibitem{HaykinCKF2009}
I.~Arasaratnam and S.~Haykin, ``Cubature {K}alman filters,'' \emph{IEEE Trans.
  Autom. Control}, vol.~54, no.~6, pp. 1254--1269, Jun. 2009.

\bibitem{RahmathullahGFS:2017}
A.~S. Rahmathullah, A.~F. Garc\'{i}a~Fern\'{a}ndez, and L.~Svensson,
  ``Generalized optimal sub-pattern assignment metric,'' in \emph{Proc. 20th
  Int. Conf. Inf. Fusion (FUSION)}, Xian, China, Jul. 2017, pp. 1--8.

\bibitem{Kay1993}
S.~Kay, \emph{Fundamentals of{ Statistical Signal Processing: Estimation
  Theory}}.\hskip 1em plus 0.5em minus 0.4em\relax Prentice Hall Signal
  Processing Series, 1993.

\end{thebibliography}

\end{document}